\documentclass[journal,final,twoside,romanappendices]{IEEEtran}

\usepackage{tcolorbox}
\usepackage[noadjust]{cite}
\usepackage[cmex10]{amsmath}
\usepackage{amssymb,amsfonts,amscd,amsthm,amsbsy}
\usepackage{graphicx}
\usepackage{stmaryrd} % for \llbracket and \rrbracket
\usepackage[mathscr]{eucal}
\usepackage{mathtools}
\usepackage{bbm}
\usepackage{xcolor}
\usepackage{algorithm}
\usepackage{algpseudocode}
\usepackage{enumerate}
\usepackage{hyperref}
\hypersetup{
  colorlinks   = true,
  urlcolor     = blue,
  linkcolor    = blue, 
  citecolor    = blue 
}

\newtheorem{theorem}{Theorem}[section]
\newtheorem{lemma}{Lemma}[section]
\newtheorem{corollary}{Corollary}[section]
\theoremstyle{remark}
\newtheorem{remark}{Remark}[section]
\newtheorem{example}{Example}[section]
\theoremstyle{definition}
\newtheorem{definition}{Definition}[section]
\usepackage{etoolbox}
\AtEndEnvironment{example}{\null\hfill$\blacksquare$}%
\AtEndEnvironment{remark}{\null\hfill$\blacksquare$}%

\let\emptyset\varnothing
\newcommand{\Fb}{\mathbbmss{F}} % field
\newcommand{\Zb}{\mathbbmss{Z}} % integers
\newcommand{\lset}{\llbracket}  % [[
\newcommand{\rset}{\rrbracket}  % ]]
\newcommand{\Cs}{\mathscr{C}}   % code
\newcommand{\Ds}{\mathscr{B}}   % code dual (this was D earlier)
\newcommand{\Csn}{\mathscr{C}_n}   % code_n
\newcommand{\Dsn}{\mathscr{B}_n} %code_Dual_n (this was Dn earlier)

\newcommand{\Zs}{\mathscr{Z}}   % zero-set
\newcommand{\Zsc}{\bar{\mathscr{Z}}} %% complement of zero-set
\newcommand{\Ss}{\mathscr{S}}   % symmetric group
\newcommand{\Gs}{\mathcal{G}}   % automorphism group
\newcommand{\Os}{\mathscr{O}}   % orbit
\newcommand{\Ps}{\mathscr{P}}   % puncturing operation
\newcommand{\wt}{w_H}   % weight
\newcommand{\supp}{\mathsf{supp}}  % support set
\newcommand{\RM}{\mathrm{RM}}   % Reed-Muller code
\newcommand{\tth}{\text{th}}
\newcommand{\p}{\pmb}           % bold font

\newcommand{\Hs}{\mathscr{H}}
\newcommand{\Tr}{{\rm Tr}}
\newcommand{\dmin}{d_{\min}}    % min distance
\newcommand{\Ks}{\mathscr{K}}   % for aliasing
\newcommand{\Dec}{\texttt{Dec}}
\newcommand{\integerbox}[1]{\lset #1 \rset}

\newenvironment{talign}
 {\let\displaystyle\textstyle\align}
 {\endalign}
\newenvironment{talign*}
 {\let\displaystyle\textstyle\csname align*\endcsname}
 {\endalign}

\title{Berman Codes: A Generalization of\\Reed-Muller Codes that Achieve BEC Capacity}

\author{Lakshmi~Prasad~Natarajan and Prasad Krishnan% <-this % stops a space
\thanks{This paper was presented in parts at the 2022 IEEE International Symposium on Information Theory (ISIT 2022), and the 2022 National Conference on Communications (NCC 2022).}%
\thanks{Lakshmi Prasad Natarajan is with the Department of Electrical Engineering, Indian Institute of Technology Hyderabad, Sangareddy 502\,284, India (email: lakshminatarajan@ee.iith.ac.in).}% <-this % stops a space
\thanks{Prasad Krishnan is with the Signal Processing and Communications Research Center, International Institute of Information Technology, Hyderabad 500\,032, India (email: prasad.krishnan@iiit.ac.in).}% <-this % stops a space
}

\begin{document}

\maketitle

\begin{abstract}
We identify a family of binary codes whose structure is similar to Reed-Muller (RM) codes and which include RM codes as a strict subclass. The codes in this family are denoted as $\Csn(r,m)$, and their duals are denoted as $\Dsn(r,m)$. The length of these codes is $n^m$, where $n \geq 2$, and $r$ is their `order'. When $n=2$, $\Csn(r,m)$ is the RM code of order $r$ and length $2^m$. 
The special case of these codes corresponding to $n$ being an odd prime was studied by Berman (1967) and Blackmore and Norton (2001). Following the terminology introduced by Blackmore and Norton, we refer to $\Dsn(r,m)$ as the \emph{Berman code} and $\Csn(r,m)$ as the \emph{dual Berman} code. We identify these codes using a recursive Plotkin-like construction, and we show that these codes have a rich automorphism group, they are generated by the minimum weight codewords, and that they can be decoded up to half the minimum distance efficiently. 
Using a result of Kumar et al. (2016), we show that these codes achieve the capacity of the binary erasure channel (BEC) under bit-MAP decoding. Furthermore, except double transitivity, they satisfy all the code properties used by Reeves and Pfister to show that RM codes achieve the capacity of binary-input memoryless symmetric channels. 
Finally, when $n$ is odd, we identify a large class of abelian codes that includes $\Dsn(r,m)$ and $\Csn(r,m)$ and which achieves BEC capacity.
\end{abstract}


% \emph{The full version of this paper, including the proofs of all the theorems, lemmas and examples, is available online~\cite{}.}

% % % % %
% include this in arXiv
% \clearpage
% \setcounter{tocdepth}{2} % do not show subsubsections
% \tableofcontents
% \clearpage

% % % % %

\section{Introduction} \label{sec:introduction}

\IEEEPARstart{R}{eed-Muller} (RM) codes \cite{Muller_IRE_54,Reed_IRE_54} form one of the important and well studied code families in coding theory, and have a rich algebraic structure. In \cite{KKMPSU_IT_17}, Kudekar et al. showed that RM codes achieve the capacity of the binary erasure channel (BEC). Furthermore, Reeves and Pfister \cite{ReP_RM_BMS_2021} have recently showed the exciting result that RM codes achieve the capacity of binary-input memoryless symmetric (BMS) channels.

In the present work, we identify a family of binary linear codes (along with its dual family) which includes the RM codes as a strict subclass. These codes are defined using a recursive construction that is similar to the Plotkin construction for RM codes. This family contains a code for each choice of three integer parameters:
% \begin{enumerate}[\((i)\)]
\begin{itemize}
\item integers $n\geq 2$ and $m\geq 1$, which determine the length of the code, and
\item an integer $r$, with $0 \leq r \leq m$, that determines the `order' of the code.
\end{itemize}
% \end{enumerate}

\noindent
We will denote the code with parameters $n,r$ and $m$ by $\Csn(r,m)$. The dual code, $\Csn(r,m)^\perp$, will be denoted by $\Dsn(r,m)$.  The length, dimension and minimum distance of $\Csn(r,m)$ are
\begin{equation} \label{eq:dual_berman_param}
\left[n^m, \, \sum_{w=0}^{r}\binom{m}{w} (n-1)^w, \, n^{m-r} \right].
\end{equation}
The dual code $\Dsn(r,m)$ has code parameters
\begin{equation} \label{eq:berman_param}
\left[n^m, \, \sum_{w=r+1}^{m}\binom{m}{w} (n-1)^w, \, 2^{r+1} \right].
\end{equation}
If we  substitute \mbox{$n=2$} in~\eqref{eq:dual_berman_param} and~\eqref{eq:berman_param} we obtain the parameters of the $r^\tth$ order RM code of length $2^m$, i.e., $\RM(r,m)$, and its dual $\RM(r,m)^\perp=\RM(m-r-1,m)$, respectively. Indeed, we will see that the code $\Cs_2(r,m)$ is identical to $\RM(r,m)$, and by duality $\Ds_2(r,m)=\RM(m-r-1,m)$.

% \IEEEpubidadjcol

\begin{figure}[t!]
\centering
\includegraphics[width=0.5\textwidth]{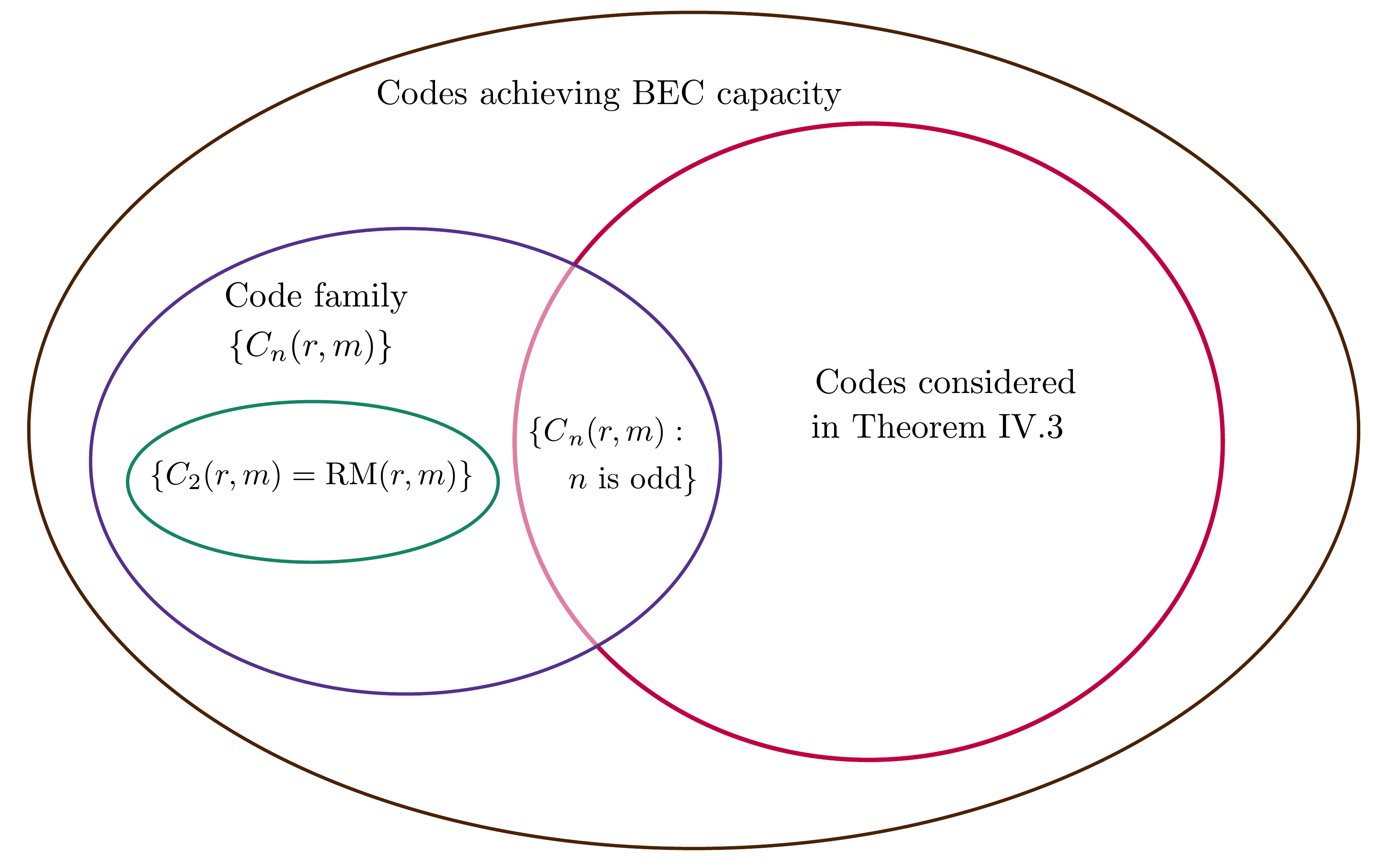}
\caption{The picture shows the relationship between the code family $\{\Csn(r,m)\}$ presented in this work, the codes considered in Theorem \ref{thm:capacity-abelian-codes}, and the Reed-Muller codes. A similar picture holds for the code family $\{\Dsn(r,m)\}$ which are dual to $\Csn(r,m)$. All the codes shown in the picture are capacity achieving for the BEC under bit-MAP decoding.}
\label{fig:inclusionofcodes}
\end{figure}

% % R0 % %
{{\em Prior Work.}
To the best of our knowledge, the subclass of the codes $\Dsn(r,m)$ corresponding to \mbox{$n=p$} being an odd prime was first introduced by Berman~\cite{Ber_Cybernetics_II_67}. Berman gave a construction of $\Ds_p(r,m)$ as an ideal in a group algebra, derived its dimension and minimum distance and showed that as $m \to \infty$ these codes can provide better minimum distance than cyclic codes for the same rate and blocklength.
Later in~\cite{BlN_IT_01} Blackmore and Norton gave a recursive group-algebraic construction of $\Csn(r,m)$, for all $n$, identified its dimension and minimum distance and analyzed its state complexity.

Blackmore and Norton refer to $\Ds_{p}(r,m)$, the code originally designed by Berman in~\cite{Ber_Cybernetics_II_67}, as the \emph{Berman code}.
We will follow this precedence, and we will refer to $\Ds_{n}(r,m)$ as the \emph{Berman code} with parameters $n,r$ and $m$, and $\Cs_{n}(r,m)$ as the \emph{dual Berman code}.

{In another direction of work \cite{MatProdCodes_Blackmore_Norton}, Blackmore and Norton introduced a generalized Plotkin-type construction of new linear codes from existing codes. These codes were termed as \textit{matrix-product} codes. A matrix-product code $\cal C$ of length $\ell N$ is generated  using $s$ component $N$-length codes \mbox{${\cal C}_i:i\in\{1,\hdots,s\}$} and a matrix $\p{A}$ of size $s\times \ell$ as follows: \mbox{${\cal C}=\{[\p c_1,\,\p c_2,\hdots,\,\p c_s ]\p{A} \colon \p c_i\in {\cal C}_i, \forall i\in\{1,\hdots,s\}\}$}, where the codewords $\p c_i$'s are interpreted as column vectors, and the $\ell N$-length codewords of the linear code ${\cal C}$ are interpreted as $N \times \ell$ matrices. The codes considered in this paper, $\Dsn(r,m)$ and $\Csn(r,m)$, do fall under the class of matrix-product codes. However, most of the results in the work \cite{MatProdCodes_Blackmore_Norton} are on matrix-product codes in which the matrix $\p{A}$ satisfies some special rank criterion (called the \emph{non-singular by columns (NSC)} property). The codes  $\Dsn(r,m)$ and $\Csn(r,m)$ presented in this work do not satisfy this property, hence the results from \cite{MatProdCodes_Blackmore_Norton} are not applicable to these codes. In \cite{HeR_JAlgebra_13}, decoding algorithms were presented for matrix-product codes satisfying some special properties. These algorithms decode such codes up to half of their minimum distance. We remark on the relevance of these results from \cite{HeR_JAlgebra_13}, at the appropriate junctures in this paper.
To the best of our knowledge, these seem to be the  previously known results on $\Dsn(r,m)$ and $\Csn(r,m)$.}
%
% A sub-class of these codes, corresponding to the case \mbox{$n = p$} with $p$ being an odd prime, was studied by Berman~\cite{Ber_Cybernetics_II_67}, and Blackmore and Norton~\cite{BlN_IT_01} using a group algebra framework.
% To the best of our knowledge, Berman~\cite{Ber_Cybernetics_II_67} introduced and investigated the code $\Ds_{p}(r,m)$ and showed that its minimum distance $2^{r+1}$ is better than cyclic codes of the same length (for large values of $m$). Later, Blackmore and Norton~\cite{BlN_IT_01} showed that the minimum distance of $\Cs_{p}(r,m)$ is $p^{m-r}$ and analyzed the state complexity of this code.

% % R0 % %
\emph{Current Work.} In this paper we study various basic properties of Berman and dual Berman codes. We first provide a simple recursive construction of these codes based on a generalization of the Plotkin $(\p{u}|\p{u}+\p{v})$ construction, and re-derive the dimension and minimum distance of these codes using only elementary methods.
Our construction allows us to identify generator and parity-check matrices of these codes that are composed of minimum weight codewords, identify a large group of code automorphisms, and provide low complexity recursive decoders that can correct up to half the minimum distance.

Naturally, we are interested in the capacity achievability of the code families $\{\Csn(r,m)\}$ and $\{\Dsn(r,m)\}$. We show that these codes achieve the capacity of the BEC under bit-MAP decoding, using a technique based on code automorphisms developed in \cite{KCP_ISIT16} (which does not require double transitivity, which was used in \cite{KKMPSU_IT_17,ReP_RM_BMS_2021}). We also present a few simulation results that illustrate the performance of these codes, in comparison with RM codes in the BEC.

For the case of $n$ being odd, we use a discrete Fourier transform (DFT) based framework to study the codes $\Csn(r,m)$ and $\Dsn(r,m)$ as abelian codes, i.e., as ideals of a group algebra $\Fb_2[G^m]$, where $G$ is any abelian group with $n$ elements. We also identify a large class of abelian codes, which includes the families \mbox{$\{\Csn(r,m):n\text{ odd}\}$} and \mbox{$\{\Dsn(r,m):n\text{ odd}\}$}, that achieves the BEC capacity. Fig.  \ref{fig:inclusionofcodes} shows the relationship between the code family $\{\Csn(r,m)\}$ presented in this work, the RM codes, the class of capacity achieving abelian codes (obtained in Theorem \ref{thm:capacity-abelian-codes} of this work), and their capacity achieving nature.
% % % % %
}

%%%%

%%%%%
\subsection{Similarity with Reed-Muller Codes}

For \mbox{$n \geq 3$}, the similarity of $\Cs_n(r,m)$ and $\Ds_n(r,m)$ with RM codes runs deep.
The likeness to RM codes was noted by Blackmore and Norton~\cite{BlN_IT_01} for the case $n$ being an odd prime.
In the current work, we define the Berman and dual Berman codes
using a recursive construction similar to the \mbox{$(\p{u}\,|\,\p{u}+\p{v})$} Plotkin construction, and we show that their puncturing properties are similar to RM~codes, they have a rich automorphism group (although they are not doubly transitive), they are generated by their minimum weight codewords, and that they can be decoded up to half the minimum distance efficiently.
Furthermore, except double transitivity, Berman and dual Berman codes satisfy all the properties of RM codes exploited by Reeves and Pfister~\cite{ReP_RM_BMS_2021} to show that RM codes achieve the capacity of BMS channels.

There are also differences with RM codes when \mbox{$n \geq 3$}.
While RM codes are either self-orthogonal or dual-containing, Berman codes have complementary duals when $n$ is odd.
The lack of double transitivity has been mentioned already.
When \mbox{$n \geq 3$}, the minimum distances of Berman and dual Berman codes grow slowly with block length $N$ compared to RM codes.
% % R0 % %
{For any fixed $n \geq 2$, and for increasing $m$, sequences of Berman (respectively dual Berman) codes with limiting rate in $(0,1)$ must have $r/m \to (n-1)/n$  (please see the discussion in Section~\ref{sec:sub:rate-berman} leading to~\eqref{eq:Cnrm_r_by_m_tends_to_mu}).}
% For any choice of $n \geq 2$ and any rate in $(0,1)$, long Berman codes and its duals have $\frac{r}{m} \approx \frac{(n-1)}{n}$ (please see the discussion in Section~\ref{sec:sub:rate-berman} leading to~\eqref{eq:Cnrm_r_by_m_tends_to_mu}).
% % % % % %
Now fixing $n$ and letting $m \to \infty$, using~\eqref{eq:dual_berman_param},~\eqref{eq:berman_param} and the fact \mbox{$r \approx m(n-1)/n$}, we see that the minimum distance $\dmin$ grows with the block length $N$ as
\begin{equation*}
\dmin\left( \Cs_n(r,m) \right) \sim N^{\frac{1}{n}}, ~
\dmin\left( \Ds_n(r,m) \right) \sim N^{\frac{(n-1)}{n \log_2 n}}.
\end{equation*}
In contrast, the minimum distance of RM codes (i.e., the case \mbox{$n=2$}) grows approximately as the square root of the block length.

\subsection{Overview of the Main Results and Organization}

We now summarize the main results and the organization of the present work {(please see Table~\ref{tbl:org})}.

\emph{Section~\ref{sec:berman_and_dual}.}
We define $\Dsn(r,m)$ and $\Csn(r,m)$ recursively using an $n$-fold generalization of the Plotkin construction.
We then identify the basic parameters of these codes, including the dimension and the minimum distance, and identify generator and parity-check matrices that are composed of minimum weight codewords of the respective codes.
% % % % R0 % % % % %
{
We also identify patterned bases for $\Dsn(r,m)$ and $\Csn(r,m)$ which illustrate the combinatorial structure of these codes
(we will use these bases in Section~\ref{sec:capacity-related} to identify some code automorphisms and show that the codes achieve BEC capacity).
}%
% % % % % % % % % % %
% % R0 version % %
{
We show that Berman and dual Berman codes can be efficiently decoded up to half the minimum distance using a technique known for matrix-product codes~\cite{HeR_JAlgebra_13} that is also similar to RM decoding~\cite{ScB_IT_95,DuS_ISIT_2000,ASY_IT_20}.
}
% % % % % % % % %

\emph{Section~\ref{sec:capacity-related}}. We focus on the code properties that Reeves and Pfister~\cite{ReP_RM_BMS_2021} relied on to prove that RM codes achieve BMS channel capacity.
{We first identify some automorphisms of Berman and dual Berman codes (Section~\ref{subsec:automorphism}).}
We show that there are multiple ways to puncture a Berman (or a dual Berman) code to a shorter length Berman (dual Berman) code (Section~\ref{sec:sub:multiple-puncturings}), and that the rate change due to this puncturing is small (Section~\ref{sec:sub:rate-berman}).
Although these codes are not doubly transitive (Section~\ref{sec:sub:double-transitivity}) we are able to show that they achieve BEC capacity (Section~\ref{sec:sub:capacity}) based on a condition on automorphisms identified by Kumar et al.~\cite{KCP_ISIT16}.

\emph{Section~\ref{sec:dft}}.
In this section we exclusively consider the case where $n \geq 3$ is an odd integer.
We use the theory of abelian codes in semi-simple group algebras~\cite{Ber_Cybernetics_II_67,Cam_Math_71,Wil_BellSys_70,KeS_ContemporaryMath_01,RaS_IT_92} to construct Berman and dual Berman codes as ideals in appropriately chosen group algebras.
Our primary tool in this section is Rajan and Siddiqi's characterization of abelian codes~\cite{RaS_IT_92} based on the discrete Fourier transform (DFT).
We begin this section with a review of abelian codes and DFT, and then introduce a convenient notation and present some results to work with the DFT for the group algebra $\Fb_2[G^m]$, where $G$ is an abelian group of order $n$.
We then construct Berman codes and their duals as abelian codes and show that this construction is equivalent to our original recursive definition of these codes when $n$ is odd.
In Appendix~\ref{app:classical_berman_codes} we show that the special case $n$ being an odd prime coincides with the codes studied by Berman~\cite{Ber_Cybernetics_II_67} and Blackmore and Norton~\cite{BlN_IT_01}.
For all odd $n$, we then identify a large family of abelian codes, that includes $\{\Dsn(r,m): n \text{ odd}\}$ and $\{\Csn(r,m): n \text{ odd}\}$, which achieves BEC capacity under bit-MAP decoding.

\emph{Section~\ref{sec:simulations}}. We present a few simulation results comparing the codes identified in this work with RM codes in the BEC. While these simulations are in no way exhaustive, we observe that in the simulation scenarios presented here the bit erasure rates (under bit-MAP decoding) of Berman code and its dual are similar to those of RM codes of comparable rate and block length, while their block erasure rates (under block-MAP decoding) are relatively worse.

We conclude this work in Section~\ref{sec:discussion}. The proofs of some of the technical results have been moved to Appendix~\ref{app:proofs}.

\begin{table}[!t]
\centering
\renewcommand{\arraystretch}{1.25}
\caption{{Organization of this paper.}}
\label{tbl:org}
\begin{tabular}{|l|l|}
\hline
\hline
Section & Title \\
\hline
\hline
{\bf II} & {\bf Berman Codes and their Duals} \\
\hline
II-A & Recursive Definition of Berman \& Dual Berman Codes \\
\hline
II-B & Dimension and Duality \\
\hline
II-C & Minimum Distance \\
\hline
II-D & Bases for $\Dsn(r,m)$ and $\Csn(r,m)$ \\
\hline
II-E & Recursive Decoding \\
\hline
\hline
{\bf III} & {\bf Capacity-Related Properties} \\
\hline
{III-A} & {Some Useful Automorphisms of  $\Dsn(r,m)$ \& $\Csn(r,m)$} \\
\hline
III-B & Puncturing $\Dsn(r,m)$ and $\Csn(r,m)$\\
\hline
III-C & Rate of  $\Dsn(r,m)$ and $\Csn(r,m)$\\
\hline
III-D & Lack of Double Transitivity \\
\hline
III-E & Achieving the BEC Capacity \\
\hline
\hline
{\bf IV} & {\bf Constructions via Discrete Fourier Transform} \\
\hline
IV-A & Background \\
\hline
IV-B & Preliminary Results \\
\hline
IV-C & Construction of $\Dsn(r,m)$ and $\Csn(r,m)$ via DFT \\
\hline
IV-D & A Family of Abelian Codes that Achieve BEC Capacity \\
\hline
\hline
{\bf V} & {\bf Simulation Results} \\
\hline
\hline
{\bf VI} & {\bf Discussion} \\
\hline
\hline
\end{tabular}
% \vspace{2mm}
% \hrule
\end{table}

\emph{Notation:} For any positive integer $\ell$, let $\lset \ell \rset$ denote the set $\{0,1,\dots,\ell-1\}$.
Let $\Zb_{\ell}=\lset \ell \rset$ be the ring where addition and multiplication are performed modulo $\ell$.
The binary field is $\Fb_2=\{0,1\}$. The empty set is $\emptyset$.
For sets $A, B$ we define $A \setminus B = \left\{a \in A : a \notin B \right\}$.
%For any vector $\p{b}$, $\supp(\p{b})$ is the support set of $\p{b}$.
%For a collection of vectors $A$ from a vector space, we denote $\mathsf{span}(A)$ the linear span of all the vectors in $A$.
The notation $\p 0$ denotes a zero-vector or a zero-matrix of appropriate size. We denote the identity matrix of size $n$ by ${\p I}_n$. For two vectors ${\p a},{\p b}$, their concatenation is denoted by $({\p a}\,|\,{\p b}).$  The dimension of a code ${\cal C}$ is denoted by $\dim({\cal C})$. The Hamming weight of a vector ${\p a}$ is denoted by $\wt({\p a})$. The minimum distance of a code ${\cal C}$ is denoted by $d_{\min}({\cal C}).$ The binomial coefficient $\binom{n}{k}$ is assumed to be $0$ if $k>n$ or if $k<0$.  We denote an $n$-length vector ${\p a}$ by its components as ${\p a}=(a_i:i\in \integerbox{n}).$ All vectors are row vectors, unless otherwise stated. The notation $(.)^T$ denotes the transpose operator. For $S\subset\integerbox{n},$ we denote the vector with components $a_i:i\in S$ as ${\p a}_S$.  If ${\p a}$ is a $n^m$-length vector for some $m\geq 1$, we also use the concatenation representation ${\p a}=({\p a}_0|{\p a}_1|\hdots|{\p a}_{n-1})$, where ${\p a}_l:l\in\integerbox{n}$ are subvectors of length $n^{m-1}$. The individual components of ${\p a}_l$ would be then denoted as $a_{l,i}:i\in\integerbox{n^{m-1}}$.
For any integer $m \geq 1$, the group of all permutations on $\lset m \rset$ is denoted as $\Ss_m$.

\section{Berman Codes and their Duals} \label{sec:berman_and_dual}
% Notation: $\p 0$ denotes a zero-vector or a zero-matrix of appropriate length. For two vectors ${\p a},{\p b}$ over vector spaces over the same field, their concatenation is denoted by $({\p a}|{\p b}).$

% REPEAT : The set $\{0,\hdots,t-1\}$ is denoted by $\integerbox{t}.$ The dimension of a code ${\cal C}$ is denoted by $\dim({\cal C})$.

% The Hamming weight of a vector ${\p a}$ is denoted by $\wt({\p a})$. The minimum distance of a code ${\cal C}$ is denoted by $d_{\min}({\cal C}).$ The binomial coefficient $\binom{n}{k}$ is assumed to be $0$ if $k>n.$  We denote a $t$-length vector ${\p a}$ by its components as ${\p a}=(a_i:i\in \integerbox{t}).$

% % R0 % %
% A class of abelian group codes was originally studied by Berman in \cite{Ber_Cybernetics_67,Ber_Cybernetics_II_67} using a group algebra framework.
% These codes include the Reed-Muller codes as a special case. We now present a recursive construction of a large class of binary codes of length $n^m$ (for some integers $n,m$), which includes the class of the Berman codes.
% Following this precedence, we continue to call our codes as Berman codes (and their duals).
{
We present a new recursive definition of Berman and dual Berman codes
% The recursive construction we present for our codes is
inspired from the Plotkin $({\p u}|{\p u}+{\p v})$ construction: the Plotkin construction involves a two-fold extension, the construction we present generalizes this (in a sense) to an $n$-fold extension for arbitrary $n$.
More general Plotkin-like constructions have been studied in the past in \cite{MatProdCodes_Blackmore_Norton,HeR_JAlgebra_13}, in which the resultant codes are called \textit{matrix-product codes}.
In the present work, our focus is on a particular Plotkin-like recursive construction that yields the Berman and dual Berman codes.
We present a broad set of results for these codes, including their dimension, minimum distance, bases, automorphism groups, and efficient decoding up to half the minimum distance.
% , and their capacity achieving nature for the BEC channel.

Please note that~\cite{Ber_Cybernetics_II_67} and~\cite{BlN_IT_01} derive the dimension and the minimum distance of $\Ds_p(r,m)$ for all odd prime $p$, and $\Csn(r,m)$ for all $n$, using a group algebra framework. In contrast, we use the new Plotkin-like construction to provide a simple derivation of the dimension and the minimum distance of all Berman and dual Berman codes.
}
% % % % % % %

\subsection{Recursive Definition of Berman \& Dual Berman Codes} \label{sec:sub:recursive_defn}

We now proceed to give the definitions of the codes presented in this work.
% % R0 % %
{These will be the primary definitions of Berman and dual Berman codes, and all other characterizations of these codes will be derived from this definition.}
% % % % % %
For some positive integers \mbox{$n\geq 2$} and $m$, for some non-negative integer $r$ such that \mbox{$0 \leq r\leq m$}, define the family of codes \mbox{$\Dsn(r,m)\subset \Fb_2^{n^m}$} recursively as follows.
%SINGLECOLVERSION
% \begin{tcolorbox}
% \begin{align*}
%     \Dsn(m,m)&\triangleq \{{\p 0}\in \Fb_2^{n^m}\}.\\
%     \Dsn(0,m)&\triangleq \{{\p c}\in \Fb_2^{n^m}\colon \sum_i c_i=0\} ~~~(\text{The single parity-check code of length } n^m).
% \end{align*}
% For $m\geq 2$ and $1\leq r\leq m-1,$
% \begin{align*}
%     \Dsn(r,m)\triangleq \{({\p v}_0|{\p v}_1|\hdots|{\p v}_{n-1})&\colon {\p v}_l\in \Dsn(r-1,m-1),
% {\forall l \in \lset n \rset,}
% \sum_{l\in\integerbox{n}}{{\p v}_l}\in \Dsn(r,m-1)\}.
% \end{align*}
% \end{tcolorbox}
%%%THEBELOWISDOUBLECOLUMNVERSION
\begin{tcolorbox}
\begin{talign*}
    &\Dsn(m,m) \triangleq \{{\p 0}\in \Fb_2^{n^m}\}.\\
    &\Dsn(0,m) \triangleq \{{\p c}\in \Fb_2^{n^m}\colon \sum_i c_i=0\}.\\
    &~~~~~~~~~~~~(\text{The single parity-check code of length } n^m).
\end{talign*}
For $m\geq 2$ and $1\leq r\leq m-1,$
\begin{talign*}
    \Dsn(r,m)\triangleq \{({\p v}_0|{\p v}_1|\hdots|&{\p v}_{n-1})\colon {\p v}_l\in \Dsn(r-1,m-1),\\
    &\sum_{l\in\integerbox{n}}{{\p v}_l}\in \Dsn(r,m-1) \}.
\end{talign*}
\end{tcolorbox}

We refer to the code $\Dsn(r,m)$ as \emph{the Berman code with parameters $n,m$, and $r$}. We similarly define the code family $\Csn(r,m)\subset \Fb_2^{n^m}$ recursively.
\begin{tcolorbox}
\begin{align*}
    &\Csn(m,m)\triangleq \Fb_2^{n^m}.\\
    &\Csn(0,m)\triangleq \{(c,\hdots,c)\in \Fb_2^{n^m}\colon c\in \Fb_2\}. \\
    & ~~~~~~~~~~~~~~~~~~~~~~(\text{The repetition code of length } n^m).
\end{align*}
%%%%
For $m\geq 2$ and $1\leq r\leq m-1,$
\begin{align*}
    \Csn(r,m)\triangleq &\{({\p u}+{\p u}_0|{\p u}+{\p u}_1|\hdots|{\p u}+{\p u}_{n-2}|{\p u}) \colon \\
  & {\p u}_l\in \Csn(r-1,m-1), {\forall l \in \lset n-1 \rset,} \\
  & {\p u}\in \Csn(r,m-1)\}.
\end{align*}
\end{tcolorbox}
We shall refer to $\Csn(r,m)$ as the \emph{dual Berman code with parameters $n,m$ and $r$} (this nomenclature will be validated in Section~\ref{subsec:dimandduality}, where we shall prove that the codes $\Dsn(r,m)$ and $\Csn(r,m)$ are dual to each other). Also, observe that when $n=2$, the code $\Cs_2(r,m)$ is defined as 
\begin{equation*}
\{({\p u}+{\p u}_0|{\p u})\colon {\p u}_0\in\Cs_2(r-1,m-1),{\p u}\in \Cs_2(r,m-1)\},
\end{equation*} 
which coincides with $\RM(r,m)$. Thus the class of codes $\Csn(r,m)$ includes the Reed-Muller codes.  Later, in Section~\ref{sec:dft}, we will also show that when $n$ is an odd prime, the code $\Dsn(r,m)$ is identical to the code designed by Berman in~\cite{Ber_Cybernetics_II_67}, and $\Csn(r,m)$ is its dual code studied by Blackmore and Norton in~\cite{BlN_IT_01}.

%TWOCOLVERSIONBELOW
% \begin{tcolorbox}
% \begin{align*}
%     \Csn(m,m)&\triangleq \Fb_2^{n^m}.\\
%     \Csn(0,m)&\triangleq \{(c,\hdots,c)\in \Fb_2^{n^m}\colon c\in \Fb_2\}\\
%&~~~\(\text{The repetition code of length } n^m).
% \end{align*}
% %%%%
% For $m\geq 2$ and $1\leq r\leq m-1,$
% \begin{align*}
%     \Csn(r,m)\triangleq &\{({\p u}+{\p u}_0|{\p u}+{\p u}_1|\hdots|{\p u}+{\p u}_{n-2}|{\p u})\\&\colon {\p u}_l\in \Csn(r-1,m-1),
%     {\p u}\in \Csn(r,m-1)\}.
% \end{align*}
% \end{tcolorbox}
%%%
\begin{example}
We give some specific examples of the codes defined above.
\begin{itemize}
\item For \mbox{$n=3$}, \mbox{$m=1$}, \mbox{$r=0$}, the code 
\begin{equation*}
\Ds_3(0,1)=\{(v_0|v_1|v_0+v_1):v_0,v_1\in\Fb_2\}
\end{equation*} 
is the single parity-check code of length $3$ and \mbox{$\Cs_3(0,1)=\{(u|u|u):u\in \Fb_2\}$} is its dual, the repetition code of length $3$.

\item By the recursive definition, the single parity check code is used as the building block along with a global parity to obtain the following code for $n=3,m=2,r=1$, 
    \begin{align*}
        \Ds_3(1,2)&=\left\{({\p v}_0|{\p v}_1|{\p v}_0+{\p v}_1):{\p v}_0,{\p v}_1\in\Ds_3(0,1)\right\},
    \end{align*}
which is equal to
    \begin{align*}
        \Big\{ (&v_{00},v_{01},v_{00}+v_{01}|v_{10},v_{11},v_{10}+v_{11}| \\
        &v_{00}+v_{10},~v_{01}+v_{11},~v_{00}+v_{01}+v_{10}+v_{11}) \colon \\
        &~~v_{ij}\in\Fb_2, \forall i,j\in\{0,1\} \Big\}.
    \end{align*}
\item The code $\Cs_3(1,2)$ is shown below as per the recursive construction. It is easy to verify that it is dual to $\Ds_3(1,2)$.
     \begin{align*}
    \Cs_3(1,2)&=\Big\{(u_{00},u_{00},u_{00}|u_{10},u_{10},u_{10}|0,0,0) \\ 
&~~~~~~~ + (u_0,u_1,u_2|u_0,u_1,u_2|u_0,u_1,u_2):\\
    &~~~~~~~~~~~~~~~~~u_{00},u_{10},u_0,u_1,u_2\in\Fb_2 \Big\}.
    \end{align*}
\end{itemize}
\end{example}

%%%%
It is clear that the codes $\Csn(r,m)$ and $\Dsn(r,m)$ are linear. We shall now obtain various properties of these codes. The techniques involved in the proofs are similar to those for $\RM$ codes, mainly involving induction on the parameter $m$.
{The parameters of $\Csn(r,m)$ were previously derived in~\cite[Remark~2.4]{BlN_IT_01}, and the parameters of $\Ds_p(r,m)$ for odd prime $p$ were derived in~\cite[Theorem~2.2]{Ber_Cybernetics_II_67}, both using a group algebra framework. In contrast, our approach is based on the recursive construction of these codes.}
The following lemma is key to the remaining results in this section.
\begin{lemma}
\label{containmentlemma}
For $1\leq r\leq m,$ we have
\begin{enumerate}
    \item $\Dsn(r,m)\subset \Dsn(r-1,m)$, and
    \item $\Csn(r-1,m)\subset \Csn(r,m)$.
\end{enumerate}
\end{lemma}
%%%%%
\begin{IEEEproof}
We prove this via induction on $m$.

\emph{Part 1}. First, we observe that the statements are true by definition for $m=1.$ We now prove the statements are true for $m\geq 2$ assuming they are true for $m-1$. Consider an arbitrary codeword of $\Dsn(r,m)$ in its concatenation representation ${\p v}=({\p v}_0|{\p v}_1|\hdots|{\p v}_{n-1})$, where the components of ${\p v}_l$ are given as \mbox{$v_{l,i}:i\in\integerbox{n^{m-1}}$}. For each \mbox{$l\in\integerbox{n},$} by definition ${\p v}_l\in \Dsn(r-1,m-1)$. Now consider the case $r=1$. Then ${\p v}_l\in \Dsn(0,m-1)$. Thus $\sum_{i\in \integerbox{n^{m-1}}}v_{l,i}=0, \forall l\in\integerbox{n}$. And hence, 
\begin{equation*}
\sum_{i\in \integerbox{n^m}}v_i=\sum_{l\in\integerbox{n}}\sum_{i\in\integerbox{n^{m-1}}}v_{l,i}=0,
\end{equation*}
which means ${\p v}\in \Dsn(0,m)$, hence proving the statement for $r=1$ for any $m$.

Now suppose \mbox{$r\geq 2$}. By induction $\Dsn(r-1,m-1)\subset \Dsn(r-2,m-1)$, thus ${\p v}_l\in  \Dsn(r-2,m-1)$. Further, $\sum_{l\in\integerbox{n}}{\p v}_l\in \Dsn(r,m-1)$ by definition, and  $\Dsn(r,m-1) \subset \Dsn(r-1,m-1)$ by induction. Hence we have $\sum_{l\in\integerbox{n}}{\p v}_l\in \Dsn(r-1,m-1).$ Thus, the codeword $\p v$ satisfies the two conditions in the definition to be a codeword of $\Dsn(r-1,m)$ which completes the proof for Part 1.

\emph{Part 2}.  The statement is true for $m=1$ by definition. We now prove for $m\geq 2$ assuming the statement is true for $m-1.$

Suppose $r=1.$ Then, using ${\p u}_l={\p 0}, \forall l\in \integerbox{n-1}$ in the codewords of $\Csn(1,m)$ gives the codewords of $\Csn(0,m).$ Thus the statement is true also for $r=1$ with any value of $m\geq 2$.

Now suppose $r\geq 2$. Consider an arbitrary codeword ${\p u}'=({\p u}+{\p u}_0|{\p u}+{\p u}_1|\hdots|{\p u}+{\p u}_{n-2}|{\p u})$ in $\Csn(r-1,m).$  Then ${\p u}\in \Csn(r-1,m-1)$ by definition and \mbox{$\Csn(r-1,m-1) \subset$} \mbox{$\Csn(r,m-1)$} by induction, which means ${\p u}\in \Csn(r,m-1)$. Similarly we can show 
\begin{equation*}
{\p u}_l\in \Csn(r-2,m-1)\subset\Csn(r-1,m-1). 
\end{equation*}
Thus we have shown that ${\p u}'\in \Csn(r,m)$ as it satisfies both  conditions in the definition of $\Csn(r,m)$. This proves Part~2.
\end{IEEEproof}
%%%%
\subsection{Dimension and Duality}
\label{subsec:dimandduality}
We obtain the dimension of the two codes and show that they are duals of each other.
\begin{lemma}
\label{dimensionlemma}
\begin{align*}
    \dim(\Dsn(r,m))&=\sum_{w=r+1}^{m}\binom{m}{w}(n-1)^w,\\
    \dim(\Csn(r,m))&=\sum_{w=0}^{r}\binom{m}{w}(n-1)^w.
\end{align*}
\end{lemma}
%%%
\begin{IEEEproof}
It is straightforward to verify the statements for $m=1.$ We now prove the statements for $m\geq 2$ assuming it is true for $m-1.$ We mainly use the two combinatorial identities: $n^m=\sum_{w=0}^m\binom{m}{w}(n-1)^w,$ and $\binom{m-1}{w-1}+\binom{m-1}{w}=\binom{m}{w}, \forall w$.

\emph{Part 1}. Again, the statements are easy to verify for $r=0,m$. Hence we assume $1\leq r\leq m-1$.
% % R0 % %
{Using the definition of $\Dsn(r,m)$, we first observe that
%ONECOLUMN
\begin{talign}
\label{eqn:dsndifferentdef}
        \Dsn(r,m) = \{(&{\p v}_0|{\p v}_1|\hdots|{\p v}_{n-2}| \sum_{l\in\integerbox{n-1}} {\p v}_l+{\p v}) \colon \nonumber \\
    & {\p v}_l\in \Dsn(r-1,m-1), {\p v}\in \Dsn(r,m-1) \}.
\end{talign}
}
% % % % % % %
%TWOCOLUMN
% \begin{align*}
%\label{eqn:dsndifferentdef}
%         \Dsn(r,m)\triangleq \{({\p v}_0&|{\p v}_1|\hdots|{\p v}_{n-2}|\hspace{-0.2cm}\sum_{l\in\integerbox{n-1}}\hspace{-0.2cm}{\p v}_l+{\p v})\colon \\
%     &{\p v}_l\in \Dsn(r-1,m-1), {\p v}\in \Dsn(r,m-1)\}.
% \end{align*}
%%%%
From the description in (\ref{eqn:dsndifferentdef}), we see that $\dim(\Dsn(r,m))$
%ONECOLUMN
\begin{align*}
    &=(n-1)\dim(\Dsn(r-1,m-1))+\dim(\Dsn(r,m-1))\\
    &=(n-1)\sum_{w=r}^{m-1}\binom{m-1}{w}(n-1)^w\\ 
    &~~~~~~~~~~~~~~~~~~~~~~~~+\sum_{w=r+1}^{m-1}\binom{m-1}{w}(n-1)^w\\
    &=\sum_{w=r+1}^{m}\binom{m-1}{w-1}(n-1)^w+\sum_{w=r+1}^{m-1}\binom{m-1}{w}(n-1)^w\\
    &=\sum_{w=r+1}^{m}\binom{m}{w}(n-1)^w,
\end{align*}
which proves Part 1.
%TWOCOLUMN
% {\small \begin{align*}
%     &\dim(\Dsn(r,m))\\
%     &=(n-1)\dim(\Dsn(r-1,m-1))+\dim(\Dsn(r,m-1))\\
%     &=(n-1)\sum_{w=r}^{m-1}\binom{m-1}{w}(n-1)^w+\sum_{w=r+1}^{m-1}\binom{m-1}{w}(n-1)^w\\
%     &=\sum_{w=r+1}^{m}\binom{m-1}{w-1}(n-1)^w+\sum_{w=r+1}^{m-1}\binom{m-1}{w}(n-1)^w\\
%     &=\sum_{w=r+1}^{m}\binom{m}{w}(n-1)^w,
% \end{align*}
% which proves Part 1).
% }

\emph{Part 2}. We observe that the statements are easy to verify for $m=1$ and for any $m$  with $r=0,m.$ Hence we consider $m\geq 2$ and $1\leq r\leq m-1$, and proceed by induction.
% % R0 % %
{From the definition of the dual Berman code $\Csn(r,m)$,} $\dim(\Csn(r,m))$
% % % % %
%ONECOLUMN
\begin{align*}
    &=(n-1)\dim(\Csn(r-1,m-1))+\dim(\Csn(r,m-1))\\
    &=(n-1)\sum_{w=0}^{r-1}\binom{m-1}{w}(n-1)^w+\sum_{w=0}^{r}\binom{m-1}{w}(n-1)^w \\
% \end{align*}
% \begin{align*}
    &=\sum_{w=1}^{r}\binom{m-1}{w-1}(n-1)^w+\sum_{w=0}^{r}\binom{m-1}{w}(n-1)^w\\
    &=\sum_{w=0}^{r}\binom{m}{w}(n-1)^w,
\end{align*}
%TWOCOLUMN
% {\small \begin{align*}
%   &\dim(\Csn(r,m))\\
%     &=(n-1)\dim(\Csn(r-1,m-1))+\dim(\Csn(r,m-1))\\
%     &=(n-1)\sum_{w=0}^{r-1}\binom{m-1}{w}(n-1)^w+\sum_{w=0}^{r}\binom{m-1}{w}(n-1)^w\\
%     &=\sum_{w=1}^{r}\binom{m-1}{w-1}(n-1)^w+\sum_{w=0}^{r}\binom{m-1}{w}(n-1)^w\\
%     &=\sum_{w=0}^{r}\binom{m}{w}(n-1)^w,
% \end{align*}
% }
which completes the proof of Part 2.
%%%%
\end{IEEEproof}
%%%%%
We now confirm that the codes $\Dsn(r,m)$ and $\Csn(r,m)$ are dual to each other.
\begin{lemma}
\label{dualitylemma}
$\Csn(r,m)^\perp=\Dsn(r,m).$
\end{lemma}
\begin{IEEEproof}
The statement holds by definition for $r=0,m$ (for any value of $m$) and thus for $m=1$ (for $r\leq 1$). Hence we now prove the statement for $m\geq 2$ and $1\leq r\leq m-1,$ assuming it is true for $m-1$.

Let ${\p v}=({\p v}_0|{\p v}_1|\hdots|{\p v}_{n-1})$ and 
\begin{align*}
{\p u}'=({\p u}+{\p u}_0|{\p u}+{\p u}_1|\hdots|{\p u}+{\p u}_{n-2}|{\p u})
\end{align*}
be any two codewords in the respective codes $\Dsn(r,m)$ and $\Csn(r,m)$. Taking their dot product we get,
\begin{align*}
    {\p u}'{\p v}^T=\sum_{l\in\integerbox{n-1}}{\p u}_l{\p v}_l^T+{\p u}\left(\sum_{l\in\integerbox{n}}{\p v}_l\right)^T.
\end{align*}
Since by the definition of the codes we have 
\begin{align*}
{\p u}_l\in\Csn(r-1,m-1) \text{ and } {\p v}_l\in \Dsn(r-1,m-1), 
\end{align*}
and further $\sum_{l\in \integerbox{n}}{\p v}_l\in\Dsn(r,m-1)$ and ${\p u}\in \Csn(r,m-1)$, by the induction hypothesis we see that ${\p u}'{\p v}^T=0$. Thus, we have $\Dsn(r,m)\subset \Csn(r,m)^\perp$. Equality follows by observing that $\dim(\Csn(r,m))=n^m-\dim(\Dsn(r,m))$, from Lemma~\ref{dimensionlemma}.
\end{IEEEproof}
\subsection{Minimum Distance}
We now obtain the minimum distance of the two codes.
\begin{lemma}
\label{mindistancelemma}
The minimum distance $d_{\min}$ of the two codes are
\begin{align*}
    d_{\min}(\Dsn(r,m))&=2^{r+1}, ~~ 0\leq r \leq m-1,\\
    d_{\min}(\Csn(r,m))&=n^{m-r},  ~~0\leq r \leq m.
\end{align*}
\end{lemma}
\begin{IEEEproof}
The claims are straightforward for $r=0,m$ and thus for $m=1$. We proceed to prove the claims for $m\geq 2$ and $1\leq r\leq m-1$ assuming they are true for $m-1.$

\emph{Part 1}. Consider an arbitrary non-zero codeword of $\Dsn(r,m)$ given by ${\p v}=({\p v}_0|{\p v}_1|\hdots|{\p v}_{n-1})$. We consider two cases.

Case (a): \textit{At least two of the ${\p v}_l$'s are non-zero.} Then $\wt({\p v})\geq 2d_{\min}(\Dsn(r-1,m-1))=2^{r+1}$ (by induction). Further, there is a codeword $({\p a}|{\p a}|{\p 0}|\hdots|{\p 0})\in\Dsn(r,m),$ where ${\p a}$ is any arbitrary non-zero codeword in $\Dsn(r-1,m-1)$ of weight $2^r$, which has weight precisely $2^{r+1}.$

Case (b): \textit{Exactly one of the ${\p v}_l$'s (say ${\p v}_0$) is non-zero}. Then clearly ${\p v}_0\in \Dsn(r,m-1)$ by definition of the code. This would mean that $r\leq m-2$, as $\Dsn(m-1,m-1)$ contains only $\p 0$. Thus, $\wt({\p v})=\wt({\p v}_0)\geq d_{\min}(\Dsn(r,m-1))=2^{r+1}$ (by induction).

The two cases are exhaustive, and thus the proof of Part 1 is completed.

\emph{Part 2}.
Consider an arbitrary non-zero codeword 
\begin{align*}
{\p u}'=({\p u}+{\p u}_0|{\p u}+{\p u}_1|\hdots|{\p u}+{\p u}_{n-2}|{\p u})\in \Csn(r,m), 
\end{align*}
where ${\p u}_l\in \Csn(r-1,m-1)$ and ${\p u}\in \Csn(r,m-1)$. Recall from Lemma \ref{containmentlemma} that $\Csn(r-1,m-1)\subset\Csn(r,m-1)$. We consider two cases.

Case (a): \emph{When} ${\p u}\in \Csn(r-1,m-1)$. In this case if ${\p u}={\p 0}$, then at least one of ${\p u}_{l}\neq {\p 0}$. Thus at least one subvector, say ${\p u}_0$, of ${\p u}'$ is non-zero in $\Csn(r-1,m-1)$. Thus, by induction, 
\begin{align*}
\wt({\p u}')\geq \wt({\p u}_0)\geq d_{\min}(\Csn(r-1,m-1))=n^{m-r}. % (by induction). 
\end{align*}
Further the codeword $({\p u}_0|{\p 0}|\hdots|{\p 0})\in \Csn(r,m)$ where ${\p u}_0$ is a minimum weight non-zero codeword in $\Csn(r-1,m-1)$ has weight precisely $n^{m-r}$.

Case (b): \emph{When} ${\p u}\in \Csn(r,m-1)\setminus \Csn(r-1,m-1)$. Then each subvector ${\p u}+{\p u}_l\in \Csn(r,m-1)\setminus \Csn(r-1,m-1)$ as well. Then $\wt({\p u}')\geq nd_{\min}(\Csn(r,m-1))=n^{m-r}$ (by induction).

This completes the proof of Part 2 and hence the lemma.
\end{IEEEproof}
%%%
\subsection{Bases for $\Dsn(r,m)$ and $\Csn(r,m)$}

{We identify two sets of bases, each for $\Dsn(r,m)$ and $\Csn(r,m)$. While the first basis follows naturally from the recursive construction of these codes and consists of minimum weight codewords, the second basis illustrates the combinatorial structure of these codes and is helpful in identifying code automorphisms.}

\subsubsection{{Natural Minimum Weight Basis}}

We define recursively a generator matrix $G_n(r,m)$  for $\Dsn(r,m)$ as follows. If $r=m$, we have nothing to define. For $r=0$, we define
$$G_n(0,m)\triangleq \begin{pmatrix}
1 & \multicolumn{3}{|c}{}\\
1 & \multicolumn{3}{|c}{{\p I}_{n^m-1}}\\
\vdots &\multicolumn{3}{|c}{}\\
 1 & \multicolumn{3}{|c}{}                              \end{pmatrix}.$$

For \mbox{$1\leq r\leq m-1$}, we define $G_n(r,m)$ recursively as in (\ref{generatorDnrecursive}).
%%%%%%
\begin{figure*}[t]
\begin{align}
\label{generatorDnrecursive}
G_n(r,m)\triangleq \begin{pmatrix}
G_n(r-1,m-1) & {\p 0}       & \hdots & {\p 0}       & G_n(r-1,m-1) \\
{\p 0}       & G_n(r-1,m-1) & \hdots & {\p 0}       & G_n(r-1,m-1) \\
{\p 0}       & {\p 0}       & \ddots & \vdots       & \vdots       \\
\vdots       & \vdots       & \ddots & G_n(r-1,m-1) & G_n(r-1,m-1) \\
{\p 0}       & {\p 0}       & \hdots & {\p 0}       & G_n(r,m-1)
\end{pmatrix}.
\end{align}
\hrule
\begin{align}
    \label{generatorCnrecursive}
H_n(r,m)\triangleq\begin{pmatrix}
H_n(r-1,m-1) & {\p 0}       & \hdots & \hdots       & {\p 0}     \\
{\p 0}       & H_n(r-1,m-1) & \ddots & \vdots       & \vdots     \\
\vdots       & \vdots       & \ddots & H_n(r-1,m-1) &  {\p 0}   \\
H_n(r,m-1)   & H_n(r,m-1)   & \hdots & H_n(r,m-1)   & H_n(r,m-1)
\end{pmatrix}.
\end{align}
\hrule
\end{figure*}
%%%%
Note that, if $r=m-1$, then the last set of rows of (\ref{generatorDnrecursive}) corresponding to $G_n(r,m-1)$ are absent. It is not difficult to show using (\ref{eqn:dsndifferentdef}) that the matrix $G_n(r,m)$ (for $1\leq r\leq m-1$) defined above generates the code $\Dsn(r,m).$

We similarly define the generator matrix $H_n(r,m)$ for $\Csn(r,m)$. We define
\begin{align*}
    &H_n(0,m) \triangleq \begin{pmatrix}1&1&\hdots&1\end{pmatrix}, (\text{all-one vector of length } n^m)\\
    &H_n(m,m) \triangleq {\p I}_{n^m},
\end{align*}
and for $1\leq r\leq m-1,$ $H_n(r,m)$ is defined in (\ref{generatorCnrecursive}). Again, using the definition of $\Csn(r,m)$, it is easy to verify the matrix $H_n(r,m)$ generates the code $\Csn(r,m)$, for $0\leq r\leq m$.

Finally, by the recursive construction, we have the following result regarding the rows of $G_n(r,m)$ and $H_n(r,m)$.
%%%%
\begin{lemma}
\label{minweightgeneratorDnCn}
For \mbox{$0\leq r\leq m-1$}, each row of the matrix $G_n(r,m)$ (respectively, $H_n(r,m)$) is a minimum weight codeword of $\Dsn(r,m)$ (respectively, $\Csn(r,m)$).
\end{lemma}
%%%%%
\begin{IEEEproof}
The proof follows by observing that $G_n(0,m)$ has minimum weight ($=2$) codewords of $\Dsn(0,m)$ as its rows, and by applying the recursive construction in (\ref{generatorDnrecursive}).

The argument for $H_n(r,m)$ is similar, as our recursive construction depends on $H_n(0,m)$ and $H_n(m,m)$ as the base cases, both of which have minimum weight codewords of the respective codes as their rows.
\end{IEEEproof}
%%%%

% % R0 version % %
{
Lemma~\ref{minweightgeneratorDnCn} shows that the minimum weight codewords of $\Dsn(r,m)$ (and $\Csn(r,m)$) span the code. Since $\Dsn(r,m)$ and $\Csn(r,m)$ are dual codes, we see that both these codes have parity-check matrices whose rows are composed of the minimum weight codewords from their respective dual codes.
This property is perhaps useful in designing low-complexity iterative decoders for these codes; such iterative decoders are available for RM codes~\cite{SHP_ISIT_18}.
}
% % % % % % % % %

% \subsubsection{Other useful generating sets}
\subsubsection{{A Patterned Basis for Berman and Dual Berman Codes}}
\label{subsubsec:usefulbasis}

In this part of the present sub-section, we show Lemma \ref{basisberman} and Lemma \ref{basisdualberman}, which give another basis for $\Dsn(r,m)$ and $\Csn(r,m)$, respectively.
% % R0 % %
{
These results further reveal the strong combinatorial structure of the respective codes. Also, we shall use the basis from Lemma \ref{basisberman} to obtain some automorphisms of the codes described in this work in Section~\ref{subsec:automorphism}.
The identification of these automorphisms will eventually allow us to prove the capacity-achieving properties of Berman codes and their duals in Section~\ref{sec:sub:capacity}.
}%
% % % % % %
Towards stating the lemma, we give some notation to work with the indices of vectors in $\Fb_2^{n^m}.$ This notation will also be used in the forthcoming section.

% Let $G$ be any set containing $n$ elements, denoted as $G=\{g_0,\hdots,g_{n-1}\}$.
{Let $G = \integerbox{n} = \{0,1,\dots,n-1\}$.}
We then identify the $n^m$ coordinates of an arbitrary vector ${\p v}\in \Fb_2^{n^m}$ using the $m$-tuples in $G^m$, i.e., ${\p v}=(v_{\p i}:{\p i}\in G^m)$.

We also write $\p v$ as a concatenation of $n$ vectors from $\Fb_2^{n^{m-1}},$ denoted by ${\p v}=({\p v}_0|\hdots|{\p v}_{n-1})$. The subvector ${\p v}_l\in \Fb_2^{n^{m-1}}$ is then identified recursively as follows.
\begin{itemize}
\item {For any ${\p i}'\in G^{m-1},$ the component of ${\p v}_l$ indexed by ${\p i}'$ is identified as $v_{l,{\p i}'}=v_{({\p i}'| l)}$ which is the component of ${\p v}$ indexed by $({\p i}'\,|\,l) \in G^m$}.
\end{itemize}
Observe that the concatenation representation can be used recursively, for instance the subvector \mbox{${\p v}_l\in\Fb_2^{n^{m-1}}$} can be written as a concatenation of $n$ subvectors from $\Fb_2^{n^{m-2}}$, and so on.

{Let us denote the support of a vector $\p{i} \in G^m$ as $\supp(\p{i}) = \{ k \in \lset m \rset : i_k \neq 0\}$. The weight or the Hamming weight of $\p{i}$ is $\wt(\p{i}) = |\supp(\p{i})|$.
We now define a partial ordering among the vectors in $G^m$. For $\p{i},\p{j} \in G^m$ we will say that `$\p{i}$ contains $\p{j}$', or equivalently, `$\p{j}$ is contained in $\p{i}$' if
\begin{equation*}
\supp(\p{j}) \subset \supp(\p{i}) \text{ and } \p{j}_{\supp(\p{j})} = \p{i}_{\supp(j)}.
\end{equation*}
That is, $\p{i}$ contains $\p{j}$ if for each $k \in \lset m \rset$ we have $j_k = 0$ or $j_k = i_k$.
We will denote this relation as $\p{i} \succeq \p{j}$ or $\p{j} \preceq \p{i}$.
Note that for any $\p{i} \in G^m$ there are exactly $2^{\wt(\p{i})}$ vectors contained by $\p{i}$, and exactly $n^{m-\wt(\p{i})}$ vectors that contain $\p{i}$.
}

We also need the following definition of a patterned-vector in $\Fb_2^{n^m}$, which will be used to show a basis for $\Dsn(r,m)$.
{For $m\geq 1$ and some ${\p i'}\in G^m$, define the vector ${\p c}_m({\p i'})\in \Fb_2^{n^m}$ as the binary vector with support set $\{ \p{i} \in G^m: \p{i} \preceq \p{i'} \}$. That is, the $n^m$ components of $\p{c}_m(\p{i'})$ are
\begin{equation*}
c_m({\p i'})_{\p i}=\begin{cases}
1 & \text{if } \p{i} \preceq \p{i'},\\
0 & \text{otherwise},
\end{cases} ~~~~~~~~\forall {\p i}\in G^m.
\end{equation*}
}
%% consisting of the entries
%% $c_m({\p i'})_{\p i}=\begin{cases}
%% 1 & \text{if } \nz({\p i})\subset \nz({\p i'}) \text{ and } {\p i}_{\nz({\p i})}={\p i'}_{\nz({\p i})},\\
%% 0 & \text{otherwise},
%% \end{cases} ~~~~~~~~\forall {\p i}\in G^m,
%% $$
%% where $\nz({\p i})\triangleq \{l\in\integerbox{m}:i_l\neq g_0\}.$ That is, the vector ${\p c}_m({\p i}')$ has $1$ exactly in those coordinates ${\p i}\in G^m$ whose non-$g_0$ locations are a subset of those in ${\p i}'$, and the values at the non-$g_0$ locations are equal in $\p i$ and ${\p i}'$. We think of the element $g_0$ as a `zero' element of $G$, in which case $\nz({\p i})$ is similar to the support of ${\p i}$.
We give an example to illustrate the definition of ${\p c}_m({\p i}')$.

\begin{example}
{Consider $m=3,n=3$, and $G=\{0,1,2\}$. Let ${\p i'}=(1,2,0)\in G^3$}.
The components of the vector ${\p c}_3({\p i'})\in \Fb_2^{27}$ are as follows.
\begin{align*}
   {c}_3({\p i'})_{\p i}=\begin{cases}
   1&\text{for } {\p i}\in\{(0,0,0),(1,0,0),(0,2,0),(1,2,0)\},\\
   0 & \text{otherwise}.
   \end{cases}
\end{align*}
\end{example}
%%%%%%

We are now ready to show a patterned basis for $\Dsn(r,m)$.
%%%
\begin{lemma}
\label{basisberman}
For $m\geq 1$, and $0\leq r\leq m-1$, consider the collection of elements in $\Fb_2^{n^m}$ given by
\begin{align}
B_{\Dsn}(r,m) &=\{{\p c}_m({\p i'})\colon {\p i'}\in G^m % \text{ such that } \nonumber \\
% &~~~~~~~~~~~~~~~~~~~~~~~~~r+1\leq {\wt({\p i}')} \leq m \}. \label{eqn4001}
\text{ and } {\wt({\p i}')} \geq r+1 \}. \label{eqn4001}
\end{align}
Then the collection $B_{\Dsn}(r,m)$ is a basis for $\Dsn(r,m)$.
\end{lemma}
\begin{IEEEproof}
We first show that the vectors in $B_{\Dsn}(r,m)$ are linearly independent. Let $B'$ be any non-empty subset of vectors from $B_{\Dsn}(r,m)$. Note that each vector in $B'$ is of the form ${\p c}_m({\p i'})$ for some unique ${\p i'}\in G^m$ with ${\wt({\p i}')} \geq r+1,$ by the construction of set $B_{\Dsn}(r,m)$.

We will show that the $\Fb_2$-sum of the vectors from $B'$ cannot be zero, which suffices to show that $B_{\Dsn}(r,m)$ is a linearly independent set of vectors.

Let ${\p c}_m({\p i}_d)\in B'$ {be} such that {$\wt({\p i}_d) \geq \wt({\p i}')$} for any ${\p c}_m({\p i}')\in B'$. Thus, ${\p c}_m({\p i}_d)$ is a maximal element in $B'$ in this sense.
Note that such a maximal element ${\p c}_m({\p i}_d)$ will always exist for any non-empty $B'\subset B_{\Dsn}(r,m).$

We observe the following by the definition of the vectors ${\p c}_m({\p i'})\in B_{\Dsn}(r,m).$
For any ${\p c}_m({\p i'})\in B',$ if $c_m({\p i'})_{{\p i}_d}=1,$ we must have {$\supp({\p i}_d)\subset \supp({\p i'})$ and ${\p i}'_{\supp({\p i}_d)}=({\p i}_d)_{\supp({\p i}_d)}$. By the maximality of ${\p c}_m({\p i}_d)$, we must have $\wt({\p i}_d)=\wt({\p i}')$}. Hence, by these observations, we must have that ${\p i}'={\p i}_d$.
%$c_m({\p i'})_{{\p i}_d}=0$ for any ${\p c}_m({\p i'})\in B'$ with ${\p i'}\neq {\p i}_d.$
Thus, the sum of vectors in $B'$ cannot be $\p 0$ (as the ${\p i}_d^\tth$ coordinate in the sum cannot be $0$). Thus, the vectors in $B_{\Dsn}(r,m)$ are linearly independent.

Also, we see that 
\begin{align*}
|B_{\Dsn}(r,m)|=\sum_{w=r+1}^m\binom{m}{w}(n-1)^w=\dim(\Dsn(r,m)).
\end{align*}
Thus, showing that $B_{\Dsn}(r,m)\subset\Dsn(r,m)$ will conclude the proof. {The rest of the proof is devoted to showing this statement.}

{Consider an arbitrary \mbox{${\p c}_m({\p i}')\in B_{\Dsn}(r,m)$}.  Note that $\wt({\p c}_m({\p i}'))=2^{\wt({\p i}')}\geq 2^{r+1},$ by definition}. Thus, ${\p c}_m({\p i}')\in \Dsn(0,m)$ has even weight. Thus the statement holds for $r=0$ for any $m$. Thus, the statement holds for $m=1$.

Now we prove the statement for \mbox{$r\geq 1, m\geq 2$} assuming it holds for \mbox{$m-1$}. Recall that we can use the concatenation representation for ${\p c}_m({\p i}')$ as ${\p c}_m({\p i}')=({\p c}_m({\p i}')_0|{\p c}_m({\p i}')_1|\hdots|{\p c}_m({\p i}')_{n-1})$. We consider two cases.

Case (a): \mbox{$m-1\in\supp({\p i}')$}.
Let \mbox{$i'_{m-1}={l'}\in G\setminus \{0\}$}.
Thus, for some \mbox{${\p i}\in G^m$}, if \mbox{$i_{m-1}\in G\setminus \{l',0\}$}, then \mbox{${\p c}_m({\p i}')_{\p i}=0$}. This means \mbox{${\p c}_m({\p i}')_l={\p 0}\in \Fb_2^{n^{m-1}}$} if \mbox{$l\notin \{l',0\}$}. Further if \mbox{$l\in \{l',0\}$}, then we can observe that \mbox{${\p c}_m({\p i}')_{l}={\p c}_{m-1}({\p i}'_{\integerbox{m-1}})\in \Fb_2^{n^{m-1}}$}, where we recall the notation \mbox{${\p i}'_{\integerbox{m-1}}=(i'_l:l\in\integerbox{m-1})$}.
As \mbox{$\supp({\p i}'_{\integerbox{m-1}})=\supp({\p i}')\setminus\{m-1\}$}, thus \mbox{$r\leq \wt({\p i}'_{\integerbox{m-1}}) \leq m-1$}, which means \mbox{${\p c}_{m-1}({\p i}'_{\integerbox{m-1}})\in B_{\Dsn}(r-1,m-1)$}.
By the induction hypothesis, we thus have ${\p c}_{m-1}({\p i}'_{\integerbox{m-1}})\in \Dsn(r-1,m-1)$. Further, $\sum_{l\in\integerbox{n}}{\p c}_m({\p i}')_l={\p c}_m({\p i}')_0+{\p c}_m({\p i}')_{l'}={\p 0}\in \Dsn(r,m-1)$. Thus the two conditions in the definition of $\Dsn(r,m)$ are satisfied, and thus ${\p c}_m({\p i}')\in \Dsn(r,m)$.

Case (b): {$m-1 \notin \supp({\p i}').$
In this case, a necessary condition for ${\p c}_m({\p i}')_{\p i}=1$  is that  $m-1\notin \supp({\p i})$. This means we have ${\p c}_m({\p i}')_l={\p 0}$ if $l\neq 0,$ and ${\p c}_m({\p i}')_l={\p c}_{m-1}({\p i}'_{\integerbox{m-1}})\in \Fb_2^{n^{m-1}}$ for $l=0$. Now, as $\supp({\p i}'_{\integerbox{m-1}})=\supp({\p i}')$, this means that $r+1\leq\wt({\p i}'_{\integerbox{m-1}})=\wt({\p i}')\leq m-1$.}
Hence, ${\p c}_{m-1}({\p i}'_{\integerbox{m-1}})\in B_{\Dsn}(r,m-1)$ and thus ${\p c}_{m-1}({\p i}'_{\integerbox{m-1}})\in\Dsn(r,m-1)$ by the induction hypothesis. By Lemma~\ref{containmentlemma} this means ${\p c}_{m-1}({\p i}'_{\integerbox{m-1}})\in \Dsn(r-1,m-1).$ It is thus clear that the two conditions in the definition of $\Dsn(r,m)$ are satisfied by the vector ${\p c}_m({\p i}').$ This concludes the proof.
\end{IEEEproof}
%%%%%
\begin{example}
For the code $\Ds_3(1,3)$, with the coordinates indexed by {$\{0,1,2\}^3$}, Lemma \ref{basisberman} shows that the following collection $B_{\Ds_3}(1,3)$ of $20$ vectors is a basis,
\begin{align*}
\bigcup_{a,b,c\in\{1,2\}} \!\!\! \{ {\p c}_3((a,b,0)),{\p c}_3((0,a,b)),{\p c}_3((a,0,b)),{\p c}_3((a,b,c)) \}.
\end{align*}
\end{example}

%%%%%
We now show a basis for the dual Berman code. We need a few more notations for the rest of this sub-section.
{
For any $\p{i'} \in G^m$ define $\p{d}_m(\p{i'}) \in \Fb_2^{n^m}$ as the vector with support $\{\p{i} \in G^m: \p{i} \succeq \p{i'}\}$. That is, the components of $\p{d}_m(\p{i'})$ are
\begin{equation*}
d_m({\p i'})_{\p i}=\begin{cases}
1&\text{if } \p{i} \succeq \p{i'} \\
0&\text{otherwise}.
\end{cases}
\end{equation*}
}

%% In other words, the vector ${\p c}_m(K,{\p i}')$ has component $1$ precisely in those coordinates ${\p i}\in G^m$, such that the locations of ${\p i}$ not in $K$ must match those in ${\p i}'$ (which are non-$g_0$), while those indexed by $K$ can have arbitrary values in $G$.

{
\begin{lemma}
\label{basisdualberman}
Consider the collection of elements in $\Fb_2^{n^m}$ given by
\begin{align} \label{eqn4002}
B_{\Csn}(r,m)=\left\{ \p{d}_m(\p{i'}): \p{i'} \in G^m \text{ and } \wt(\p{i'}) \leq r \right\}.
\end{align}
The collection $B_{\Csn}(r,m)$ is a basis for $\Csn(r,m)$.
\end{lemma}
}
\begin{IEEEproof}
We first note that 
\begin{equation*}
|B_{\Csn}(r,m)|=\sum_{w=0}^{r} \binom{m}{w}(n-1)^{w}=\dim(\Csn(r,m)). 
\end{equation*}
We now show that the vectors in $B_{\Csn}(r,m)$ are linearly independent.  To do this, we show that the sum of vectors in any non-empty subset $B'\subset B_{\Csn}(r,m)$ is non-zero.

Let $\p{d}_m(\p{i}_d) \in B'$ be such that for any $\p{d}_m(\p{i}') \in B'$ we have $\wt(\p{i}_d) \leq \wt(\p{i}')$. Note that for any non-empty $B'$ such a minimal element $\p{i}_d$ exists.
Now suppose that the $\p{i}_d^\tth$ component of $\p{d}_m(\p{i}') \in B'$ is equal to $1$. This implies $\p{i}_d \succeq \p{i}'$ and $\wt(\p{i}_d) \geq \wt(\p{i})$. By the minimality of $\p{i}_d$, we have $\p{i}' = \p{i}_d$. Thus, the only vector in $B'$ whose $\p{i}_d^\tth$ component is non-zero is $\p{d}_m(\p{i}_d)$. Hence, the sum of the vectors in $B'$ is non-zero.
This implies that $B_{\Csn}(r,m)$ is linearly independent.

We now show that $B_{\Csn}(r,m)\subset\Csn(r,m)$. Firstly, it is straightforward to check that the statement holds for $r=0,m$ for any $m$. Thus, the statement is also true for $m=1$. Now we will prove the statement for $m\geq 2$, \mbox{$1\leq r\leq m-1$}.
Let $\p{d}_m(\p{i}') \in B_{\Csn}(r,m)$. We will use the concatenation representation $\p{d}_m(\p{i}') = (\p{d}_0|\cdots|\p{d}_{n-1})$, where $\p{d}_l \in \Fb_2^{n^{m-1}}$ for all $l \in \lset m \rset$. We consider two cases.

Case (a): $i'_{m-1} = 0$. In this case, note that for any $\p{j} \in G^m$ we have $\p{j} \succeq \p{i}'$ if and only if $\p{j}_{\lset m-1 \rset} \succeq \p{i'}_{\lset m-1 \rset}$. Hence,
\begin{equation*}
{d}_m(\p{i}')_{\p{j}} =
\begin{cases}
1 & \text{ if } \p{j}_{\lset m-1 \rset} \succeq \p{i'}_{\lset m-1 \rset}, \\
0 & \text{ otherwise}.
\end{cases}
\end{equation*}
Thus, we observe that $\p{d}_l = \p{d}_{m-1}\left(\p{i}'_{\lset m-1 \rset} \right)$ for all $l \in \lset n \rset$. Since $i'_{m-1}=0$, we have $\wt(\p{i}'_{\lset m-1 \rset}) \leq r$. By induction hypothesis we have $\p{d}_0=\cdots=\p{d}_{n-1} \in \Csn(r,m-1)$. Thus, $\p{d}_m(\p{i}')$ satisfies the properties of a codeword of $\Csn(r,m)$.

Case (b): $i'_{m-1} = l' \in G \setminus \{0\}$. In this case we have
\begin{equation*}
{d}_m(\p{i}')_{\p{j}} =
\begin{cases}
1 & \text{ if } \p{j}_{\lset m-1 \rset} \succeq \p{i'}_{\lset m-1 \rset} \text{ and } j_{m-1}=l', \\
0 & \text{ otherwise}.
\end{cases}
\end{equation*}
This implies that $\p{d}_l = 0$ for all $l \neq l'$, and $\p{d}_{l'} = \p{d}_{m-1}(\p{i}'_{\lset m-1 \rset})$. Since $i'_{m-1} \neq 0$, we have $\wt(\p{i}'_{\lset m-1 \rset}) \leq r-1$. Hence, by induction hypothesis, $\p{d}_{l'} = \p{d}_{m-1}(\p{i}'_{\lset m-1 \rset}) \in \Csn(r-1,m-1)$. Clearly, $\p{d}_m(\p{i}')$ satisfies the properties of a codeword in $\Csn(r,m)$.
This concludes the proof.
\end{IEEEproof}

% % % % % BEGIN Remark on the bases % %

The bases presented in Lemmas~\ref{basisberman} and~\ref{basisdualberman} for the Berman code and its dual are related to each other, and arise as rows and columns, respectively, of a single $n^m \times n^m$ binary matrix.
% For ease of exposition, let us say that a tuple $\p{j} \in G^m$ is \emph{contained} in the tuple $\p{i} \in G^m$ if $\nz(\p{j}) \subset \nz(\p{i})$ and $\p{j}_{\nz(\p{j})} = \p{i}_{\nz(\p{j})}$.
{To see this, note that for any choice of $\p{i}, \p{j} \in G^m$, the $\p{j}^\tth$ component of the vector $\p{c}_m(\p{i})$ has value ${c}_m(\p{i})_{\p{j}} = 1$ if and only if $\p{j} \preceq \p{i}$. Again, the $\p{i}^\tth$ component of $\p{d}_m(\p{j})$ is equal to $1$ if and only if $\p{j} \preceq \p{i}$.}
Hence, we have
\begin{equation*}
c_m(\p{i})_{\p{j}} = d_m(\p{j})_{\p{i}}
=
\begin{cases}
1, \text{ if } \p{j} \preceq \p{i},\\
0, \text{ otherwise },
\end{cases}
\text{ for all } \p{i}, \p{j} \in G^m.
\end{equation*}
This observation leads us to define a binary $n^m \times n^m$ matrix $\p{A}_m = [A_m(\p{i},\p{j})]$ whose rows and columns are indexed by tuples from $G^m$, with its entry in row $\p{i}$ and column $\p{j}$ defined as
\begin{equation*}
A_m(\p{i},\p{j})
=
\begin{cases}
1, \text{ if } \p{j} \preceq \p{i},\\
0, \text{ otherwise }.
\end{cases}
\end{equation*}
We see that $\p{c}_m(\p{i})$ is the $\p{i}^\tth$ row of $\p{A}_m$ and $\p{d}_m(\p{j})$ is the $\p{j}^\tth$ column of $\p{A}_m$.
% Using this with Lemmas~\ref{basisberman} and~\ref{basisdualberman},
We thus arrive at the following result.

\begin{corollary} \label{cor:bases}
For any $m \geq 1$ and $0 \leq r \leq m$,
\begin{enumerate}
\item the rows of $\p{A}_m$ indexed by $\{\p{i} : \wt(\p{i}) \geq r+1 \}$ (or equivalently, the rows of $\p{A}_m$ with Hamming weight at least $2^{r+1}$) form a basis for $\Dsn(r,m)$, and
\item the columns of $\p{A}_m$ indexed by $\{\p{j}: \wt(\p{j}) \leq r\}$ (equivalently, the columns of $\p{A}_m$ with Hamming weight at least $n^{m-r}$) form a basis for $\Csn(r,m)$.
\end{enumerate}
\end{corollary}
\begin{IEEEproof}
These claims follow immediately from Lemmas~\ref{basisberman} and~\ref{basisdualberman} and the facts $\wt(\,\p{c}_m(\p{i})\,) = 2^{\wt(\p{i})}$ and $\wt(\, \p{d}_m(\p{j}) \,) = n^{m - \wt(\p{j})}$.
\end{IEEEproof}

The matrix $\p{A}_m$ can be defined recursively using $\p{A}_{m-1}$.
To show this recursion, we impose the following colexicographic order on the elements of $G^m$.
For $m=1$, we impose the {natural order $0 < 1 < \cdots < {n-1}$.}
For $m \geq 2$, and distinct $\p{i},\p{i}' \in G^m$, we {define} $\p{i} < \p{i}'$ if and only if 
\begin{equation*}
% \p{i} < \p{i}' \text{ if and only if either } 
i_{m-1} < i'_{m-1}, \text{ or } i_{m-1} = i'_{m-1} \text{ and } \p{i}_{\lset m-1 \rset} < \p{i}'_{\lset m-1 \rset}.
\end{equation*}
We assume that the rows and columns of $\p{A}_m$ are indexed by the elements of $G^m$ in the ascending order (from left to right for columns, and top to bottom for rows).
Now using 
% the facts $A_m(\p{i},\p{j}) = A_{m-1}(\p{i}_{\lset m-1 \rset},\p{j}_{\lset m-1 \rset})$ if $j_{m-1} \in \{0,i_{m-1}\}$ and $A_m(\p{i},\p{j}) = 0$ otherwise.
\begin{align*}
A_m(\p{i},\p{j})
=
\begin{cases}
A_{m-1}(\p{i}_{\lset m-1 \rset},\p{j}_{\lset m-1 \rset}), &\text{if } j_{m-1} \in \{0,i_{m-1}\},\\
0, &\text{otherwise},
\end{cases}
\end{align*}
we arrive at the recursive structure of $\p{A}_m$ given below
\begin{align*}
\p{A}_m &=
\begin{pmatrix}
\p{A}_{m-1} & \p{0} & \p{0} & \cdots & \p{0} \\
\p{A}_{m-1} & \p{A}_{m-1} & \p{0} & \cdots & \p{0} \\
\p{A}_{m-1} & \p{0} & \p{A}_{m-1} & \cdots & \p{0} \\
\vdots & \vdots &  & \ddots & \vdots \\
\p{A}_{m-1} & \p{0} & \p{0} & \cdots & \p{A}_{m-1}
\end{pmatrix},
\text{ and} \\
\p{A}_1 &=
\begin{pmatrix}
1 & {0} & {0} & \cdots & {0} \\
1 & 1 & {0} & \cdots & {0} \\
1 & {0} & 1 & \cdots & {0} \\
\vdots & \vdots &  & \ddots & \vdots \\
1 & {0} & {0} & \cdots & 1
\end{pmatrix} \in \Fb_2^{n \times n}.
\end{align*}
Note that $\p{A}_m = \p{A}_1^{\otimes m}$ where the exponent $\otimes m$ denotes the $m$-fold Kronecker product of a matrix with itself.
When $n=2$, part~2 of Corollary~\ref{cor:bases} yields the well known characterization of RM codes in terms of the columns of $\begin{pmatrix}1 & 0 \\ 1 & 1 \end{pmatrix}^{\otimes m}$.

% % % % % END Remark on the bases % %

% % % % % Decoding BEGINS % % % % %

\subsection{Recursive Decoding} \label{sec:sub:majority-decoding}

% % R0 % %
{
The recursive structure of the codes $\Dsn(r,m)$ and $\Csn(r,m)$ allow us to perform efficient bounded distance decoding up to half the minimum distance using techniques similar to RM decoding~\cite{ScB_IT_95,DuS_ISIT_2000}.
For dual Berman codes, we use a decoder known for matrix-product codes~\cite[Section~4]{HeR_JAlgebra_13}.
% \pk{Please check from here}
Our decoder for Berman codes is similar to the idea presented in~\cite[Section~3]{HeR_JAlgebra_13} for decoding $(\p{u}|\p{u} + \p{v})$ Plotkin construction, however not identical to it. The idea of this algorithm is to first decode $\p{v}$ by cancelling the effect of $\p{u}$, and then decode the two independent copies of $\p{u}$ in the codeword to obtain two estimates of $\p{u}$, and finally choose one of these two estimates of $\p{u}$ via minimum distance decoding.

Since we use these decoders recursively, a key step in our process is to show that the outputs of these decoders are valid codewords irrespective of the number of errors introduced in the channel.
We also present complexity analyses for both the decoders. These two aspects of our presentation here are novel compared to \cite{HeR_JAlgebra_13}.
}
% % % % %

We will assume that $\p{v} \in \Fb_2^{n^m}$ is a codeword transmitted through a noisy binary-input binary-output channel and $\p{y} \in \Fb_2^{n^m}$ is the channel output. The number of errors introduced by the channel is the Hamming distance between $\p{v}$ and $\p{y}$, which is
\begin{equation*}
d(\p{v},\p{y}) = \left| \left\{ i \in \lset n^m \rset~:~v_{i} \neq y_{i} \right\} \right|.
\end{equation*}
% We consider the scenario where $d(\p{v},\p{y})$ is less than half the minimum distance of the code, and provide algorithms to decode $\p{y}$ to the transmitted codeword $\p{v}$.

\subsubsection{Recursive Decoding for $\Cs_n(r,m)$}

For $r=0$ and $r=m$, we perform minimum distance decoding of the channel output to the codebook $\Cs_n(r,m)$.
% % R0 version % %
{
When $1 \leq r \leq m-1$, the decoder $\Dec_{\Csn,r,m}$ for $\Csn(r,m)$ will use $\Dec_{\Csn,r,m-1}$ and $\Dec_{\Csn,r-1,m-1}$ as subroutines.
In this case, we will assume that the transmitted codeword is
\begin{equation*}
\p{v} = (\p{u}|\cdots|\p{u}|\p{u}) + (\p{u}_0|\cdots|\p{u}_{n-2}|\p{u}_{n-1})
\end{equation*}
where $\p{u} \in \Csn(r,m-1)$, $\p{u}_0,\dots,\p{u}_{n-2} \in \Csn(r-1,m-1)$ and $\p{u}_{n-1}=\p{0}$.
The decoder will first decode $\p{u}_0,\dots,\p{u}_{n-2}$, remove their effect in the channel output $\p{y}$, and then decode $\p{u}$.
% We prove the correctness of the decoder (by induction on the parameter $m$) when the number of errors introduced in the channel is less than $n^{m-r}/2$.

% The key property used in our induction argument will be the fact that the output of the decoder is a valid codeword for any channel output (irrespective of the number of errors introduced in the channel).

% For the boundary cases $r=0$ (the repetition code) and $r=m$ (the universe code), $\Dec_{\Csn,r,m}$ is the minimum distance decoder.
It is clear that for the boundary cases $r=0$ and $r=m$ (the repetition code and the universe code), the output of $\Dec_{\Csn,r,m}$ is a valid codeword for any channel output, and the decoder outputs the correct codeword if the number of channel errors is less than half the minimum distance.
Let us now assume that $m \geq 2$, and $1 \leq r \leq m-1$.
We will denote the channel output by $\p{y}=(\p{y}_0|\cdots|\p{y}_{n-1})$.
% Now consider the first stage of the decoder where we decode $\p{u}_0,\dots,\p{u}_{n-2}$.
For each $l \in \lset n-1 \rset$,
% note that $\p{u}_l$ is a codeword of $\Csn(r-1,m-1)$ which has minimum distance $n^{m-r}$. To
we decode $\p{u}_l$ using $\Dec_{\Csn,r-1,m-1}(\p{\tilde{y}}_l)$ where
\begin{equation*}
\p{\tilde{y}}_l \triangleq \p{y}_l + \p{y}_{n-1} % = \p{u}_l + \left( (\p{y}_l + \p{u}_l) + (\p{y}_{n-1} + \p{u}) \right)
\end{equation*}
is a noisy version of $\p{u}_l$. % and satisfies $d(\p{u}_l,\p{\tilde{y}}_l) < n^{m-r}/2$, which is half the distance of $\Csn(r-1,m-1)$.
% We will decode $\p{u}_l$ using $\Dec_{\Csn,r-1,m-1}(\p{\tilde{y}}_l)$.
If the number of channel errors in $\p{y}$ is less than $n^{m-r}/2$, then the effective number of channel errors in $\p{\tilde{y}}_l$ is also less than $n^{m-r}/2$, which is half the minimum distance of $\Csn(r-1,m-1)$.
By induction, if $\p{y}$ contains less than $n^{m-r}/2$ errors then the output of $\Dec_{\Csn,r-1,m-1}(\p{\tilde{y}}_l)$ is correct (that is, equal to $\p{u}_l$), otherwise $\Dec_{\Csn,r-1,m-1}(\p{\tilde{y}}_l)$ will be some codeword of $\Csn(r-1,m-1)$.
}

% Hence, by induction on $m$, we can assume that $\Dec_{\Csn,r-1,m-1}(\p{\tilde{y}}_l)$ will successfully decode $\p{u}_l$ for each $l \in \lset n-1 \rset$.

In the next stage, we decode $\p{u}$ via $\Dec_{\Csn,r,m-1}(\p{y}_l + \p{u}_l)$ for $l \in \lset n \rset$, where $\p{u}_{n-1} = \p{0}$.
% and $\Dec_{\Csn,r,m-1}(\p{y}_{n-1})$.
% These $n$ subroutine calls will each return $n^{m-r-1}$ votes for every bit of $\p{u}$. Combining these votes will yield $n^{m-r}$ votes for each bit of $\p{u}$.
Now consider the scenario where the number of channel errors in $\p{y}$ less than $n^{m-r}/2$. We know that the receiver has correctly decoded $\p{u}_0,\dots,\p{u}_{n-2}$. Hence, $\p{y}_l + \p{u}_l$ is a noisy version of $\p{u}$, and the total number of channel errors in $\p{y}$ is
\begin{equation*}
\sum_{l \in \lset n \rset} d(\p{y}_l + \p{u}_l, \p{u})<\frac{n^{m-r}}{2}.
\end{equation*}
Hence, there will be at least once choice of $l$ such that 
\begin{equation*}
d(\p{y}_l + \p{u}_l, \p{u}) < n^{m-r-1}/2, 
\end{equation*}
and by induction, $\Dec_{\Csn,r,m-1}(\p{y}_l + \p{u}_l)=\p{u}$ for this $l$.
Our decoder will look for an $l$ such that
\begin{equation*}
\sum_{k \in \lset n \rset} d(\Dec_{\Csn,r,m-1}(\p{y}_l + \p{u}_l),\p{y}_k + \p{u}_k) < \frac{n^{m-r}}{2}
\end{equation*}
and use $\Dec_{\Csn,r,m-1}(\p{y}_l + \p{u}_l)$ as the decoded value of $\p{u}$ if such an $l$ exists. If more than one such $l$ exists, each of these choices of $l$ will yield the correct value of $\p{u}$ (this follows from the fact that $\Dec_{\Csn,r,m-1}(\p{y}_l + \p{u}_l) \in \Csn(r,m-1)$, and the minimum distance of the concatenation of $\Csn(r,m-1)$ and the $n$-length repetition code is $n^{m-r}$).
If no such $l$ exists (this could happen when the number of errors in $\p{y}$ is at least $n^{m-r}/2$), the decoder uses $\Dec_{\Csn,r,m-1}(\p{y}_{n-1})$ (which, even if incorrect, will still belong to $\Csn(r,m-1)$) as the decoded value of $\p{u}$.
Finally, the decoder will output $(\p{u}|\p{u}|\cdots|\p{u}) + (\p{u}_0|\cdots|\p{u}_{n-2}|\p{0})$, which belongs to $\Csn(r,m)$ irrespective of the number of errors in the channel output $\p{y}$.

% The terminal cases for the recursive calls are $r=0$ and $r=m$. Since $\Csn(0,m)$ is the repetition code, the $n^m$ components of $\p{y}$ directly yield $n^m$ votes for the single transmitted message bit $v_0$.
% For the case $r=m$, we note that $\Csn(m,m)=\Fb_2^{n^m}$ and has distance $n^{m-m}=1$. Each bit $v_i$ of $\p{v}$ receives exactly one vote, which is dependent on the corresponding channel output $y_i$.

\begin{center}
\begin{algorithm}
\caption{$\Dec_{\Csn,r,m}$ algorithm for decoding $\Csn(r,m)$}
{\bf Input}: Channel output $\p{y} \in \Fb_2^{n^m}$

{\bf Output}: A codeword $\p{v} \in \Csn(r,m)$

% \vspace*{0.05in}
\begin{algorithmic}[1]

\If{$r=0$}

% \For{each $i \in \lset n^m \rset$}

% \State{Count one vote for ${v}_0$ in favor of $y_i$}

% \EndFor

% \Return{$n^m$ votes collected for message bit $v_0$}

\State{$\p{v} \gets \p{0}$}

\If{$\wt(\p{y}) > n^{m}/2$}

\State{$\p{v} \gets (1~1~\cdots~1)$}

\EndIf

\Return{$\p{v}$}

\EndIf

\If{$r=m$}

\State{$\p{v} \gets \p{y}$}

\Return{$\p{v}$}

% \For{each $i \in \lset n^m \rset$}

% \State{Count one vote for $v_i$ in favor of $y_i$}

% \EndFor

% \Return{the one vote collected for each bit of $\p{v}$}

\EndIf

\If{$1 \leq r \leq m-1$}

\For{each $l \in \lset n-1 \rset$}

\State{$\p{\tilde{y}}_l \gets \p{y}_l + \p{y}_{n-1}$}

% \State{Decode $\p{u}_l$ using $\Dec_{\Csn,r,m}(\p{\tilde{y}}_l)$}

\State{$\p{u}_l \gets \Dec_{\Csn,r-1,m-1}(\p{\tilde{y}}_l)$}

\State{$\p{y'}_l \gets \p{y}_l + \p{u}_l$}

\EndFor

\State{$\p{u}_{n-1} \gets \p{0}$}

\State{$\p{y'}_{n-1} \gets \p{y}_{n-1}$}

\For{each $l \in \lset n \rset$}

% \State{$\p{y'}_l \gets \p{y}_l + \p{u}_l$}

\State{$\p{u} \gets \Dec_{\Csn,r,m-1}(\p{y'}_l)$}

\If{$\sum_{k \in \lset n \rset} d(\p{u},\p{y'}_k) < n^{m-r}/2$}

\State{{\bf break}}

\EndIf

\EndFor

% \State{Collect $n^{m-r-1}$ votes for each bit of $\p{u}$ via $\Dec_{\Csn,r,m}(\p{y}_{n-1})$}

% \For{$l \in \lset n-1 \rset$}

% \State{Collect $n^{m-r-1}$ more votes for each bit of $\p{u}$ via $\Dec_{\Csn,r,m}(\p{y}_{l} + \p{u}_l)$}

% \EndFor

\State{$\p{v} \gets (\p{u}|\p{u}|\cdots|\p{u}) + (\p{u}_0|\p{u}_1|\cdots|\p{u}_{n-1})$}

\Return{$\p{v}$}

\EndIf

\end{algorithmic}
\end{algorithm}
\end{center}

The decoder $\Dec_{\Csn,r,m}$ is summarized in Algorithm~1. We now derive an upper bound on the complexity of this decoder.
We will denote the complexity of this algorithm by $g(r,m)$ and derive an upper bound of the form $b_m n^m$ on $g(r,m)$, where $b_m$, $m \geq 1$, is a constant.
Clearly, $g(0,m),g(m,m) \leq n^m$, and $b_1=1$ is a valid choice for upper bounding $g(0,1)$ and $g(1,1)$.
Now consider an arbitrary $m \geq 2$ and $1 \leq r \leq m-1$.
The complexity of steps 11--15 is at the most 
\begin{align*}
n(2n^{m-1} + g(r-1,m-1)).
\end{align*}
Steps 18--23 incur at the most $n(g(r,m-1) + n^m)$ computations. Finally, step~24 uses less than $n^m$ XORs.
Hence, we have
\begin{align*}
g(r,m) &\leq 3n^m + n^{m+1} + n(g(r-1,m-1) + g(r,m-1)) \\
&\leq (3+n+2b_{m-1}) n^m.
\end{align*}
Hence, we choose $b_1=1$ and $b_m = (3+n+2b_{m-1})$ to arrive at the bound $g(r,m) \leq b_m n^m$ for all $r$ and $m$.
The solution to this linear recursion is $b_m=2^{m-1}(n+4) - (n+3)$.
Thus, we have the bound
\begin{equation*}
g(r,m) \leq b_m n^m \leq 2^{m} \frac{(n+4)}{2} n^m,
\end{equation*}
which is $O(N^{1 + \log_n 2})$, where $N$ is the code length.

\subsubsection{Recursive Decoding for $\Dsn(r,m)$}

% Our decoder for $\Dsn(r,m)$ is defined recursively using the decoders of $\Dsn(r,m-1)$ and $\Dsn(r-1,m-1)$.

% % R0 % % %
{
We will denote the decoding function for $\Dsn(r,m)$ as $\Dec_{\Dsn,r,m}$.
}
% % % % % %
% Similar to the decoder for $\Csn(r,m)$, the key property of $\Dec_{\Dsn,r,m}$ is that its output is always a codeword of $\Dsn(r,m)$ and this output is the transmitted codeword if the number of channel errors is less than $2^r$.
%
For the boundary cases $r=0$ (the single parity-check code) and $r=m$ (the all-zero code), we perform minimum distance decoding.
We will now assume $1 \leq r \leq m-1$.
While decoding the channel output $\p{y}$ to $\Dsn(r,m)$, we are searching for the transmitted codeword $\p{v}=(\p{v}_0|\cdots|\p{v}_{n-1})$ with each $\p{v}_{l} \in \Dsn(r-1,m-1)$ and $\p{v}_{\rm sum} \triangleq \sum_{l}\p{v}_{l} \in \Dsn(r,m-1)$.
Let us denote the channel output as $\p{y}=(\p{y}_0|\cdots|\p{y}_{n-1})$.
We have assumed that the number of channel errors is less than $2^r$. Hence, we observe that the distance between $\p{y}_{\rm sum} \triangleq \sum_{l}\p{y}_{l}$ and $\p{v}_{\rm sum}$ is less than $2^r$.
When $r \leq m-2$, since the minimum distance of $\Dsn(r,m-1)$ is $2^{r+1}$, we observe that (by using an induction argument on the parameter $m$) we can use $\Dec_{\Dsn,r,m-1}(\p{y}_{\rm sum})$ to decode $\p{v}_{\rm sum}$.
When $r = m-1$, we know that $\Dsn(m-1,m-1)=\{\p{0}\}$, and hence, $\p{v}_{\rm sum} = \p{0}$.

We now assume that the receiver knows $\p{v}_{\rm sum}$. Observe that for each $l \in \lset n-1 \rset$, we now have two noisy versions of $\p{v}_l$, which are
\begin{equation*}
\p{y}_l \text{ and } \p{\tilde{y}}_l \triangleq \p{v}_{\rm sum} + \sum_{k \neq l} \p{y}_k.
\end{equation*}
Since the total number of channel errors in $\p{y}$ is less than $2^r$, at least one of $\p{y}_l$ and $\p{\tilde{y}}_l$ is within a distance of $2^{r-1} - 1$ to $\p{v}_l$. By induction, we then see that at least one of
\begin{equation*}
\p{\hat{v}}_l(0) \triangleq \Dec_{\Dsn,r-1,m-1}(\p{y}_l), \,
\p{\hat{v}}_l(1) \triangleq \Dec_{\Dsn,r-1,m-1}(\p{\tilde{y}}_l),
\end{equation*}
is equal to $\p{v}_l$ and the other belongs to $\Dsn(r-1,m-1)$.
We thus see that $\p{v}$ belongs to the below collection of $2^{n-1}$ vectors if the number of channel errors in $\p{y}$ is less than $2^r$,
\begin{align*}
\mathscr{L} \triangleq \Big\{ & \Big( \p{\hat{v}}_0(a_0) \Big| \cdots \Big| \p{\hat{v}}_{n-2}(a_{n-2}) \Big| \p{v}_{\rm sum} + \sum_{l \in \lset n-1 \rset} \p{\hat{v}}_l(a_l)  \Big) : \\
&~~~~~~~~~~~~~~~~~~~(a_0,\dots,a_{n-2}) \in \{0,1\}^{n-1}  \Big\}.
\end{align*}
Note that every vector in $\mathscr{L}$ is a codeword from $\Dsn(r,m)$ irrespective of the number of channel errors in $\p{y}$.
This is because, by induction, we have that $\p{v}_{\rm sum} \in \Dsn(r,m-1)$, each $\p{\hat{v}}_l(a_l) \in \Dsn(r-1,m-1)$ and the structure of the vectors in $\mathscr{L}$ follows the definition of $\Dsn(r,m)$.
Thus, we deduce that performing a minimum distance decoding among the vectors in $\mathscr{L}$ will return the correct codeword if the number of errors is less than $2^r$, and this will return some codeword of $\Dsn(r,m)$ otherwise.
This recursive decoding technique for $\Dsn(r,m)$ is summarized in Algorithm~2.

\begin{center}
\begin{algorithm}
\caption{$\Dec_{\Dsn,r,m}$ algorithm for decoding $\Dsn(r,m)$}
{\bf Input}: Channel output $\p{y} \in \Fb_2^{n^m}$

{\bf Output}: A codeword $\p{v} \in \Dsn(r,m)$

% \vspace*{0.05in}
\begin{algorithmic}[1]

\If{$r=m$}

\State{$\p{v} \gets \p{0}$}

\Return{$\p{v}$}

\EndIf

\If{$r=0$}

\State{$\p{v} \gets \p{y}$}

\If{$\wt(\p{v})$ is odd}

\State{Flip the first bit of $\p{v}$}

\EndIf

\Return{$\p{v}$}

\EndIf

\If{$1 \leq r \leq m-1$}

\State{$\p{y}_{\rm sum} \gets \sum_{l \in \lset n \rset} \p{y}_{l}$}

\State{$\p{v}_{\rm sum} \gets \Dec_{\Dsn,r,m-1}(\p{y}_{\rm sum})$}

\For{each $l \in \lset n-1 \rset$}

\State{$\p{\tilde{y}}_l \gets \p{v}_{\rm sum} + \p{y}_{\rm sum} + \p{y}_l$}

\State{$\p{\hat{v}}_l(0) \gets \Dec_{\Dsn,r-1,m-1}(\p{y}_l)$}

\State{$\p{\hat{v}}_l(1) \gets \Dec_{\Dsn,r-1,m-1}(\p{\tilde{y}}_l)$}

\EndFor

\State{$\p{v} \gets$ Minimum distance decode $\p{y}$ to $\mathscr{L}$}

\Return{$\p{v}$}

\EndIf

\end{algorithmic}
\end{algorithm}
\end{center}

We now derive an upper bound on the complexity of this algorithm.
Let $f(r,m)$ denote the number of computations incurred by $\Dec_{\Dsn,r,m}$.
We see that $f(0,m) \leq n^m$ and $f(m,m)=0$. We will derive a loose upper bound of the form
\begin{equation} \label{eq:dec_rm:dn:1}
f(r,m) \leq c_m n^m, \text{ for all } r \in \lset m+1 \rset,
\end{equation}
where the constants $c_m$, $m \geq 1$, are appropriately chosen.
Clearly, $f(r,1) \leq n$ for all $r \in \lset 2 \rset$, and hence, we choose $c_1=1$.

We now derive a recursive upper bound on $f(r,m)$ when \mbox{$1 \leq r \leq m-1$}.
We see that step~11 incurs $n^m$ computations, step~12 uses $f(r,m-1)$ computations, steps~13-17 have complexity at the most $n(2n^{m-1} + 2f(r-1,m-1))$, and step~18 uses $2^{n-1} n^m$ computations.
Computing all the vectors in $\mathscr{L}$ takes at the most $2^{n-1}n^m$ further computations.
Combining these terms we arrive at the following upper bound
\begin{align*}
f(r,m) &\leq n^m + f(r,m-1) + 2^{n-1} n^m + 2^{n-1}n^m \\ 
&~~~~~~~~~~~~~~~~~+ n\left(2f(r-1,m-1) + 2n^{m-1} \right) \\
&\leq (3 + 2^{n})n^m + c_{m-1}n^{m-1} + 2c_{m-1}n^m \\
&\leq \left(3 + 2^{n} + c_{m-1}\left(2 + \frac{1}{n}\right) \right) n^m.
\end{align*}
Thus, we choose $c_m = 3 + 2^{n} + c_{m-1}(2 + \frac{1}{n})$ to satisfy~\eqref{eq:dec_rm:dn:1}, with $c_1=1$.
This is a linear recurrence with constant coefficients, and its solution
$c_m$ grows with $m$ as $O\left(\left(2+\frac{1}{n}\right)^m \right)$.
Thus, the complexity of $\Dec_{\Dsn,r,m}$ is at the most $c_m n^m$, which is $O\left(N^{1 + \log_{n}\left(2 + \frac{1}{n} \right)}\right)$ since $N = n^m$.

% % % % % Decoding ENDS % % % %

\section{Capacity-Related Properties} \label{sec:capacity-related}

Reeves and Pfister~\cite{ReP_RM_BMS_2021} recently showed that Reed-Muller (RM) codes achieve the capacity of binary-input memoryless symmetric (BMS) channels.  They rely on the following properties of the RM codes in their work.
% We denote the RM code of length $2^m$ and order $r$ as $\RM(r,m)$.
\begin{enumerate}
\item For every $R^* \in (0,1)$ there exists a sequence of RM codes of increasing block lengths with rates converging to $R^*$.
\item RM codes are transitive (that is, for any chosen pair of coordinates, there is an automorphism that maps the first coordinate to the second), and more importantly, RM codes are doubly transitive (that is, for any three distinct coordinates $i,j,k$, there is a code automorphism that fixes $i$ and sends $j$ to $k$).
\item For $1 \leq k \leq m$ and $r \leq m-k$, we can puncture $\RM(r,m+k)$ to obtain $\RM(r,m)$. Further, there are multiple ways to do this puncturing. In particular, suppose we use the elements of $\Fb_2^{m+k}$ to represent the coordinates of the codewords in $\RM(r,m+k)$.
For any $H \subset \Fb_2^{m+k}$ and any vector $\p{v}=(v_{\p{i}}: \p{i} \in \Fb_2^{m+k}) \in \Fb_2^{2^{m+k}}$, define the puncturing operation $\Ps_H$ as $\Ps_H(\p{v}) = (v_{\p{i}}:\p{i} \in H)$.
Let $H_1 = \Fb_2^{m} \times \{0\}^k$ and $H_2 = \Fb_2^{m-k} \times \{0\}^k \times \Fb_2^{k}$. Then,
\begin{align*}
\Ps_{H_1}\left( \RM(r,m+k) \right) &= \RM(r,m), \\
\Ps_{H_2}\left( \RM(r,m+k) \right) &= \RM(r,m), \text{ and } \\
\Ps_{H_1 \cap H_2} \left( \RM(r,m+k) \right) &= \RM(r,m-k),
\end{align*}
see Lemma~8 of~\cite{ReP_RM_BMS_2021}.
\item For long RM codes the rate change due to puncturing is small. Let us denote the rate of $\RM(r,m)$ as $R_{\RM}(r,m)$. For any $r \leq m$ and $k \geq 1$, we have
\begin{equation*}
0 \leq R_{\RM}(r,m) - R_{\RM}(r,m+k) \leq \frac{3k+4}{5\sqrt{m}}.
\end{equation*}
Reeves and Pfister derive this bound and exploit the fact that this rate change is $O(k/\sqrt{m})$.
\end{enumerate}

In this section we show that $\Dsn(r,m)$ and $\Csn(r,m)$ satisfy all these properties except double transitivity when $n \geq 3$.
The proof that these codes are not doubly transitive is based on the observation in \cite{Von_Thesis_ETH} that the product of the minimum distances of a code and its dual is much smaller than the block length.

Since $\Dsn(r,m)$ and $\Csn(r,m)$ are not doubly transitive, we can not use~\cite[Theorem~20]{KKMPSU_IT_17} for capacity-achievability in the BEC.
Fortunately, these codes do satisfy the weaker sufficient conditions put forth in~\cite[Theorem~19]{KCP_ISIT16} for a code sequence to achieve the BEC capacity.
(In fact, for all odd $n$, we are able to show that a broader class of abelian codes are capacity-achieving in the BEC under bit-MAP decoding, see Section~\ref{sec:dft}).

% % % R0 % % % % %
%%%%%%%%
\subsection{Some Useful Automorphisms of $\Dsn(r,m)$ and $\Csn(r,m)$}
\label{subsec:automorphism}
In this sub-section, we obtain some automorphisms of the code $\Dsn(r,m)$.
Note that since the automorphism groups of a code and its dual are the same, these apply to the code $\Csn(r,m)$ also.
{
These automorphisms will be used to show capacity achievability in the BEC, and to analyse the puncturing properties of Berman and dual Berman codes.
}%

We use the scheme in Section \ref{subsubsec:usefulbasis} that identifies the coordinates of vectors in $\Fb_2^{n^m}$ with $G^m$ (for any set $G$ with $n$ elements) for specifying the automorphisms of the code $\Dsn(r,m)$.
\begin{lemma}
\label{lemma:group-permutationautomorphism}
Let $\sigma$ be any permutation of the set $G$. The following permutation $\pi_{m-1,\sigma}$ on the $m$-tuples in $G^m$ is an automorphism of $\Dsn(r,m)$,
\begin{align*}
    \pi_{m-1,\sigma}\colon (i_0,\hdots,i_{m-2},i_{m-1})\mapsto (i_0,\hdots,i_{m-2},\sigma(i_{m-1})).
\end{align*}
\end{lemma}
%%%%
\begin{IEEEproof}
Let ${\p v}=({\p v}_0|\hdots|{\p v}_{n-1})$ be an arbitrary codeword in $\Dsn(r,m)$. We want to show that the vector ${\p v}'$, with coordinates $ v'_{\p i}=v_{\pi_{m-1,\sigma}({\p i})}$, lies in $\Dsn(r,m)$.

To see this, observe that if we write ${\p v}'$ as $({\p v}'_0|\hdots|{\p v}'_{n-1})$, then by the notation developed in Section \ref{subsubsec:usefulbasis},  for any $l\in\integerbox{n}$ we have that ${\p v}'_l={\p v}_{l'}$,
{for precisely that unique $l'$ such that $\sigma({l'})=l$}. Thus, the subvectors of ${\p v}'$ are precisely the same as those in ${\p v}$, only their positions are permuted. Thus ${\p v}'$ satisfies the two conditions in the definition of $\Dsn(r,m)$. Hence ${\p v}'\in \Dsn(r,m)$, which completes the proof.
\end{IEEEproof}
%%%
\begin{lemma}
\label{lemma:exchangeautomorphism}
For any $t\in\integerbox{m-1}$, the permutation $\beta_t$ on $G^m$ defined below is an automorphism of $\Dsn(r,m)$ and $\Csn(r,m)$.
\begin{align*}
    \beta_t\colon (i_0,\hdots,i_{m-1})\mapsto (i_0,\hdots,i_{t-1},i_{m-1},i_{t+1},\hdots,i_{m-2},i_t).
\end{align*}
\end{lemma}
%%%%
\begin{IEEEproof}
Clearly, the statement is true for $r=m$. Hence we assume $r\leq m-1$. Recalling the definition of the set $B_{\Dsn}(r,m)$ in (\ref{eqn4001}), to show the lemma, it is sufficient to show that for each ${\p c}_m({\p i}')\in B_{\Dsn}(r,m),$ the permuted vector ${\p c}'$ defined below is also in $B_{\Dsn}(r,m).$
\begin{align}
    \label{eqn0101}
    c'_{\p i}=c_m({\p i}')_{\beta_t({\p i})},~\forall {\p i}\in G^m.
\end{align}
We shall in fact prove that ${\p c}'={\p c}_m(\beta_t({\p i}'))
).$
The proof will then be complete as $\beta_t$ is a {one-to-one} map.

Firstly we observe that the following two statements are equivalent for any ${\p i}\in G^m$, because $\beta_t$ is a self-inverse permutation.
\begin{itemize}
\item {$\beta_t(\p{i}) \preceq \p{i}'$, i.e., $\supp(\beta_t({\p i})) \subset \supp({\p i}')$ and $\beta_t({\p i})_{\supp(\beta_t({\p i}))}={\p i}'_{\supp(\beta_t({\p i}))}$, are both true.}
\item {$\p{i} \preceq \beta_t(\p{i}')$, i.e., $\supp({\p i}) \subset \supp(\beta_t({\p i}'))$ and ${\p i}_{\supp({\p i})}=\beta_t({\p i}')_{\supp({\p i})}$, are both true.}
\end{itemize}

{Now, using the above equivalence and~\eqref{eqn0101}, we have,}
%ONECOLUMN
\begin{align}
\label{eqn0102}
c'_{\p i}=
\begin{cases}
1&\text{if } {\p{i} \preceq \beta_t(\p{i}')}\\
0&\text{otherwise},
\end{cases}
\end{align}
%TWOCOLUMN
% \begin{align}
% \label{eqn0102}
% c'_{\p i}=
% \begin{cases}
% 1&\text{if } \nz({\p i})\subset \nz(\beta_t({\p i}')) \text{ and }
% \\&{\p i}_{\nz({\p i})}=\beta_t({\p i}')_{\nz({\p i})}\\
% 0&\text{otherwise },
% \end{cases}
% \end{align}
for all ${\p i}\in G^m$. Clearly, by (\ref{eqn0102}), we see that ${\p c}'$ is precisely the vector ${\p c}_m(\beta_t({\p i}')).$ Since {$\wt(\beta_t({\p i}'))=\wt({\p i}')\geq r+1,$} we have that ${\p c}_m(\beta_t({\p i}'))\in B_{\Dsn}(r,m)$ as well. This completes the proof.
\end{IEEEproof}

We now summarize the results from the above two lemmas. We use $\Ss_m$ to denote the symmetric group of degree $m$, i.e., $\Ss_m$ is the group of all permutations on $\integerbox{m}$.

\begin{theorem} \label{thm:berman-all-automorphisms}
Let $\sigma_0,\dots,\sigma_{m-1}$ be any permutations of the set $G$, and let $\gamma \in \Ss_m$. The following permutations on $G^m$ are automorphisms of $\Dsn(r,m)$ and $\Csn(r,m)$,
\begin{align*}
(i_0,\dots,i_{m-1}) &\to (\sigma_0(i_0),\dots,\sigma_{m-1}(i_{m-1})), \\
(i_0,\dots,i_{m-1}) &\to (i_{\gamma(0)},\dots,i_{\gamma(m-1)}).
\end{align*}
\end{theorem}
%%%%%
\begin{IEEEproof}
It is sufficient to show that the permutations in the statement are automorphisms of $\Dsn(r,m)$, because of the duality of $\Dsn(r,m)$ and $\Csn(r,m)$.

\emph{Part 1}. For any $t\in\integerbox{m-1}$, the permutation
\begin{align*}
(i_0,\dots,i_{m-1}) \to (i_0,\dots,i_{t-1},\sigma_t(i_t),i_{t+1},\hdots,i_{m-1}),
\end{align*}
is identical with the composition $\beta_t\pi_{m-1,\sigma_t}\beta_t$, where $\pi_{m-1,\sigma_t}$ and $\beta_t$ are as defined in Lemma \ref{lemma:group-permutationautomorphism} and Lemma \ref{lemma:exchangeautomorphism} respectively. Thus, the permutation 
\begin{align}
\label{eqn:auto1}
(i_0,\dots,i_{m-1}) &\to (\sigma_0(i_0),\dots,\sigma_{m-1}(i_{m-1})),
\end{align}
is identical with the composition 
\begin{equation*}
\pi_{m-1,\sigma_{m-1}}\left(\prod\limits_{t\in\integerbox{m-1}} \, \beta_t \, \pi_{m-1,\sigma_t} \, \beta_t \right).
\end{equation*} 
Since the set of automorphisms of $\Dsn(r,m)$ form a group under composition, by Lemmas \ref{lemma:group-permutationautomorphism} and \ref{lemma:exchangeautomorphism}, we have that (\ref{eqn:auto1}) is an automorphism of $\Dsn(r,m)$.

\emph{Part 2}. It is known (see for example, \cite{Heap_permutation_via_interchanges}) that any permutation $\gamma\in \Ss_m$ can be generated by a composition of transpositions, where a transposition refers to a permutation which interchanges one element of $\integerbox{m}$ with another, and leaves the other elements as is. Observe that $\beta_t$ as in Lemma \ref{lemma:exchangeautomorphism} is precisely the transposition that interchanges $t$ with $m-1$. Further, a transposition in $\Ss_m$ that interchanges two distinct elements $t_1,t_2\in \integerbox{m-1}$ can be obtained as the composition  $\beta_{t_2}\beta_{t_1}\beta_{t_2}$. This completes the proof, following Lemma \ref{lemma:exchangeautomorphism} and because the automorphisms of $\Dsn(r,m)$ form a group.
\end{IEEEproof}

% % % % % % % % %

\subsection{Puncturing $\Dsn(r,m)$ and $\Csn(r,m)$} \label{sec:sub:multiple-puncturings}

% % R0 % %
{We intend to show that there are multiple ways to puncture $\Csn(r,m+k)$ to obtain $\Csn(r,m)$, and similarly, puncture $\Dsn(r+k,m+k)$ to $\Dsn(r,m)$.}
% % % % % %
The key observation will be the fact that $\Csn(r,m)$ can be punctured to get $\Csn(r,m-1)$ (similarly, puncturing $\Dsn(r,m)$ yields $\Dsn(r-1,m-1)$).

We first set up some terminology to work with puncturing operation.
Recall the notation from Section~\ref{subsubsec:usefulbasis} for representing the coordinates of codewords using tuples $\p{i} = (i_0,\dots,i_{m-1}) \in G^m$ where {$G=\lset n \rset$}.
For any set $\Ks \subset \lset m \rset$ and any $\p{b} \in G^{|\Ks|}$, we say that `$H \subset G^m$ is the \emph{direct-product subset} of $G^m$ corresponding to $\Ks$ and $\p{b}$' if
\begin{equation*}
H = \left\{ \p{i} \in G^m: \p{i}_{\Ks} = \p{b} \right\},
\end{equation*}
where $\p{i}_{\Ks} = (i_l: l \in \Ks)$ is a sub-vector of $\p{i}$.
For any such $H$, we define the puncturing operation $\Ps_H$ as follows
\begin{equation*}
\Ps_H(\p{v}) = \Ps_H\left(\, (v_{\p{i}}:\p{i} \in G^m) \, \right) = (v_{\p{i}}:\p{i} \in H).
\end{equation*}
That is, $\Ps_H$ discards all the coordinates not in $H$.
The size of $H$ and the length of the vector $\Ps_H(\p{v})$ are both equal to $n^{m - |\Ks|}$.
For instance, with our usual concatenated representation $\p{v} = (\p{v}_0|\cdots|\p{v}_{n-1})$, we see that {$\p{v}_l = (v_{\p{i}}:i_{m-1} = l)$.
Hence, if $H$ is the direct-product subset corresponding $\Ks = \{m-1\}$ and $b=l$, then $\Ps_H(\p{v}) = \p{v}_l$.}

\begin{lemma} \label{lem:puncture-simple}
Let $\Ks=\{m-1\}$, {$l \in G$, and $H = G^{m-1} \times \{l\}$} be the corresponding direct-product subset of $G^m$. Then,
\begin{enumerate}
\item $\Ps_H\left(\Csn(r,m) \right) = \Csn(r,m-1)$ if $r \leq m-1$,
\item $\Ps_H\left(\Dsn(r,m) \right) = \Dsn(r-1,m-1)$ if $r \geq 1$.
\end{enumerate}
\end{lemma}
\begin{IEEEproof}
We will use the code definitions to prove this result. We know that $\Ps_H\left( (\p{v}_0|\cdots|\p{v}_{n-1}) \right) = \p{v}_l$.
From the definition of $\Csn(r,m)$ and the fact $\Csn(r-1,m-1) \subset \Csn(r,m-1)$, we see that $\Ps_H(\Csn(r,m)) = \Csn(r,m-1)$, irrespective of the value of $l$.
The proof for $\Dsn(r,m)$ is similar.
\end{IEEEproof}

We can easily generalize the results of Lemma~\ref{lem:puncture-simple} to include arbitrary direct-product subsets of $G^m$.

\begin{theorem} \label{thm:puncture}
Let \mbox{$\Ks \subset \lset m \rset$}, $\p{b} \in G^{|\Ks|}$ and $H$ be the corresponding direct-product subset of $G^m$. Then
\begin{enumerate}
\item $\Ps_H\left(\Csn(r,m) \right) = \Csn(r,m-|\Ks|)$ if $r \leq m-|\Ks|$,
\item $\Ps_H\left(\Dsn(r,m) \right) = \Dsn(r-|\Ks|,m-|\Ks|)$ if $r \geq |\Ks|$.
\end{enumerate}
\end{theorem}
\begin{IEEEproof}
We will prove only Part~1, the proof of Part~2 uses the same ideas.

Let us first consider the case $|\Ks|=1$. Suppose $\Ks = \{k\}$ and {$b=l$}.
Let $\gamma \in \Ss_m$ be such that $\gamma(k) = m-1$, \mbox{$\gamma(m-1) = k$} and $\gamma$ fixes all other points in $\lset m \rset$.
We note that the map $\Ps_H$ is equal to the composite of the permutation $\gamma$ applied on the coordinates $G^m$ followed by the puncturing operation {$\Ps_{G^{m-1} \times \{l\}}$}. Since $\gamma$ is an automorphism of $\Csn(r,m)$ (see Theorem~\ref{thm:berman-all-automorphisms}) we have
\begin{align*}
\Ps_H(\Csn(r,m)) &= {\Ps_{G^{m-1} \times \{l\}}} \Big( \, \gamma \big( \Csn(r,m) \big) \,\Big) \\
&= {\Ps_{G^{m-1} \times \{l\}}}\big(\Csn(r,m)\big) \\
&= \Csn(r,m-1),
\end{align*}
where the last step follows from Lemma~\ref{lem:puncture-simple}.

The case $|\Ks| \geq 2$ now simply follows by induction.
To see this consider $\Ks = \{k_1,k_2\}$ with $k_1 > k_2$ and $\p{b} \in G^2$ consisting of elements {${l_1}$ and ${l_2}$}.
Let $H_1$ be the direct-product subset of $G^{m}$ corresponding to $\{k_1\}$ and {${l_1}$}, and $H_2$ be the direct-product subset of $G^{m-1}$ corresponding to $\{k_2\}$ and {${l_2}$}.
We notice that $\Ps_H$ is the composite of the functions $\Ps_{H_1}$ and $\Ps_{H_2}$. Hence,
\begin{align*}
\Ps_H(\Csn(r,m)) 
&= \Ps_{H_2} \Big( \Ps_{H_1} \big(\Csn(r,m)\big) \Big) \\
&= \Ps_{H_2} \big(\Csn(r,m-1)\big)\\
&= \Csn(r,m-2).
\end{align*}
The proof for other values of $|\Ks|$ is similar.
\end{IEEEproof}

We can puncture $\Csn(r,m+k)$, $1 \leq k \leq m$, in multiple ways to obtain $\Csn(r,m-k)$.
Let $H_1$ and $H_2$ be direct-product subsets of $G^{m+k}$ such that
\begin{align*}
|H_1| = |H_2| = n^m \text{ and } |H_1 \cap H_2| = n^{m-k}.
\end{align*}
Note that $H_1 \cap H_2$ is also a direct-product subset of $G^{m+k}$.
For instance, we could have {$H_1=G^m \times \{0\}^k$ and $H_2=G^{m-k} \times \{0\}^k \times G^k$. In this case, $H_1 \cap H_2 = G^{m-k} \times \{0\}^{2k}$}.
Applying Theorem~\ref{thm:puncture}, we have
\begin{align*}
\Ps_{H_1}\left( \Csn(r,m+k) \right) &= \Csn(r,m), \\
\Ps_{H_2}\left( \Csn(r,m+k) \right) &= \Csn(r,m), \text{ and} \\
\Ps_{H_1 \cap H_2}\left( \Csn(r,m+k) \right) &= \Csn(r,m-k).
\end{align*}
A similar property holds for $\Dsn(r,m)$ also. Let $1 \leq k \leq r$, and let $H_1$ and $H_2$ be as defined above, then
\begin{align*}
\Ps_{H_1}\left( \Dsn(r+k,m+k) \right) &= \Dsn(r,m), \\
\Ps_{H_2}\left( \Dsn(r+k,m+k) \right) &= \Dsn(r,m), \text{ and} \\
\Ps_{H_1 \cap H_2}\left( \Dsn(r+k,m+k) \right) &= \Dsn(r-k,m-k).
\end{align*}

\subsection{Rate of $\Dsn(r,m)$ and $\Csn(r,m)$} \label{sec:sub:rate-berman}

The rate of $\Cs_n(r,m)$ is
\begin{equation*}
R_n(r,m) \triangleq \frac{\sum_{w=0}^{r} \binom{m}{w} (n-1)^w}{n^m},
\end{equation*}
which is equal to the fraction of vectors in {$G^m$} with weight at the most $r$.
Let $X_0,\dots,X_{m-1}$ be independent and identically distributed Bernoulli random variables with \mbox{$\Pr(X_k = 1) = (n-1)/n$} for all $k \in \lset m \rset$.
The mean and variance of $X_k$ are
\begin{equation*}
\mu = \frac{n-1}{n} \text{ and } \sigma^2 = \frac{n-1}{n^2},
\end{equation*}
respectively.
Define $Y_m = X_0 + \cdots + X_{m-1}$, which follows binomial distribution.
We observe that
\begin{align*}
R_n(r,m) &= \sum_{w=0}^{r} \binom{m}{w} \left( \frac{n-1}{n} \right)^w \, \left( \frac{1}{n} \right)^{m-w} \\
&= \Pr\left( Y_m \leq r \right) \\
&= \Pr\left( \frac{\sum_{k \in \lset m \rset} \left(X_k - \mu\right)}{\sqrt{m \sigma^2}} \leq \frac{r - m \mu}{\sqrt{m \sigma^2}}\right).
\end{align*}
Note that $(X_k-\mu)/\sigma$ has zero mean and unit variance, and the distribution of $(X_k-\mu)/\sigma$ is completely determined by the value of $n$. The \emph{Berry-Esseen inequality}~\cite{KoS_SIAM_10} can be used to approximate the distribution function of $\sum_k (X_k - \mu)/(\sqrt{m \sigma^2})$ with that of the standard Gaussian.
Let 
\begin{equation*}
Q(x) = \int_{x}^{\infty} \frac{1}{\sqrt{2\pi}} \exp\left(-\frac{t^2}{2}\right) {\rm d}t
\end{equation*}
be the tail probability of the standard Gaussian distribution.
The Berry-Esseen inequality guarantees that there exists a constant $\kappa > 0$, that depends only on $n$, such that
\begin{equation} \label{eq:berry-esseen}
\left| R_n(r,m) - \left( 1 - Q\left( \frac{r - m \mu}{\sqrt{m \sigma^2}} \right) \right)  \right| \leq \frac{\kappa}{\sqrt{m}}.
\end{equation}
Let $R^* \in (0,1)$ and let $\{\Csn(r_l,m_l)\}$ be a sequence of codes with $m_l \to \infty$ and rates converging to $R^*$.
From~\eqref{eq:berry-esseen}, we have
\begin{equation*}
r_l = m_l \mu + Q^{-1}\left( 1 - R^* - o(1) \right)\sqrt{m_l \sigma^2}.
\end{equation*}
Observe that
\begin{equation} \label{eq:Cnrm_r_by_m_tends_to_mu}
\frac{r_l}{m_l} \to \mu, \text{ as } l \to \infty.
\end{equation}

We also note that $1-R_n(r,m)$ is the rate of $\Dsn(r,m)$. Hence, from~\eqref{eq:berry-esseen}, we deduce that, for a fixed $n$, the rate of $\Dsn(r,m)$ is
\begin{equation*}
Q\left( \frac{r - m \mu}{\sqrt{m \sigma^2}} \right) + O\left( \frac{1}{\sqrt{m}} \right).
\end{equation*}

\subsubsection{Rate Change from Puncturing}

We now compare the rates of $\Csn(r,m)$ and $\Csn(r,m+k)$ where $k \geq 1$.
We recognize that $\Csn(r,m)$ can be obtained from $\Csn(r,m+k)$ via puncturing where a fraction $1 - \frac{1}{n^k}$ of the code symbols are removed.
Hence $R_n(r,m) - R_n(r,m+k)$ is the increase in the rate due to this puncturing operation.
We first observe that $R_n(r,m) - R_n(r,m+k)$ equals
\begin{equation*}
\Pr(X_0 + \cdots + X_{m-1} \leq r) - \Pr(X_0 + \cdots + X_{m+k-1} \leq r).
\end{equation*}
This implies $R_n(r,m) - R_n(r,m+k) \geq 0$.
Using~\eqref{eq:berry-esseen} it is easy to show that this rate change is $O(k/\sqrt{m})$.

\begin{lemma} \label{lem:rate-change:dual_Berman}
For any $k \geq 1$ and $0 \leq r \leq m$,
\begin{equation*}
0 \leq  R_n(r,m) - R_n(r,m+k) \leq \frac{2\kappa}{\sqrt{m}} + \frac{k}{\sqrt{m}}\left(\frac{1 + \mu}{\sqrt{8\pi \sigma^2}} \right).
\end{equation*}
\end{lemma}
\begin{IEEEproof}
Using~\eqref{eq:berry-esseen} we observe that the rate difference \mbox{$R_n(r,m) - R_n(r,m+k)$} is upper bounded by
\begin{align*}
&~~~ \frac{\kappa}{\sqrt{m}} + \frac{\kappa}{\sqrt{m+k}} + Q\left(\frac{r-(m+k)\mu}{\sigma\sqrt{m+k}}\right) - Q\left( \frac{r-m\mu}{\sigma\sqrt{m}} \right) \\
&\leq \frac{2\kappa}{\sqrt{m}} + \int\displaylimits_{\frac{r}{\sigma\sqrt{m+k}} - \frac{\mu\sqrt{m+k}}{\sigma}}^{\frac{r}{\sigma\sqrt{m}} - \frac{\mu\sqrt{m}}{\sigma}} \frac{1}{\sqrt{2\pi}} \exp\left(-\frac{t^2}{2}\right) {\rm d}t \\
&\leq \frac{2\kappa}{\sqrt{m}} + \! \frac{1}{\sqrt{2\pi}}\left( \frac{r}{\sigma\sqrt{m}} - \frac{r}{\sigma\sqrt{m+k}} + \frac{\mu\sqrt{m+k} - \mu\sqrt{m}}{\sigma} 
% - \frac{\mu\sqrt{m}}{\sigma} 
\right) \!.
\end{align*}
We now use the AM-GM inequality $\sqrt{m (m+k)} \leq m + \frac{k}{2}$, i.e., $\sqrt{m+k} - \sqrt{m} \leq \frac{k}{2\sqrt{m}}$, to further loosen the above upper bound to
\begin{align*}
&~~~\frac{2\kappa}{\sqrt{m}} + \frac{r}{\sigma\sqrt{2\pi m (m+k)}}\frac{k}{2\sqrt{m}} + \frac{\mu}{\sigma\sqrt{2\pi}}\frac{k}{2\sqrt{m}}\\
&\leq \frac{2\kappa}{\sqrt{m}} + \frac{k}{\sqrt{m}}\left(\frac{1 + \mu}{\sqrt{8\pi \sigma^2}} \right),
\end{align*}
where we have used the fact $r \leq m \leq \sqrt{m(m+k)}$.
\end{IEEEproof}

The following similar result holds for the family of codes $\{\Dsn(r,m)\}$.

\begin{lemma} \label{lem:rate-change:Berman}
For any $k \geq 1$, the rate of $\Dsn(r,m)$ is greater than or equal to the rate of $\Dsn(r+k,m+k)$, and the difference in these rates is upper bounded by
\begin{equation*}
\frac{2\kappa}{\sqrt{m}} + \frac{k}{\sqrt{m}}\frac{(\mu + 2)}{\sqrt{8\pi\sigma^2}}.
\end{equation*}
\end{lemma}
\begin{IEEEproof}
Proof is similar to the proof of Lemma~\ref{lem:rate-change:dual_Berman} and is available in Appendix~\ref{app:proofs}.
\end{IEEEproof}

\subsection{Lack of Double Transitivity} \label{sec:sub:double-transitivity}

It is well known that RM codes (the case $n=2$) are doubly transitive. In the rest of this sub-section we will assume that $n \geq 3$.

We now show that for long block lengths, $\Dsn(r,m)$ and $\Csn(r,m)$ are not doubly transitive. We will rely on the following property of doubly transitive codes~\cite[Theorem~E.9]{Von_Thesis_ETH}, viz., for any doubly transitive code
\begin{equation*}
(\dmin-1)(\dmin^\perp - 1) \geq N-1,
\end{equation*}
where $N$ is the block length of the code, $\dmin$ is the minimum distance of the code and $\dmin^\perp$ is the minimum distance of its dual code.

Let $R^* \in (0,1)$ and let $\{\Csn(r_l,m_l)\}$ be a sequence of codes with $m_l \to \infty$ and rates converging to $R^*$.
From~\eqref{eq:Cnrm_r_by_m_tends_to_mu} we have $r_l = m_l \left(\mu + o(1) \right)$.
% From~\eqref{eq:berry-esseen}, we have
% \begin{equation*}
% r_l = m_l \mu + Q^{-1}\left( 1 - R^* - o(1) \right)\sqrt{m_l \sigma^2}.
% \end{equation*}
% Observe that
% \begin{equation*}
% \frac{r_l}{m_l} \to \mu, \text{ as } l \to \infty.
% \end{equation*}
Now consider the product \mbox{$(\dmin-1)(\dmin^\perp-1)$} for the code $\Cs_n(r_l,m_l)$ (or equivalently for the code $\Ds_n(r_l,m_l)$). We have
\begin{align*}
(\dmin-1)(\dmin^\perp-1) &\leq n^{m_l - r_l} \times 2^{r_l + 1} \\
&= n^{m_l} \times 2 \times \left( \frac{2}{n} \right)^{r_l} \\
&= n^{m_l} \times 2 \times \left( \frac{2}{n} \right)^{m_l(\mu + o(1))}.
\end{align*}
Since $n \geq 3$, we deduce that
\begin{align*}
(\dmin-1)(\dmin^\perp-1) < n^{m_l} - 1
\end{align*}
for all sufficiently large $l$.
Hence, we have proved

\begin{lemma}
Let $n \geq 3$, $R^* \in (0,1)$ and $\{\Csn(r_l,m_l)\}$ be a sequence of codes with increasing block lengths and rates converging to $R^*$. For every sufficiently large $l$, the code $\Csn(r_l,m_l)$ does not have a doubly transitive automorphism group.
\end{lemma}

\subsection{Achieving the BEC Capacity} \label{sec:sub:capacity}

Kumar, Calderbank and Pfister~\cite[Theorem~19]{KCP_ISIT16} use code automorphisms to provide a sufficient condition for a code to achieve the capacity of BEC under bit-MAP decoding.
This condition is less demanding than requiring double transitivity, which was the property used in~\cite{KKMPSU_IT_17} to prove RM codes achieve BEC capacity.
To use this result on a sequence of codes with increasing block lengths, we require the following
\begin{enumerate}
\item the rates of the sequence of codes must converge to a value in $(0,1)$,
\item each code in this sequence must be transitive,
\item for each code in the sequence, the orbits of the coordinates under a subgroup of automorphisms (those automorphisms that fix an arbitrarily chosen coordinate) must be sufficiently large.
\end{enumerate}
We will now apply this result to the family of codes $\{\Csn(r,m)\}$. A similar result holds for $\{\Dsn(r,m)\}$.

\subsubsection{Code Sequence with Converging Rate} \label{sub:sub:sec:converging_rate}
For a given $n \geq 2$ and $R^* \in (0,1)$, consider a sequence of codes $\{\Csn(r_l,m_l)\}$ with $m_l \to \infty$ and
\begin{equation*}
r_l = m_l \mu + Q^{-1}(1 - R^*) \sqrt{m_l \sigma^2}  + o(\sqrt{m_l}).
\end{equation*}
Using~\eqref{eq:berry-esseen} we note that the rate $R_n(r_l,m_l) \to R^*$ as $m_l \to \infty$.
Hence, for any $R^* \in (0,1)$ there exists a sequence of $\Csn(r,m)$ codes with increasing block lengths and rates converging to $R^*$.

\subsubsection{Transitivity}
We now use Theorem~\ref{thm:berman-all-automorphisms} to observe that for any choice of parameters $n,r,m$, the code $\Csn(r,m)$ is transitive. We will use the notation introduced in Section~\ref{subsubsec:usefulbasis} for the coordinates of a codeword.
Consider any choice of coordinates $\p{i},\p{j} \in G^m$. We need to show that there is a code automorphism that maps $\p{i}$ to $\p{j}$.
Let $\sigma_0,\dots,\sigma_{m-1}$ be permutations of the set $G$ such that
\begin{equation*}
\sigma_0(i_0) = j_0,\dots,\sigma_{m-1}(i_{m-1}) = j_{m-1}.
\end{equation*}
Applying Theorem~\ref{thm:berman-all-automorphisms} for this choice of $\sigma_0,\dots,\sigma_{m-1}$ shows that $\Csn(r,m)$ is indeed transitive.

\subsubsection{Orbits Under a Subgroup of Automorphisms}

{Let $\Gs_0$ be the subgroup of automorphisms of $\Csn(r,m)$ that fixes the coordinate $\p{0} \in G^m$.}
We want a lower bound on the size of the orbits of \mbox{$\p{i} \in G^m \setminus \{\p{0}\}$} under the action of $\Gs_0$, which is
\begin{equation*}
\Os_{r,m}(\p{i}) {\triangleq} \left\{ \pi(\p{i}) : \pi \in \Gs_0 \right\}.
\end{equation*}
We will identify a subset of $\Os_{r,m}(\p{i})$ to obtain this lower bound.

Consider any $\p{i} \neq \p{0}$ and any $\p{j} \in G^m$ such that {$\supp(\p{i}) = \supp(\p{j})$}. There exist $m$ permutations of $G$, $\sigma_0,\dots,\sigma_{m-1}$, such that
\begin{equation*}
\sigma_l(i_l) = j_l \text{ and } {\sigma_l(0) = 0} \text{ for all } l \in \lset m \rset.
\end{equation*}
Using Theorem~\ref{thm:berman-all-automorphisms}, we see that the map $(k_0,\dots,k_{m-1}) \to (\sigma_0(k_0),\dots,\sigma_{m-1}(k_{m-1}))$ is an automorphism of $\Csn(r,m)$ that fixes $\p{0}$ and sends $\p{i}$ to $\p{j}$.
We thus conclude that $\Os_{r,m}(\p{i})$ contains all $\p{j}$ such that {$\supp(\p{j}) = \supp(\p{i})$}.

Now, for a given $\p{i} \neq \p{0}$, consider any $\p{j}$ such that {$\wt(\p{j}) = \wt(\p{i})$}. Clearly, there exists a permutation $\gamma \in \Ss_m$ such that {$\supp(\p{j}) = \supp(\gamma(\p{i}))$}.
From our argument in the previous paragraph, $\p{j} \in \Os_{r,m}\left( \gamma(\p{i}) \right)$.
Since $\gamma$ is a code automorphism that fixes $\p{0}$, we see that $\gamma(\p{i})$  itself belongs to $\Os_{r,m}\left(\p{i}\right)$, and therefore, $\p{j} \in \Os_{r,m}(\p{i})$.
We have thus showed that for any $\p{i} \neq \p{0}$,
% \begin{equation*}
{$\Os_{r,m}(\p{i}) \supset \{ \p{j} \in G^m: \wt(\p{j}) = \wt(\p{i}) \}$.}
% \end{equation*}
Hence,
\begin{equation*}
{\left| \Os_{r,m}(\p{i}) \right| \geq \binom{m}{\wt(\p{i})} (n-1)^{\wt(\p{i})}.}
\end{equation*}
Note that for any $n \geq 3$, and any $\p{i} \in G^m \setminus \{\p{0}\}$, we have
\begin{equation*}
\left| \Os_{r,m}(\p{i}) \right| \geq 2m.
\end{equation*}

We are now ready to apply~\cite[Theorem~19]{KCP_ISIT16}. Let $n \geq 3$. Consider a sequence of codes $\{\Csn(r_l,m_l)\}$ with $m_l \to \infty$ and rates converging to $R^* \in (0,1)$. All the codes in this sequence are transitive, and they satisfy
\begin{equation*}
\min_{\p{i} \in G^{m_l} \setminus \{\p{0}\}} \left| \Os_{r_l,m_l}(\p{i}) \right| \to \infty \text{ as } l \to \infty.
\end{equation*}
These are precisely the sufficient conditions identified in~\cite{KCP_ISIT16} to guarantee that this sequence of codes has a vanishing bit erasure probability under bit-MAP decoding in the BEC for any channel erasure probability $\epsilon < (1 - R^*)$.
Since a similar result holds for $\{\Dsn(r,m)\}$, we have thus proved the following.
%%%%%%
\begin{theorem} \label{thm:berman_codes_achieve_BEC_capacity}
Consider any BEC with capacity $C\in(0,1)$. For any $n\geq 3$ and any $R\in[0,C)$, there exists a sequence of codes from the family $\{\Csn(r,m)\}$ with strictly increasing blocklengths with rates converging to $R$ and the bit erasure probability under bit-MAP decoding converging to zero. The same statement holds for the family $\{\Dsn(r,m)\}$ as well.
\end{theorem}

% \begin{theorem}
% For any $n \geq 3$, the code families $\{\Csn(r,m)\}$ and $\{\Dsn(r,m)\}$ achieve the capacity of the BEC under bit-MAP decoding.
% \end{theorem}

It is well known that the above result is true for $n=2$, i.e., for RM codes. In particular, we know that RM codes achieve BEC capacity under both bit-MAP and block-MAP decoding~\cite{KKMPSU_IT_17}.

\section{Constructions via\\ Discrete Fourier Transform} \label{sec:dft}

A number of good binary error correcting codes are abelian codes, that is, they are ideals in commutative group algebras. Examples include BCH codes, quadratic residue codes, and RM codes~\cite{Ber_Cybernetics_67,Ber_Cybernetics_II_67,Wil_BellSys_70,Cam_Math_71,macwilliams1977theory,KeS_ContemporaryMath_01,Assmus_1996}.
When the order of the abelian group is odd (the order of the group is relatively prime with the characteristic of $\Fb_2$), the corresponding group algebra is semi-simple.
It is well known that abelian codes constructed from semi-simple group algebras can be characterized using discrete Fourier transform (DFT) over finite fields~\cite{Cam_Math_71,Blahut_Cambridge,RaS_IT_92}.

% % R0 version % %
{
Unless specified otherwise, throughout this section we will assume that $n \geq 3$ is an odd integer.
% % We will use the transform domain characterization of abelian codes of Rajan and Siddiqi~\cite{RaS_IT_92} to identify $\Csn(r,m)$ and $\Dsn(r,m)$ as ideals in appropriate group algebras.
% % Further, with the help of the DFT tool, we will identify a large class of abelian codes (containing $\Csn(r,m)$ and $\Dsn(r,m)$) that achieve the capacity of the BEC under bit-MAP decoding.
% % During this process, we will also show that when $n$ is an odd prime, $\Dsn(r,m)$ is identical to the code designed by Berman in~\cite{Ber_Cybernetics_II_67}, and $\Csn(r,m)$ is its dual code studied by Blackmore and Norton in~\cite{BlN_IT_01}.
%
In this section we first review the necessary background on abelian codes and DFT, we then derive some preliminary results related to DFT, and provide a construction of $\Csn(r,m)$ and $\Dsn(r,m)$ for odd $n$ using the transform domain approach.
Finally, we identify a large family of abelian codes that achieve BEC capacity in Theorem~\ref{thm:capacity-abelian-codes}. This new family includes codes outside the ones constructed in Section~\ref{sec:berman_and_dual}; please see Fig.~\ref{fig:inclusionofcodes} in Section~\ref{sec:introduction}.
}
% % % % % % % %

\subsection{Background} \label{sec:sub:background}

Throughout this section we will assume that $(G,\oplus)$ is any finite abelian group of odd order, with \mbox{$|G| \geq 3$}.
Please be aware that $|G|$ will replace the role played by the parameter $n$ in the previous sections.
We denote the identity element of $G$ by $0$.
Since $G$ is finite abelian, it is a direct product of cyclic groups~\cite{jacobson2009basic}.
Let the number of cyclic groups in this decomposition be $s$ and let their orders be $m_0,\dots,m_{s-1}$, i.e., 
\begin{equation*}
G = \Zb_{m_0} \times \cdots \times \Zb_{m_{s-1}}.
\end{equation*}
Since $|G|$ is odd, so are $m_0,\dots,m_{s-1}$.
We will treat each element \mbox{$i \in G$} as an $s$-tuple
\begin{equation*}
i = \left( i[0], \, i[1], \cdots, \, i[s-1] \right) \in \Zb_{m_0} \times \cdots \times \Zb_{m_{s-1}}.
\end{equation*}
% % R0 version % %
{Addition and subtraction of $i,j \in G$ is performed component-wise, that is,
% The sum of two elements $i,j \in G$ is performed component-wise, that is,
for each $\ell \in \lset s \rset$
\begin{align*}
(i \oplus j)[\ell] &= (i[\ell] + j[\ell]) \mod m_\ell, \text{ and } \\ 
(i \ominus j)[\ell] &= (i[\ell] - j[\ell]) \mod m_\ell.
\end{align*}
% We denote the sum of $i$ and the additive inverse of $j$ as $i \ominus j$, i.e.,  $(i \ominus j)[\ell] = (i[\ell] - j[\ell]) \mod m_\ell$.
}
% % % % % % % % % %

\subsubsection{Group Algebra \& Abelian Codes}

The group algebra $\Fb_2[G]$ is
% \begin{equation*}
$\textstyle \left\{ \sum_{i \in G} a_i X^i~:~a_i \in \Fb_2 \right\}$,
% \end{equation*}
where each element $a$ of the group algebra is a formal sum $\sum_{i \in G} a_i X^i$.
% % R0 % % % % %
{The sum and product of two elements $a$ and $b$ is
\begin{align*}
 &a + b = \sum_{i \in G} a_i X^i + \sum_{i \in G} b_i X^i = \sum_{i \in G} (a_i + b_i) X^i, \\
 &ab = \left( \sum_{i \in G} a_i X^i \right) \cdot \left( \sum_{j \in G} b_j X^j \right)
% &= \sum_{i \in G} \sum_{j \in G} a_ib_j X^{i \oplus j} \nonumber \\
= \sum_{i \in G} \left( \sum_{k \in G} a_k b_{i \ominus k} \right) X^i. \nonumber
\end{align*}
}
% where the addition $a_i + b_i$ is over $\Fb_2$.
% The product $ab$ of two elements $a, b \in \Fb_2[G]$ is
% \begin{align}
% \left( \sum_{i \in G} a_i X^i \right) \cdot \left( \sum_{j \in G} b_j X^j \right)
% &= \sum_{i \in G} \sum_{j \in G} a_ib_j X^{i \oplus j} \nonumber \\
% &= \sum_{i \in G} ( \sum_{k \in G} a_k b_{i \ominus k} ) X^i. \nonumber
% \end{align}
% The multiplication operation in $\Fb_2[G]$ is a convolution of the length-$|G|$ binary vectors $(a_i:i \in G)$ and $(b_i:i \in G)$.
{Since $G$ is abelian,
% this multiplication operation is commutative, and
$\Fb_2[G]$ is a commutative ring.
}

An abelian code $\Cs$ is an ideal in the group algebra $\Fb_2[G]$, that is, $(\Cs,+)$ is an additive subgroup of $\Fb_2[G]$ and for any $a \in \Cs$ and $b \in \Fb_2[G]$ we have $ab \in \Cs$.
Observe that $\Cs$ yields a length-$|G|$ binary linear code
\begin{equation*}
\{(a_i:i \in G)~:~a \in \Cs\} \subset \Fb_2^{|G|}.
\end{equation*}
We will make no distinction between an ideal $\Cs$ and the binary linear code associated with this ideal.
%% It is well known that BCH codes, quadratic residue codes and Reed-Muller codes\footnote{For Reed-Muller codes, the appropriate group algebra is modular and is based on $G=(\Zb_2^m,+)$ which is not of odd order.} can be constructed as ideals in appropriate group algebras~\cite{Ber_Cybernetics_67,Ber_Cybernetics_II_67,Wil_BellSys_70,Cam_Math_71,KeS_ContemporaryMath_01}.
We make the following simple observation here.

\begin{lemma} \label{lem:abelian-codes-transitive}
Abelian codes are transitive.
\end{lemma}
\begin{IEEEproof}
Let $\Cs$ be an ideal in $\Fb_2[G]$, and \mbox{$j,k \in G$}. We want to show that there exists an automorphism of $\Cs$ that maps the coordinate $k$ to $j$.
% % R0 % %
Let $a = \sum_{i \in G}a_i X^i$ and $c = \sum_{i \in G} c_i X^i$ be such that $a \in \Cs$ and
\begin{equation*}
c = X^{j \ominus k}a= \sum_{i \in G} a_i X^{i \oplus j \ominus k}.
\end{equation*}
% % % % %
% For any \mbox{$a \in \Cs$}, consider \mbox{$c = X^{j \ominus k}a= \sum_{i \in G} a_i X^{i \oplus j \ominus k}$}.
Note that \mbox{$c_{i \oplus j \ominus k} = a_i$}. Hence, $c$ is obtained by permuting the coordinates of $a$, and this permutation sends $k$ to $j$, i.e., $c_j=a_k$.
Since $\Cs$ is an ideal {and $a \in \Cs$, we have} $c = X^{j \ominus k}a \in \Cs$.
Hence, the map $a \to X^{j \ominus k}a$ is a code automorphism that maps $k$ to $j$.
\end{IEEEproof}

\subsubsection{Discrete Fourier Transform}

% % R0 % %
{
We now recall the definition of conjugacy classes and then review the DFT-based approach for studying abelian codes in semi-simple group algebras. Please refer to~\cite{RaS_IT_92} and the references therein for more information.
}
% % % % % % % %
% % This approach uses the DFT to explicitly characterize all the ideals in $\Fb_2[G]$.
% % This will be the primary tool used in this section.
% % We will first recall the notion of \emph{conjugacy classes} and then review the DFT.

{
% {\em Conjugacy Classes:}
For any integer $\ell \geq 0$ and any $i \in G$, we use $\ell i$ to denote the $\ell$-fold sum $(i \oplus i \oplus \cdots \oplus i)$.
The \emph{conjugacy class} of $G$ containing the element \mbox{$i \in G$} is the set
\begin{equation*}
\Gamma_i = \{i,2i, 2^2i,\dots,2^{\ell-1}i\}
\end{equation*}
where $\ell=|\Gamma_i|$ is the smallest integer such that $2^\ell i = i$.
% Such an integer $\ell$ exists since $2$ is not a divisor of $|G|$.
% For any $i,j \in G$, either $\Gamma_i=\Gamma_j$ or $\Gamma_i \cap \Gamma_j = \emptyset$.
% Thus,
The distinct conjugacy classes of $G$ partition $G$.
% We note that \mbox{$\Gamma_i = \Gamma_j$} if and only if \mbox{$i \in \Gamma_j$} (or equivalently, $j \in \Gamma_i$).
The conjugacy class of the identity element $\Gamma_0$ is $\{0\}$, and this is the only conjugacy class with size equal to $1$.
}
\begin{lemma} \label{lem:G-non-zero-conjugacy-class}
If $i \in G \setminus \{0\}$ then $|\Gamma_i| \geq 2$.
\end{lemma}
\begin{IEEEproof}
Since \mbox{$i \neq 0$}, \mbox{$2i = i \oplus i \neq i$}, and hence, \mbox{$|\Gamma_i| \geq 2$}.
\end{IEEEproof}

We will denote the set of all distinct conjugacy classes of $G$ by $\Lambda$. %
% Let $\rep: \Lambda \to G$ be any function that allows us to pick a representative element from a conjugacy class, i.e., for any conjugacy class $\Gamma$, $\rep(\Gamma)$ is an element of $\Gamma$.
% For example, $\rep(\Gamma)$ can be the `smallest' element in $\Gamma$ under the lexicographic ordering of elements in $\Zb_{m_0} \times \cdots \times \Zb_{m_{s-1}}$.
%
% Observe that if $\Gamma \in \Lambda$ and $j = \rep(\Gamma)$ then $\Gamma = \Gamma_j$.
Note that $\sum_{\Gamma \in \Lambda} |\Gamma| = |G|$ since the distinct conjugacy classes partition $G$.

\begin{example} \label{ex:conj_class_Z3xZ3}
We will use \mbox{$G = \Zb_3 \times \Zb_3$} to illustrate the definitions introduced in this sub-section. Each element $i \in G$ is a $2$-tuple $(i[0]~~i[1])$. The distinct conjugacy classes of $G$ are
\begin{align*}
&\{(0~0)\},~
\{(0~1), (0~2)\}, ~ \{(1~0), (2~0)\}, \\
&\{(1~1), (2~2)\}, ~ \{(1~2), (2~1)\}.
\end{align*}
In this case $\Lambda$ contains $5$ classes, and all classes except $\Gamma_0= \{(0~0)\}$ have size $2$.
%% A possible choice of $\rep$, based on the lexicographic ordering, is as follows
%% \begin{align*}
%% &\rep\left(\{(0~0)\}\right) = (0~0)\\
%% &\rep\left(\{(0~1), (0~2)\}\right) = (0~1), ~ \rep\left(\{(1~0), (2~0)\}\right) = (1~0) \\
%% &\rep\left(\{(1~1), (2~2)\}\right) = (1~1), ~ \rep\left(\{(1~2), (2~1)\}\right) = (1~2).
%% \end{align*}
\end{example}

We will be interested in subsets of $G$ that are unions of conjugacy classes.
{
Note that $Z \subset G$ is a union of conjugacy classes if and only if
% there exists a $\Lambda' \subset \Lambda$ such that $Z = \cup_{\Gamma \in \Lambda'}\Gamma$.
% Alternatively, $Z$ is a union of conjugacy classes if and only if
for each $j \in Z$ we have $2j \in Z$.
}

% {\em Discrete Fourier Transform:} % \label{sec:background:sub:DFT}
The DFT for $\Fb_2[G]$ is based on the decomposition $G=\Zb_{m_0} \times \cdots \times \Zb_{m_{s-1}}$.
Let $\Fb_q$ be a finite extension of $\Fb_2$ that contains the primitive $m_{\ell}^\tth$ root of unity for every $\ell \in \lset s \rset$. Let $\alpha_0,\dots,\alpha_{s-1}$ be elements of $\Fb_q^*$ with multiplicative orders $m_0,\dots,m_{s-1}$, respectively.
The DFT is an $\Fb_2$-linear injective map $\Phi: \Fb_2[G] \to \Fb_q[G]$.
We will denote the DFT of $a = \sum_{i \in G}a_i X^i \in \Fb_2[G]$ by
\begin{equation*}
\Phi(a) = A = \sum_{j \in G} A_j X^j \in \Fb_q[G].
\end{equation*}
The DFT is defined as follows
\begin{align*}
A_j = \sum_{i \in G} \alpha_0^{i[0] \cdot j[0]}\,\alpha_1^{i[1] \cdot j[1]} \cdots \alpha_{s-1}^{i[s-1] \cdot j[s-1]} \, a_i,
\end{align*}
where $i[\ell] \cdot j[\ell]$ is the product of two elements in the ring $\Zb_{m_\ell}$.
If $A \in \Phi(\Fb_2[G])$ we can obtain $a = \Phi^{-1}(A)$ through the inverse DFT
\begin{equation*}
a_i = \sum_{j \in G} \alpha_0^{-i[0] \cdot j[0]}\,\alpha_1^{-i[1] \cdot j[1]} \cdots \alpha_{s-1}^{-i[s-1] \cdot j[s-1]} \, A_j.
\end{equation*}
We will often refer to the elements $a$ and $A$ as sequences existing in the `time-domain' and the `transform-domain', respectively.
We now recall some key properties of the DFT.

{
\emph{The Convolution Property.}
Let $a,b \in \Fb_2[G]$ and $c = ab$. If $A,B,C$ are the images of $a,b,c$, respectively, under the DFT then $C_j = A_jB_j$ for all $j \in G$.
% Hence convolution in time-domain is equivalent to component-wise product in the transform-domain.

\emph{The Modulation Property.}
If $a_ib_i = c_i$ for all $i \in G$, then $AB = C$ in $\Fb_q[G]$, where $A,B,C$ are images of $a,b,c$, respectively, under the DFT.
% Component-wise product in time-domain is equivalent to convolution in the transform-domain.
% The proof of this property is similar to that of the convolution property above.
% Since~\cite{RaS_IT_92} does not include a proof of this fact, we provide a proof in Appendix~\ref{app:proofs} for completeness.

\emph{The Reversal Property.}
Let $a,b \in \Fb_2[G]$ be such that $a_i = b_{0 \ominus i}$ for all $i \in G$.
% That is, $a$ is obtained via a `reversal' permutation on the coordinates of $b$.
Then $A$ and $B$ satisfy $A_j = B_{0 \ominus j}$ for all $j \in G$.
% i.e., reversals in time-domain and transform-domain are equivalent.

\emph{The Conjugate Symmetry Property.}
% The image of $\Fb_2[G]$ under the DFT is a proper subset of $\Fb_q[G]$.
An element $A \in \Fb_q[G]$ belongs to $\Phi(\Fb_2[G])$ if and only if
% \begin{equation*}
$A_j^2 = A_{2j}$ for all $j \in G$. This implies that $A_j$ belongs to a subfield of $\Fb_q$ of size $2^{|\Gamma_j|}$.
% \end{equation*}
% If $A \in \Phi(\Fb_2[G])$ then the value of $A_j$ completely determines the value of $A_k$ for all $k \in \Gamma_j$.
% In particular, if $k = 2^l j$ then $A_k = A_j^{2^l}$.
}

\subsubsection{Characterization of Abelian Codes}

% % R0 % %
% There are two direct consequences of the conjugate symmetry property. Assume that $a \in \Fb_2[G]$ and $A = \Phi(a)$.
% The first consequence is that, since $j = 2^{|\Gamma_j|} j$, we have $A_j = A_j^{2^{|\Gamma_j|}}$.
% Hence, for each $j \in G$, $A_j$ belongs to a subfield of $\Fb_q$ of size $2^{|\Gamma_j|}$.
% Second, \mbox{$A_j = 0$} if and only if \mbox{$A_k = 0$} for all \mbox{$k \in \Gamma_j$}. Hence, the set of coordinates $j$ for which $A_j=0$ is always a union of conjugacy classes.
% % % % % % %
Let $Z \subset G$ be a union of conjugacy classes. Define
\begin{equation*}
\Cs(Z) \triangleq \left\{ a \in \Fb_2[G]~:~A_j = 0 \text{ for all } j \in Z \right\}.
\end{equation*}
Using the convolution property of DFT it is easy to show that $\Cs(Z)$ is an ideal in $\Fb_2[G]$. We will say that $Z$ is the \emph{zero-set} of the abelian code $\Cs(Z)$.
A key structural property of semi-simple abelian group algebras is that every ideal is of the form $\Cs(Z)$ for some choice of $Z$.

\begin{theorem}
(DFT Characterization of Abelian Codes~\cite{RaS_IT_92}.)
Let $G$ be an abelian group of odd order and $\Cs$ be any ideal in $\Fb_2[G]$. There exists a $Z \subset G$, which is a union of conjugacy classes, such that $\Cs=\Cs(Z)$.
\end{theorem}

The \emph{non-zeros} of $\Cs(Z)$ is $\bar{Z} = G \setminus Z$. If $j \in \bar{Z}$ then there exists an $a \in \Cs(Z)$ such that $A_j \neq 0$. The dimension of $\Cs(Z)$ over $\Fb_2$ is $|\bar{Z}| = |G| - |Z|$, see~\cite{RaS_IT_92}.

{
% An ideal of $\Fb_2[G]$ is \emph{minimal} if it is non-trivial and if it contains no non-trivial ideal as a proper subset.
Any abelian code is a direct sum of a collection of \emph{irreducible codes}, which are minimal ideals of $\Fb_2[G]$~\cite{Ber_Cybernetics_II_67,Wil_BellSys_70,RaS_IT_92}.
% There is a one-to-one correspondence between the conjugacy classes of $G$ and the minimal ideals of $\Fb_2[G]$. Let $\Gamma$ be a conjugacy class of $G$. The minimal ideal corresponding to the class $\Gamma$ is the abelian code whose non-zeros are precisely $\Gamma$, that is $\Cs(G \setminus \Gamma)$.
% In general,
If $Z$ is a union of conjugacy classes of $G$, then $\Cs(Z)$ is the direct sum of the following irreducible codes: $\Cs(G \setminus \Gamma)$, where $\Gamma \subset G \setminus Z$ and $\Gamma$ is a conjugacy class.
% \begin{equation*}
% \Cs(Z) = \sum_{\substack{\Gamma \subset G \setminus Z \\ \Gamma \text{ is a conjugacy class}}} \!\!\!\! \Cs(G \setminus \Gamma).
% \end{equation*}
% The dimension of the minimal ideal corresponding to $\Gamma$ is $|\Gamma|$.
}

\subsection{Preliminary Results} \label{sec:sub:berman-prelim}

Let \mbox{$G^m = G \times \cdots \times G$} be the $m$-fold product of a group $G$ with itself.
We will define two families of codes $\Ds_G(r,m)$ and $\Cs_G(r,m)$ as ideals in the group algebra $\Fb_2[G^m]$.
Later in this section we will show that these codes are indeed identical to $\Dsn(r,m)$ and $\Csn(r,m)$, where $n=|G|$.
In this sub-section, we will introduce some notation to work with the group algebra $\Fb_2[G^m]$ and its DFT and present some preliminary results related to abelian codes from $\Fb_2[G^m]$.

As in Section~\ref{sec:sub:background}, let $G=\Zb_{m_0} \times \cdots \times \Zb_{m_{s-1}}$.
We will represent the elements of $G^m$ as vectors and denote them using bold font. If $\p{i} \in G^m$, then $\p{i}=(i_0,\dots,i_{m-1})$ where $i_k \in G$ for each $k \in \lset m \rset$.
Recall that since $i_k \in G$, we treat $i_k$ itself as an $s$-tuple $(i_k[0],~\cdots~,i_k[s-1]) \in \Zb_{m_0} \times \cdots \times \Zb_{m_{s-1}}$.
The identity element of $G^m$ is $\p{0} = (0,\dots,0)$ where $0$ is the identity element of $G$.
The sum of two elements $\p{i},\p{j} \in G^m$ is $\p{i} \oplus \p{j} = (i_0 \oplus j_0,\dots,i_{m-1} \oplus j_{m-1})$.

The group algebra $\Fb_2[G^m]$ is $\left\{ \sum_{\p{i} \in G^m} a_{\p{i}} X^{\p{i}}~:~a_{\p{i}} \in \Fb_2 \right\}$.
Let $\Fb_q$, $\alpha_0,\dots,\alpha_{s-1}$ be as defined in Section~\ref{sec:sub:background}.
Applying the definition of DFT from Section~\ref{sec:sub:background} to the group $G^m$ (instead of $G$) we observe that the DFT $\Phi$ is map from $\Fb_2[G^m]$ into $\Fb_q[G^m]$. If $\Phi(a) = A = \sum_{\p{j} \in G^m} A_{\p{j}}X^{\p{j}}$ then we have
\begin{align*}
A_{\p{j}} &= \sum_{\p{i} \in G^m} \prod_{k \in \lset m \rset } \alpha_0^{i_k[0] \cdot j_k[0]}\,\alpha_1^{i_k[1] \cdot j_k[1]} \cdots \alpha_{s-1}^{i_k[s-1] \cdot j_k[s-1]} \, a_{\p{i}} \\
&= \sum_{\p{i} \in G^m} \left( \prod_{k \in \lset m \rset } \prod_{\ell \in \lset s \rset} \alpha_{\ell}^{i_k[\ell] \cdot j_k[\ell]} \right) a_{\p{i}}.
\end{align*}
The inverse DFT is given by
\begin{equation} \label{eq:IDFT-Gm}
a_{\p{i}} = \sum_{\p{j} \in G^m} \left( \prod_{k \in \lset m \rset } \prod_{\ell \in \lset s \rset} \alpha_{\ell}^{-i_k[\ell] \cdot j_k[\ell]} \right) A_{\p{j}}.
\end{equation}

We will need the following result about the conjugacy classes of $G^m$.
Define the {Hamming weight}, or simply the {weight}, $\wt(\p{i})$ of $\p{i} \in G^m$ as the number of non-zero components in the vector $\p{i}$, i.e., \mbox{$\wt(\p{i}) = |\{k~|~i_k \neq 0\}|$}.

\begin{lemma} \label{lem:conj_class_same_weight}
Let $\Gamma$ be a conjugacy class of $G^m$. The weights of all the vectors in $\Gamma$ are equal.
\end{lemma}
\begin{IEEEproof}
Observe that for every \mbox{$\p{i} \in \Gamma$} there exists a \mbox{$\p{j} \in \Gamma$} such that \mbox{$\p{i} = 2\p{j} = \p{j} \oplus \p{j}$}. Hence, it is enough to show that $\wt(\p{j}) = \wt(2\p{j})$ for all $\p{j} \in G^m$.
Note that 
\begin{equation*}
2\p{j} = (j_0 \oplus j_0,\dots,j_{m-1} \oplus j_{m-1}).
\end{equation*}
If $j_k \neq 0$ and $j_k \oplus j_k = 0$, then $j_k$ has order $2$ in $G$. Since $|G|$ is odd this is impossible. Thus, $j_k \oplus j_k =0$ if and only if $j_k = 0$. This completes the proof.
\end{IEEEproof}

We say that \mbox{$W \subset G^m$} is a \emph{weight class of weight} $w$ if $W$ is the set of all vectors in $G^m$ of weight $w$.
We now consider ideals $\Cs(Z)$ where $Z$ is a union of weight classes from $G^m$.
Using Lemma~\ref{lem:conj_class_same_weight} we observe that such a $Z$ is a union of conjugacy classes and hence $\Cs(Z)$ is indeed an ideal.

\begin{lemma} \label{lem:time-reversal}
Let $Z \subset G^m$ be a union of weight classes. The code $\Cs(Z)$ is closed under time-reversal permutation, that is, if $a \in \Cs(Z)$ and if $c \in \Fb_2[G^m]$ is such that $c_{\p{i}} = a_{\p{0} \ominus \p{i}}$ for all $\p{i}$, then $c \in \Cs(Z)$.
\end{lemma}
\begin{IEEEproof}
From the reversal property of DFT we know that $C_{\p{j}} = A_{\p{0} \ominus \p{j}}$ for all $\p{j}$.
Since $Z$ is a union of weight classes and $\wt(\p{0} \ominus \p{j}) = \wt(\p{j})$ we observe that $\p{j} \in Z$ if and only if $\p{0} \ominus \p{j} \in Z$.
We conclude that $C_{\p{j}} = 0$ for all $\p{j} \in Z$.
%% Since $\wt(\p{0} \ominus \p{j}) = \wt(\p{j})$ we see that $C_{\p{j}} = 0$ for all $\p{j}$ such that $\wt(\p{j}) \geq r+1$.
This completes the proof.
\end{IEEEproof}

We are now ready to identify $\Cs(Z)^\perp$ when $Z$ is a union of weight classes.
\begin{lemma} \label{lem:dual-abelian-codes-weight-classes}
Let $Z\subset G^m$ be a union of weight classes. The ideal with zero-set $G^m \setminus Z$, i.e., $\Cs(G^m \setminus Z)$, is the dual code of $\Cs(Z)$.
\end{lemma}
\begin{IEEEproof}
% For brevity, we will use $\Zs$ and $\Zsc$ to denote $\Zs(r,m)$ and $\Zsc(r,m)$, respectively.
% Let $\Cs(\Zsc)$ be the abelian code with $\Zsc$ as the zero-set.
% We need to show that $\Cs(\Zsc)$ is the dual of $\Cs_G(r,m)=\Cs(\Zs)$.
Let $\bar{Z} = G^m \setminus Z$. Note that $\bar{Z}$ is a union of weight classes.
The dimension of $\Cs(\bar{Z})$ is $|G^m| - |\bar{Z}| = |Z|$, which is equal to the dimension of $\Cs(Z)^\perp$. Hence, to complete the proof, it is sufficient to show that every codeword in $\Cs(\bar{Z})$ is orthogonal to every codeword in $\Cs(Z)$.

We first observe that $cb = 0$ in $\Fb_2[G^m]$ if $c \in \Cs(Z)$ and $b \in \Cs(\bar{Z})$. This follows from applying the convolution property of DFT and observing that for any $\p{j} \in G^m$ either $C_{\p{j}}=0$ or $B_{\p{j}} = 0$.
Since $cb=0$, the $\p{0}^\tth$ component of the product $cb$ is
% \begin{equation*}
$\sum_{\p{i} \in G^m} c_{\p{0} \ominus \p{i}} \, b_{\p{i}} = 0$
% \end{equation*}
for any $c \in \Cs(Z)$ and $b \in \Cs(\bar{Z})$.

We are now ready to complete the proof. Let $a \in \Cs(Z)$ and $b \in \Cs(\bar{Z})$. Let $c$ be the time-reversal of $a$.
From Lemma~\ref{lem:time-reversal}, we know that \mbox{$c \in \Cs(Z)$}, and hence, \mbox{$cb=0$}. This implies $\sum_{\p{i}} c_{\p{0} \ominus \p{i}} b_{\p{i}} = 0$. Since $c_{\p{0} \ominus \p{i}} = a_{\p{i}}$, we have $\sum_{\p{i}} a_{\p{i}} b_{\p{i}} = 0$.
\end{IEEEproof}

\subsection{Construction of $\Ds_n(r,m)$ and $\Cs_n(r,m)$ via DFT} \label{sec:sub:berman_def}

We will now define two codes $\Ds_G(r,m)$ and $\Cs_G(r,m)$ as ideals in $\Fb_2[G^m]$.
% We start by defining the dual Berman code $\Cs_G(r,m)$. We identify $\Cs_G(r,m)$ using its zero-set.
%
Let
\begin{equation*}
\Zs(r,m) = \{\p{j} \in G^m~:~\wt(\p{j}) \geq r+1\}
\end{equation*}
be the set of all elements in $G^m$ with weight greater than $r$.
The complement of $\Zs(r,m)$ is
\begin{equation*}
\Zsc(r,m) = G^m \setminus \Zs(r,m) = \left\{ \p{j} \in G^m~:~\wt(j) \leq r \right\}.
\end{equation*}
%
% From Lemma~\ref{lem:conj_class_same_weight} we deduce that both $\Zs(r,m)$ and $\Zsc(r,m)$ are unions of conjugacy classes in $G^m$.
% Hence, they can be used as zero-sets to define abelian codes.
Note that both $\Zs(r,m)$ and $\Zsc(r,m)$ are unions of weight classes.

\begin{definition}
The code $\Cs_G(r,m)$ is the ideal in $\Fb_2[G^m]$ with zero-set $\Zs(r,m)$, and $\Ds_G(r,m)$ is the dual code of $\Cs_G(r,m)$.
\end{definition}

Clearly, $\Cs_G(r,m)$ has block length $|G|^m$ and dimension
\begin{equation*}
|G^m| - |\Zs(r,m)| = |\Zsc(r,m)| = \sum_{w=0}^{r} \binom{m}{w} \left( |G| -1 \right)^w.
\end{equation*}
Since $|G^m| = \sum_{w=0}^{m} \binom{m}{w}(|G|-1)^w$, we deduce that the dimension of $\Ds_G(r,m)$ is
\begin{equation*}
\sum_{w=r+1}^{m} \binom{m}{w}(|G|-1)^w,
\end{equation*}
which is equal to $|\Zs(r,m)|$.

The following properties are direct consequences of the above definition.

\begin{lemma}
Let $0 \leq r_1 \leq r_2 \leq m$.
\begin{enumerate}
\item $\Cs_G(r_1,m) \subset \Cs_G(r_2,m)$.
\item $\Cs_G(0,m)$ is the repetition code of length $|G|^m$.
\item $\Cs_G(m,m) = \Fb_2[G^m]$.
\end{enumerate}
\end{lemma}
\begin{IEEEproof}
\emph{Part 1}. Let $a \in \Cs_G(r_1,m)$. Then $A_{\p{j}} = 0$ if $\wt(\p{j}) \geq r_1 + 1$. This implies $A_{\p{j}} = 0$ for all $\p{j}$ with $\wt(\p{j}) \geq r_2 + 1$. Hence, $a \in \Cs_G(r_2,m)$.

\emph{Part 2}. Note that $\Zsc(0,m) = \{\p{0}\}$. Hence, $a \in \Cs_G(0,m)$ if and only if $A_{\p{j}} = 0$ for all $\p{j} \neq \p{0}$ and $A_{\p{0}}$ satisfies the conjugate symmetry property. That is, $A_{\p{0}}^2 = A_{2 \cdot \p{0}} = A_{\p{0}}$, or equivalently, $A_{\p{0}} \in \Fb_2$. Using the inverse DFT, we deduce that $a_{\p{i}} = A_{\p{0}}$ for all $\p{i} \in G^m$. This proves that $\Cs_G(0,m)$ is the repetition code.

\emph{Part 3}. Since the zero-set $\Zs(m,m) = \emptyset$, we have $\Cs_G(m,m)=\Fb_2[G^m]$.
\end{IEEEproof}

The following results on $\Ds_G(r,m)$ are consequences of Lemmas~\ref{lem:time-reversal} and~\ref{lem:dual-abelian-codes-weight-classes}.

\begin{corollary}
For any $0 \leq r \leq m$, we have
\begin{enumerate}
\item $\Ds_G(r,m) = \Cs(\Zsc(r,m))$.
\item Both $\Cs_G(r,m)$ and $\Ds_G(r,m)$ are closed under time-reversal permutation.
\item $\Cs_G(r,m)$ and $\Ds_G(r,m)$ are complementary duals.
\end{enumerate}
\end{corollary}
\begin{IEEEproof}
The proofs of Parts 1 and 2 follow directly from Lemmas~\ref{lem:dual-abelian-codes-weight-classes} and~\ref{lem:time-reversal}, respectively.

\emph{Part 3.} Let $a \in \Cs_{G}(r,m) \cap \Ds_G(r,m)$. Consider $A = \Phi(a)$. If $\wt(\p{j}) \leq r$ then $A_{\p{j}}=0$ since $a \in \Ds_G(r,m)$. If $\wt(\p{j}) \geq r + 1$ then $A_{\p{j}}=0$ since $a \in \Cs_G(r,m)$. Hence, $A=0$ and $a=0$. Thus, $\Cs_{G}(r,m) \cap \Ds_G(r,m) = \{0\}$.
\end{IEEEproof}

We note that codes with complementary duals are known to be useful in cryptography~\cite{CaG_Springer_15}.

\subsubsection{Relation to the Classical Berman Codes}

The subclass of codes corresponding to $G=\Zb_p$ with $p$ being an odd prime was introduced and studied by Berman~\cite{Ber_Cybernetics_II_67}, and Blackmore and Norton~\cite{BlN_IT_01}. Both~\cite{Ber_Cybernetics_II_67} and~\cite{BlN_IT_01} provide a time-domain description of $\Ds_{\Zb_p}(r,m)$ and $\Cs_{\Zb_p}(r,m)$.
In Appendix~\ref{app:classical_berman_codes} we recall the original description of these codes from~\cite{BlN_IT_01} and show that this is identical to our transform-domain description for the case $G=\Zb_p$.

\subsubsection{Relation to $\Ds_n(r,m)$ and $\Cs_n(r,m)$}

We will use a technique from~\cite{BlN_IT_01} to identify a recursive structure for $\Cs_G(r,m)$, and use this structure to show that $\Cs_G(r,m)$ and $\Ds_G(r,m)$ are in fact the same as $\Cs_n(r,m)$ and $\Ds_n(r,m)$ for any choice of $G$ with $|G|=n$.
This technique was used in~\cite{BlN_IT_01} to show that the minimum distance of $\Cs_{\Zb_p}(r,m)$ is $p^{m-r}$ for any odd prime $p$.
The key difference between our work and~\cite{BlN_IT_01} is that we rely on the DFT tool for our analysis, and our results hold for any odd-ordered abelian group $G$.

Towards obtaining the recursive structure we will assume an ordering on the index set $G^m$ of the coordinates of any $a \in \Fb_2[G^m]$. Suppose the elements of $G$ are $\{g_0,\dots,g_{|G|-1}\}$ with $g_0=0$. We represent $a$ as
\begin{equation*}
a = \left(a_0~|~a_1~|~\cdots~|~a_{|G|-1}\right)
\end{equation*}
where $a_{\ell} \in \Fb_2[G^{m-1}]$ for each $\ell \in \lset\, |G| \,\rset$. We will use the standard notation
\begin{equation*}
a_{\ell} = \sum_{\p{i'} \in G^{m-1}} a_{\ell,\p{i'}} X^{\p{i'}}, \text{ where } a_{\ell,\p{i'}} \in \Fb_2,
\end{equation*}
and relate $a_0,\dots,a_{|G|-1}$ to $a$ as follows
\begin{equation*}
a = \sum_{\ell \in \lset \, |G| \,\rset} ~ \sum_{\p{i'} \in G^{m-1}} a_{\ell,\p{i'}} \, X^{(\p{i'}|g_{\ell})}.
\end{equation*}
Equivalently, if $a = \sum_{\p{i} \in G^m} a_{\p{i}} X^{\p{i}}$, and if for a given $\p{i} \in G^m$ we have $\p{i} = (i_0,\dots,i_{m-2},g_{\ell})$ for some choice of $\ell$, then $a_{\ell,(i_0,\dots,i_{m-2})} = a_{\p{i}}$.

\begin{lemma} \label{lem:direct-sum-dual-berman}
(Recursive structure of $\Cs_G(r,m)$.)
For any odd ordered abelian group $G$ and any \mbox{$1 \leq r \leq m-1$}, we have
\begin{align*}
\Cs_G(r,m) = \{ &(u+u_0|u+u_1|\cdots|u+u_{|G|-2}|u)~: \\ 
& u_l \in \Cs_G(r-1,m-1), u \in \Cs_G(r,m-1) \}.
\end{align*}
\end{lemma}
\begin{IEEEproof}
The full proof is available in Appendix~\ref{app:proofs}. The proof uses two fundamental properties of the DFT, viz., \emph{(i)}~the DFT of the indicator function of the subgroup $G^{m-1} \times \{0\}$ is the indicator function of $\{0\}^{m-1} \times G$; and \emph{(ii)}~the characterization of the DFT of the periodic sequence $a=(b~|~\cdots~|~b)$ in terms of the DFT of its fundamental copy $b$.
Both these results are also included as part of the proof of this lemma.

An intermediate step in the proof is to identify two subcodes $\Cs_1$ and $\Cs_2$ of $\Cs_G(r,m)$, and show that $\Cs_G(r,m)$ is a direct sum of $\Cs_1$ and $\Cs_2$.
A similar idea was used in~\cite{BlN_IT_01} to derive the minimum distance of $\Cs_{\Zb_p}(r,m)$.
Please see Appendix~\ref{app:proofs} for the complete proof.
\end{IEEEproof}

%% \begin{remark} \label{rem:RM_Berman_recursive_structure}
%% We note that the structure of Berman codes given in Lemma~\ref{lem:recursive-berman} is similar to the recursive $(u~|~u+v)$ construction of RM codes.
%% Recall that $\RM(r,m)$ is
%% \begin{align*}
%% \{(u~|~u + v)~|~u \in \RM(r,m-1), v \in \RM(r-1,m-1)\}.
%% \end{align*}
%% Using the fact $\RM(r-1,m-1) \subset \RM(r,m-1)$, we see that $\RM(r,m)$ can be equivalently described as
%% \begin{align*}
%% &\{(v_0~|~v_1)~|~v_0,v_1 \in \RM(r,m-1),\\
%% &~~~~~~~~~~~~~~~~~~~v_0+v_1 \in \RM(r-1,m-1)\}.
%% \end{align*}
%% Describing in terms of dual codes, and using $\RM(m-r-1,m) = \RM(r,m)^\perp$, we see that $\RM(r,m)^\perp$ is equal to
%% \begin{align*}
%% &\{(v_0~|~v_1)~|~v_0,v_1 \in \RM(r-1,m-1)^\perp,\\
%% &~~~~~~~~~~~~~~~~~~~v_0+v_1 \in \RM(r,m-1)^\perp\}.
%% \end{align*}
%% This has the same form as the recursive structure of Berman codes (Lemma~\ref{lem:recursive-berman}), where $|G|=2$ and the role of $\Cs_G(r,m)^\perp$ is filled by $\RM(r,m)^\perp$.
%% % This is similar to Lemma~\ref{lem:recursive-berman} where in $\RM(r,m)$, $\RM(r,m-1)$ and $\RM(r-1,m-1)$ take the roles of $\Cs_G(r,m)^\perp$, $\Cs_G(r-1,m-1)^\perp$ and $\Cs_G(r,m-1)^\perp$, respectively, and we have $|G|=2$.
%% \end{remark}

The recursive construction of $\Cs_G(r,m)$ in Lemma~\ref{lem:direct-sum-dual-berman} ultimately uses codes of the form $\Cs_G(0,m')$ and $\Cs_G(m',m')$, for some choices of $m'$, as its building blocks.
Note that $\Cs_G(0,m')$ and $\Cs_G(m',m')$ are the repetition code and the universe code of length $|G|^{m'}$, and hence, both these codes depend on $G$ only via $|G|$.
By using an induction argument, we conclude that the construction shown in Lemma~\ref{lem:direct-sum-dual-berman} depends on the choice of $G$ only through its order $|G|$.
Furthermore, we observe that this recursion is identical to the definition of $\Cs_n(r,m)$ if $n=|G|$.
Using this observation together with the facts $\Cs_G(r,m)^\perp=\Ds_G(r,m)$ and $\Csn(r,m)^\perp = \Dsn(r,m)$ immediately leads us to the following main result of this sub-section.

\begin{corollary} \label{cor:berman_dft_construction}
Let \mbox{$n \geq 3$} be an odd integer and let $G$ be any abelian group of order $n$. Then for all \mbox{$0 \leq r \leq m$},
\begin{equation*}
\Cs_G(r,m) = \Csn(r,m) \text{ and } \Ds_G(r,m)= \Dsn(r,m).
\end{equation*}
\end{corollary}

\subsection{A Family of Abelian Codes that Achieve BEC Capacity} \label{sec:sub:automorphisms}

We will now study a family of abelian codes that includes $\Cs_G(r,m)$ and $\Ds_G(r,m)$.
The family of codes that we are interested in are ideals $\Cs(Z)$ in $\Fb_2[G^m]$ with the property that the zero-set $Z$ is closed under certain permutations of $G^m$.
We detail these permutations next.

\subsubsection{Two Classes of Permutations on $G^m$}

We will first need the result that the map $i \to 2i$ is a permutation on $G$. If there exist $i, i' \in G$ such that $2i = 2i'$ then we have $2(i - i') = 0$. Since there is no element in $G$ of order $2$ (because $|G|$ is odd), we conclude that $i = i'$. Hence, $i \to 2i$ is a permutation on $G$.
We observe that for each \mbox{$j \in G$} there exists a unique element in $G$, which we will denote as $2^{-1}j$, such that $j = 2 \left( 2^{-1}j \right)$, i.e., $j = 2^{-1}j \oplus 2^{-1}j$.

For each $k \in \lset m \rset$, let $\pi_{k}$ be the permutation
\begin{equation*}
(i_0,\dots,i_{m-1}) \to (i_0,\dots,i_{k-1},2i_k,i_{k+1},\dots,i_{m-1}).
\end{equation*}
The inverse permutation $\pi_k^{-1}$ is
\begin{equation} \label{eq:inverse-pi-k}
(j_0,\dots,j_{m-1}) \to (j_0,\dots,j_{k-1},2^{-1}j_k,j_{k+1},\dots,j_{m-1}).
\end{equation}

Note that the permutation $i_k \to 2i_k$ on $G$ fixes $0$, hence, we see that \mbox{$i_k \neq 0$} if and only if \mbox{$2i_k \neq 0$}. Thus, both $\pi_k$ and $\pi_k^{-1}$ are weight preserving permutations on $G^m$, that is $\wt(\pi_k(\p{i})) = \wt(\p{i})$ for all $\p{i} \in G^m$.

Note that each \mbox{$\gamma \in \Ss_m$} is a permutation on $\lset m \rset$. We let $\gamma \in \Ss_m$ act on $G^m$ as follows
\begin{equation*}
\left( j_0,\dots,j_{m-1} \right) \to \left( j_{\gamma(0)},\dots,j_{\gamma(m-1)} \right).
\end{equation*}
With this group action we view $\gamma \in \Ss_m$ as a permutation on $G^m$. The inverse permutation is $\gamma^{-1}$, which is the inverse of $\gamma$ in $\Ss_m$.
Similar to $\pi_k$, we see that $\gamma$ is also a weight preserving permutation.

The following two lemmas show the interplay between the permutations $\gamma$ and $\pi_k$ and the DFT.

\begin{lemma} \label{lem:pi-k-DFT}
Let $k \in \lset m \rset$ and $a \in \Fb_2[G^m]$, and let $b \in \Fb_2[G^m]$ be the sequence obtained by applying $\pi_k$ on the coordinates of $a$, i.e., $b_{\pi_k(\p{i})} = a_{\p{i}}$ for all $\p{i} \in G^m$. Then
\begin{equation*}
B_{\p{j}} = A_{\pi_k^{-1}(\p{j})} \text{ for all } \p{j} \in G^m.
\end{equation*}
\end{lemma}
\begin{IEEEproof}
Viewing $j_k \in G = \Zb_{m_0} \times \cdots \Zb_{m_{s-1}}$ as an $s$-tuple we have
\begin{equation*}
2^{-1}j_k = (2^{-1}j_k[0], \, \dots, \, 2^{-1}j_k[s-1]),
\end{equation*}
where $2^{-1}j_k[\ell]$ is the product of $j_k[\ell]$ with the multiplicative inverse of $2$ in the ring $\Zb_{m_\ell}$ (since $m_\ell$ is odd, $2$ is a unit in $\Zb_{m_\ell}$).
For any $i_k[\ell], j_k[\ell] \in \Zb_{m_{\ell}}$, we observe that
\begin{equation} \label{eq:2_inv_ring_product}
2i_k[\ell] \cdot 2^{-1}j_k[\ell] = i_k[\ell] \cdot j_k[\ell].
\end{equation}

% Let $a,b \in \Fb_2[G]$ be such that $b_{\pi_k(\p{i})} = a_{\p{i}}$ for all $\p{i}$, that is, $b$ is the sequence obtained by applying the permutation $\pi_k$ on the coordinates of $a$.
% Note that $b_{\p{i}} = a_{\pi_k^{-1}(\p{i})}$.
Since $b_{\pi_k(\p{i})} = a_{\p{i}}$, the DFT of $b$ at $\p{j}$ is
\begin{align*}
B_{\p{j}} = \sum_{\p{i} \in G^m}  \left( \prod_{\substack{k' \in \lset m \rset \\ k' \neq k}} \prod_{\ell \in \lset s \rset} \alpha_{\ell}^{i_{k'}[\ell] \cdot j_{k'}[\ell]} \cdot \prod_{\ell \in \lset s \rset} \alpha_{\ell}^{2i_k[\ell]\cdot j_k[\ell]} \right)  a_{\p{i}}.
\end{align*}
Using the inverse permutation~\eqref{eq:inverse-pi-k} and the property~\eqref{eq:2_inv_ring_product}, we immediately deduce that $B_{\pi_k^{-1}(\p{j})} =A_{\p{j}}$ for all $\p{j} \in G^m$.
\end{IEEEproof}

\begin{lemma} \label{lem:gamma-dft}
Let $\gamma \in \Ss_m$, and $a,b \in \Fb_2[G^m]$ be such that $b_{\gamma(\p{i})} = a_{\p{i}}$ for all $\p{i} \in G^m$.
Then $B_{\p{j}} = A_{\gamma^{-1}(\p{j})}$ for all $\p{j} \in G^m$.
\end{lemma}
\begin{IEEEproof}
We note that
\begin{equation*}
B_{\p{j}} = \sum_{\p{i} \in G^m}  \left( \prod_{k \in \lset m \rset} \prod_{\ell \in \lset s \rset} \alpha_{\ell}^{i_{\gamma(k)}[\ell] \cdot j_{k}[\ell]} \right)  a_{\p{i}}.
\end{equation*}
Clearly,
\begin{align*}
B_{\p{j}} = \sum_{\p{i} \in G^m}  \left( \prod_{k \in \lset m \rset} \prod_{\ell \in \lset s \rset} \alpha_{\ell}^{i_{k}[\ell] \cdot j_{\gamma^{-1}(k)}[\ell]} \right)  a_{\p{i}} = A_{\gamma^{-1}(\p{j})}.
\end{align*}
This completes the proof.
\end{IEEEproof}

\subsubsection{A Family of Abelian Codes from $\Fb_2[G^m]$}

The codes studied in this sub-section are precisely the ideals $\Cs(Z) \subset \Fb_2[G^m]$ whose zero-set $Z$ is closed under all the below permutations
\begin{equation*}
\pi_k, k \in \lset m \rset \text{ and all } \gamma \in \Ss_m.
\end{equation*}
% From Lemma~\ref{lem:conj_class_same_weight}, we observe that such a $Z$ is a union of conjugacy classes, and hence, $\Cs(Z)$ is indeed an ideal.
% Note that $\Cs_G(r,m)$ and $\Ds_G(r,m)$ belong to this family of codes.

\begin{example} \label{ex:new_zero_sets}
%% To illustrate properties~\eqref{eq:gamma-req} and~\eqref{eq:pi-k-req}, we
Consider the group $G=\Zb_{15}$ and use $m=2$. In the additive group $\Zb_{15}=\{0,1,\dots,14\}$, the set $\{5,10\}$ forms a conjugacy class, i.e., this set is closed under multiplication by $2$ in the ring $\Zb_{15}$. Now consider
\begin{align*}
Z_1 = \left\{ (0,5), (0,10), (5,0), (10,0)  \right\} \subset \Zb_{15}^2.
\end{align*}
This is a union of two conjugacy classes $\{(0,5), (0,10)\}$ and $\{(5,0), (10,0) \}$.
Further, $Z_1$ is closed under the permutation $(j_0~j_1) \to (j_1~j_0)$. Clearly, $Z_1$ is a zero-set of an ideal in $\Fb_2[\Zb_{15}^2]$ and is closed under $\pi_0$, $\pi_1$ and all $\gamma \in \Ss_2$.

Another such choice of zero-set is
\begin{equation*}
Z_2 = \{(5,5), (5,10), (10,5), (10,10)\} \subset \Zb_{15}^2.
\end{equation*}
% % R0 % %
{
We observe that neither $Z_1$ nor $Z_2$ is the zero-set of $\Cs_{\Zb_{15}}(r,2)$ or $\Ds_{\Zb_{15}}(r,2)$ for any choice of $r$.
Further, the dimensions of these codes do not match with the dimensions of $\Csn(r,m)$ and $\Dsn(r,m)$ for $n=15$, $m=2$ and any choice of $r$.
Hence these codes are not Berman or dual Berman codes.
}
% % % % % %
\end{example}

\begin{remark} \label{rem:zero-sets_berman_automorphisms}
% We remark that if $W$ is a weight class in $G^m$, then it is closed under $\pi_0,\dots,\pi_{m-1}$ and $\gamma \in \Ss_m$. To see this, note that $\pi_k$ and $\gamma$ are weight preserving maps and $W$ is the set of all tuples in $G^m$ of a given weight.
% This immediately implies that
% % R0 % % %
{
Any zero-set that is a union of weight classes is closed under the permutations $\{\pi_0,\dots,\pi_{m-1}\}$ and $\Ss_m$.
Therefore, the zero-sets of $\Ds_G(r,m)$ and $\Cs_G(r,m)$ belong to this class of zero-sets.}
% % % % % % %
\end{remark}

We will now analyse some automorphisms of this family of abelian codes. Unless otherwise specified, we will assume that $Z$ is a zero-set closed under the actions of all permutations in $\{\pi_k: k \in \lset m \rset \} \cup \Ss_m$.
Since $\Cs(Z)$ is an ideal in a group algebra, it is transitive, see Lemma~\ref{lem:abelian-codes-transitive}.
From the proof of Lemma~\ref{lem:abelian-codes-transitive} we recognize that for each choice of $\p{k} \in G^m$ the following permutation of the set of coordinates $G^m$ is an automorphism of $\Cs(Z)$
\begin{equation*}
 \p{i} \to \p{i} \oplus \p{k},
\end{equation*}
which corresponds to multiplication of a codeword $a \in \Cs(Z)$ by the element $X^{\p{k}} \in \Fb_2[G^m]$.
We now identify two more classes of automorphisms of $\Cs(Z)$.

\begin{theorem} \label{thm:abelian-codes-automorphism}
Let \mbox{$Z \subset G^m$} be a zero-set that is closed under the actions of $\pi_0,\dots,\pi_{m-1}$ and all $\gamma \in \Ss_m$.
For each \mbox{$k \in \lset m \rset$}, and for each $\gamma \in \Ss_m$, the permutations $\pi_{k}$ and $\gamma$ applied on the coordinates $\p{i} \in G^m$ of a codeword $a = \sum_{\p{i} \in G^m} a_{\p{i}} X^{\p{i}} \in \Fb_2[G^m]$ are automorphisms of $\Cs(Z)$.
\end{theorem}
\begin{IEEEproof}
Let $a \in \Cs(Z)$ and $b$ be the sequence obtained by applying $\pi_k$ on the coordinates of $a$.
We know that $A_{\p{j}}=0$ for all $\p{j} \in Z$.
We want to show that $B_{\p{j}} = 0$ for all $\p{j} \in Z$.
Towards this, observe that $\pi_k(Z) = Z$, and hence, $\pi_k^{-1}(\p{j}) \in Z$.
Now using Lemma~\ref{lem:pi-k-DFT}, we have $B_{\p{j}} = A_{\pi_k^{-1}(\p{j})} = 0$ for all $\p{j} \in Z$. Hence, $b \in \Cs(Z)$.

The proof for $\gamma \in \Ss_m$ is similar and uses Lemma~\ref{lem:gamma-dft}.
\end{IEEEproof}

\begin{remark}
The proof of the $\pi_k$ part of Theorem~\ref{thm:abelian-codes-automorphism} mainly relies on the fact that the integer $2$ is a unit in all the rings $\Zb_{m_0},\dots,\Zb_{m_{s-1}}$.
These results in Lemma~\ref{lem:pi-k-DFT} and Theorem~\ref{thm:abelian-codes-automorphism} will hold if the role played by $2$ is taken by any integer that is a common unit of $\Zb_{m_0},\dots,\Zb_{m_{s-1}}$.
\end{remark}

\subsubsection{Achieving the Capacity of BEC}

Our proof for capacity achievability relies on the sufficient condition identified by Kumar et al.~\cite{KCP_ISIT16},
{and is similar to the proof of the capacity-achievability of Berman and dual Berman codes given in Section~\ref{sec:sub:capacity}.
}

{
% Let $\Gs$ be the automorphism group of $\Cs(Z)$.
Let $\Gs_0$ be the subgroup of automorphisms for which $\p{0}$ is a fixed point.
% Of main interest to us are the orbits of the non-zero points of $G^m$ under the action of $\Gs_0$.
% We will now derive a lower bound on the sizes of these orbits.
For each $\p{i} \in G^m \setminus \{\p{0}\}$ define
\begin{equation*}
\Os(\p{i}) = \left\{ \pi(\p{i})~:~\pi \in \Gs_0\right\},
\end{equation*}
to be the orbit of $\p{i}$ under the action of $\Gs_0$.
From Theorem~\ref{thm:abelian-codes-automorphism} we know that $\pi_k$, $k \in \lset m \rset$, and $\gamma \in \Ss_m$ are code automorphisms that are weight preserving on $G^m$,
% In particular, these automorphisms have $\p{0}$ as a fixed point,
and hence, they belong to $\Gs_0$.
}
We observe that the automorphisms $\pi_k,\pi_k^2,\dots$ acting on $\p{i} \in G^m$ generate the set of vectors
\begin{equation*}
\left\{\, \left(i_0,\dots,i_{k-1},j,i_{k+1},\dots,i_{m-1}\right)~~:~~j \in \Gamma_{i_k} \,\right\},
\end{equation*}
where $\Gamma_{i_k}$ is the conjugacy class of $i_k$ in the group $G$.
Considering the actions of all possible powers and products of $\pi_k$, $k \in \lset m \rset$, and $\gamma \in \Ss_m$, we conclude that $\Os(\p{i})$ includes
\begin{equation} \label{eq:orbit_berman_codes_includes}
\bigcup_{\gamma \in \Ss_m} \left\{ \, (j_0,\dots,j_{m-1}) \in G^m : j_{\gamma(k)} \in \Gamma_{i_k} \text{ for all } k \in \lset m \rset \, \right\}.
\end{equation}
Using the fact \mbox{$|\Gamma_{i_k}| \geq 2$} if \mbox{$i_k \in G \setminus \{0\}$} (see Lemma~\ref{lem:G-non-zero-conjugacy-class}), we observe that the size of~\eqref{eq:orbit_berman_codes_includes}, and hence $|\Os(\p{i})|$ is lower bounded by
\begin{equation*}
\binom{m}{\wt(\p{i})} \, 2^{\wt(\p{i})}.
\end{equation*}
% The minimum of $|\Os(\p{i})|$ over all non-zero $\p{i}$ plays a role in determining the performance of $\Cs(Z)$ in the BEC~\cite{KCP_ISIT16}.
We note that
\begin{equation} \label{eq:orbit-lower-bound}
\min_{\p{i} \in G^m \setminus \{\p{0}\}} |\Os(\p{i})| \geq \min_{w \in \{1,\dots,m\}} \binom{m}{w}2^w = 2m.
\end{equation}

\begin{theorem} \label{thm:capacity-abelian-codes}
Let $G$ be any odd-ordered abelian group with $|G| \geq 3$ and $\{m_l\}$ be a sequence of positive integers with $m_l \to \infty$. If for each $l$, $Z_l \subset G^{m_l}$ is a union of conjugacy classes satisfying the following conditions
\begin{enumerate}
\item $Z_l$ is closed under the actions of $\pi_0,\dots,\pi_{m_l-1}$ and $\Ss_{m_l}$, and
\item the rates of the sequence of abelian codes $\{\Cs(Z_l)\}$ converges to $R^*\in(0,1)$,
\end{enumerate}
then the bit erasure probability of the sequence of codes $\{\Cs(Z_l)\}$ converges to zero under bit-MAP decoding on any BEC with channel erasure probability $\epsilon\in[0,1-R^*)$.
Further, for any $R^* \in (0,1)$ there exists a sequence of abelian codes satisfying the above conditions.
\end{theorem}
\begin{IEEEproof}
% % % R0 % % % %
{
The proof of the first part of the theorem is similar to the proof of Theorem~\ref{thm:berman_codes_achieve_BEC_capacity}. It relies on~\cite[Theorem~19]{KCP_ISIT16} and uses the fact that the codes $\Cs(Z_l)$ are transitive, and via~\eqref{eq:orbit-lower-bound}, uses the observation $\min_{\p{i} \in G^{m_l} \setminus \{\p{0}\}} |\Os_l(\p{i})| \to \infty$ and $l \to \infty$, where $\Os_l(\p{i})$ is the orbit of $\p{i}$ under the action of the subgroup $\Gs_0$ of automorphisms of the code $\Cs(Z_l)$.
}
% % % % % % % %

% We observe that each code $\Cs(Z_l)$ is an ideal, and hence, transitive (see Lemma~\ref{lem:abelian-codes-transitive}).
% From Theorem~\ref{thm:abelian-codes-automorphism} and~\eqref{eq:orbit-lower-bound}, we have
% \begin{equation*}
% \min_{\p{i} \in G^{m_l} \setminus \{\p{0}\}} |\Os_l(\p{i})| \geq 2m_l
% \end{equation*}
% where $\Os_l(\p{i})$ is the orbit of $\p{i}$ under the action of the subgroup $\Gs_0$ of automorphisms of the code $\Cs(Z_l)$.
% Clearly,
% \begin{equation*}
% \min_{\p{i} \in G^{m_l} \setminus \{\p{0}\}} |\Os_l(\p{i})| \to \infty \text{ as } l \to \infty.
% \end{equation*}
% Now using~\cite[Theorem~19]{KCP_ISIT16}, we conclude that in any BEC with channel erasure probability less than $1-R^*$, the bit erasure probabilities of the sequence of codes $\{\Cs(Z_l)\}$ under bit-MAP decoding converges to $0$.

{
Towards proving the second part of the theorem, note that the family of abelian codes closed under $\{\pi_0,\dots,\pi_{m_l-1}\} \cup \Ss_{m_l}$ is large, and includes the Berman and dual Berman codes with $n=|G|$ as a strict subset (see Example~\ref{ex:new_zero_sets}, Remark~\ref{rem:zero-sets_berman_automorphisms} and Corollary~\ref{cor:berman_dft_construction}).
Hence, to complete the proof it is enough to show that there exists a sequence of dual Berman codes $\{\Csn(r_l,m_l)\}$ with rates converging to $R^*$. However, this was already shown as part of the proof of Theorem~\ref{thm:berman_codes_achieve_BEC_capacity}.
}
% From Theorem~\ref{thm:berman_codes_achieve_BEC_capacity} we know that for any choice of $R^* \in (0,1)$ there exists a sequence of dual Berman codes $\{\Csn(r_l,m_l)\}$ with rates converging to $R^*$. Choosing $n=|G|$, and using Corollary~\ref{cor:berman_dft_construction}, we observe that $\Csn(r_l,m_l)$ is an abelian code in $\Fb_2[G^{m_l}]$, and its zero-set $Z_l \subset G^{m_l}$ is a union of weight classes.
% From the discussion in Remark~\ref{rem:zero-sets_berman_automorphisms}, we observe that this zero-set $Z_l$ is closed under the actions of $\pi_0,\dots,\pi_{m_l-1}$ and $\Ss_{m_l}$.
% This completes the proof of the second part of this theorem.
\end{IEEEproof}

% % R0 version % %
{
The sequence of capacity-achieving codes identified in Theorem~\ref{thm:capacity-abelian-codes} are ideals in $\Fb_2[G^m]$. For instance, if $G$ is the cyclic group of order $n$, then the corresponding group algebra $\Fb_2[G^m]$ is isomorphic to 
\begin{equation*}
\Fb_2[X_0,\dots,X_{m-1}] / \langle X_0^n-1,\dots,X_{m-1}^n-1 \rangle, 
\end{equation*}
where $X_0,\dots,X_{m-1}$ are variables. To obtain a code sequence, we consider the sequence of algebras for increasing values of $m$ keeping $n$ fixed. That is, we obtain codes of increasing blocklengths by increasing the number of variables in the polynomial ring. Note that $m$ is the dimension of the DFT used for spectral analysis.

We can compare our construction with the approach used in~\cite[Corollary~21]{KCP_ISIT16} which identifies a family of capacity-achieving cyclic codes for the BEC. These codes from~\cite{KCP_ISIT16} are ideals in $\Fb_2[X_0] / \langle X_0^n - 1 \rangle$, where $X_0$ is a variable and $n$ equals the blocklength. A code sequence is obtained by considering a carefully identified sequence of blocklengths $n$ (in particular, each $n$ should be a divisor of $2^p-1$ for some choice of prime $p$). In contrast to our approach, the codes in~\cite{KCP_ISIT16} use $m=1$ variable only, and a code sequence is obtained by considering increasing values of $n$. In other words, while the Fourier-domain construction of our code sequence uses $m$-dimensional DFT of length $n$ per dimension (with increasing values of $m$, and fixing the value of $n$), the spectral description of the codes from~\cite[Corollary~21]{KCP_ISIT16} uses $1$-dimensional DFT of length $n$ (with increasing values of $n$ while fixing $m=1$).
}
% % % % % % % % %

\section{Simulation Results} \label{sec:simulations}

We present a few simulation results to get a glimpse of the performance of the codes identified in this work in comparison with the RM codes in the BEC.
For a given code $\Cs \subset \Fb_2^{n^m}$, we assume that the transmitted codeword $c = (c_i: i \in \lset n^m \rset)$ is picked uniformly at random from the codebook. The code symbols $c_i \in \{0,1\}$ are transmitted sequentially through the BEC with erasure probability $\epsilon$.
The channel output corresponding to the input $c_i$ is $y_i$, i.e., $\Pr(y_i=c_i) = 1-\epsilon$ and $\Pr(y_i = ?) = \epsilon$.
We use $y=(y_{{i}}~:~{i} \in \lset n^m \rset)$ to denote the channel output sequence, and
$y_{\sim {i}}$ to denote the collection of all channel outputs except $y_{{i}}$.
Please note that $c_{{i}}$ and $y_{{i}}$ are random variables.

The EXIT function for the ${i}^\tth$ transmitted bit is
% \begin{equation*}
$h_{{i}}(\epsilon) = \Hs(c_{{i}}|y_{\sim {i}})$,
% \end{equation*}
where $\Hs(\cdot|\cdot)$ is the conditional entropy. The average EXIT function is
\begin{equation*}
h(\epsilon) = \frac{1}{n^m} \sum_{{i} \in \lset n^m \rset} h_{{i}}(\epsilon).
\end{equation*}
All the codes considered in this section are transitive. For transitive codes we have \mbox{$h(\epsilon) = h_{{i}}(\epsilon)$} for all ${i}$~\cite{KKMPSU_IT_17}.
The average bit erasure probability under bit-MAP decoding satisfies $P_b = \epsilon h(\epsilon)$~\cite{KKMPSU_IT_17}.
%% \begin{align*}
%% P_b &= \frac{1}{|G|^m}\sum_{\p{i} \in G^m} \Hs(c_{\p{i}}|y) \\
%% &= \frac{1}{|G|^m} \sum_{\p{i} \in G^m} \Pr(y_{\p{i}}=?) \Hs(c_{\p{i}}|y_{\sim \p{i}}) \\
%% &= \frac{1}{|G|^m} \sum_{\p{i} \in G^m} \epsilon h_{\p{i}}(\epsilon) \\
%% &= \epsilon h(\epsilon).
%% \end{align*}
For transitive codes \mbox{$P_b = \epsilon h_{{i}}(\epsilon)$} for every \mbox{${i} \in \lset n^m \rset$}.
We perform bit-MAP decoding for exactly one of the transmitted bits, viz., $c_{{0}}$, and numerically estimate the value of $h_{{0}}(\epsilon)$. We then obtain the bit erasure probability as $P_b = \epsilon h_{{0}}(\epsilon)$.

% At small and medium block lengths it is troublesome to identify an abelian code from $\Fb_2[G^m]$ whose rate $R$ and block length $N$ are close to those of an RM code.
We have tried to compare codes with reasonably close rates and lengths.
To account for the difference in the rates we use $\epsilon - (1-R)$, instead of $\epsilon$, as the horizontal axis in our plots. Note that $\epsilon - (1-R)$ is the difference between the actual channel erasure probability and the capacity limit (i.e., $1-R$ is the highest possible channel erasure probability that any code of rate $R$ can withstand).

The codes considered in our simulations use \mbox{$n=3$}, or equivalently the abelian group \mbox{$G=\Zb_3$}. This abelian group belongs to the family of cyclic groups $(\Zb_p,+)$ of odd prime order $p$, with $2$ being a primitive root modulo $p$.
For such values of $p$, we explicitly identify the generator matrices of all abelian codes obtained from $\Fb_2[\Zb_p^m]$ in Appendix~\ref{app:gen_matrix_prime_group}.

\begin{figure}[!t]
\centering
\includegraphics[width=3.0in]{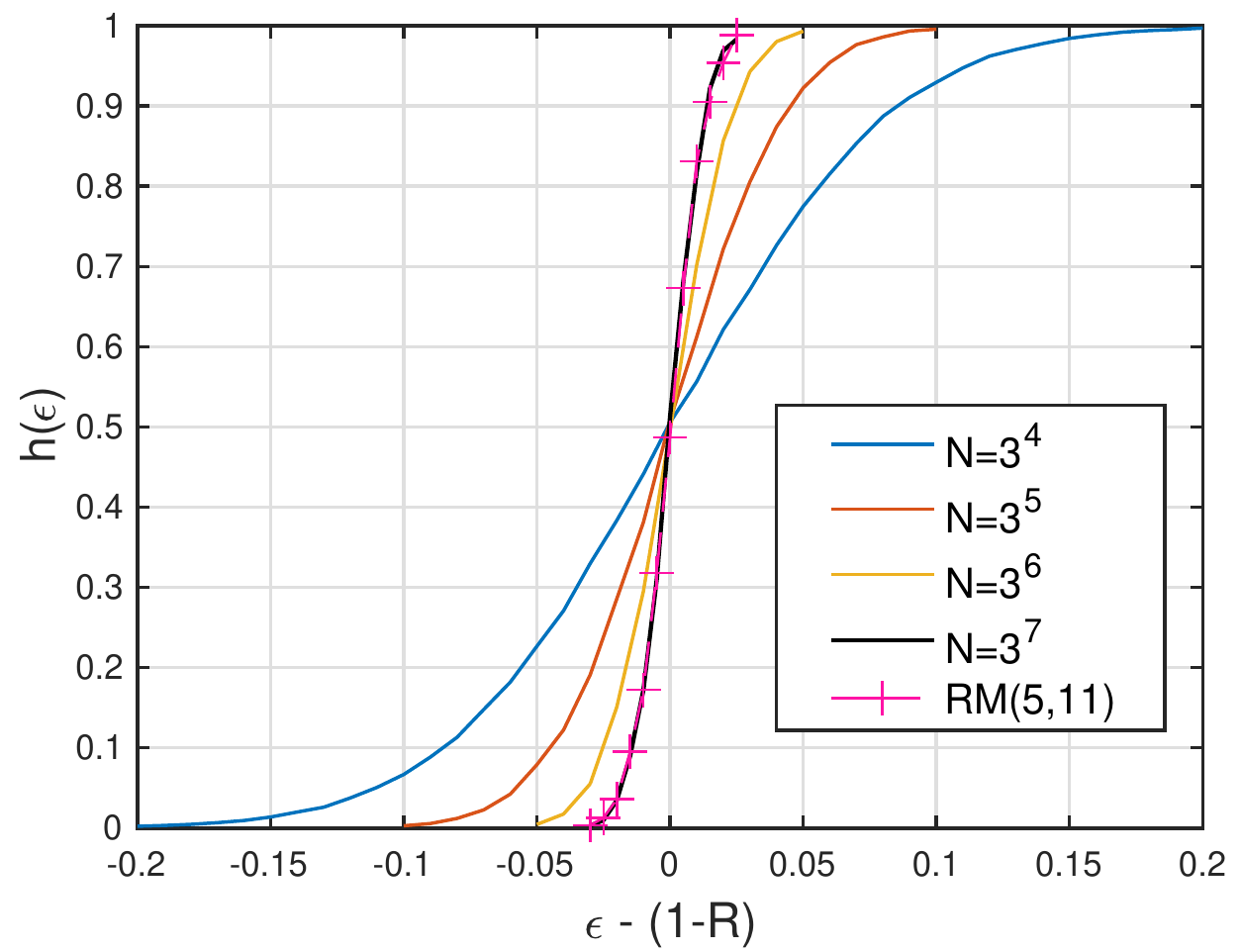}
\caption{EXIT functions of abelian codes from $\Fb_2[\Zb_3^m]$, with $m=4,5,6,7$, whose zero-set is the collection of all vectors of odd weight from $\Zb_3^m$. The EXIT function of rate $1/2$ RM code of length $2^{11}$ is also shown.}
\label{fig:threshold}
\end{figure}

\begin{table}[!t]
\renewcommand{\arraystretch}{1.25}
\centering
\caption{Basic parameters of the codes compared in Fig.~\ref{fig:threshold}.}
\begin{tabular}{|c|c|c|c|}
\hline
\hline
Code & Length & Dimension & Rate \\
\hline
\hline
Ideal in $\Fb_2[\Zb_3^4]$ & $81$ & $41$ & $0.5062$ \\
\hline
Ideal in $\Fb_2[\Zb_3^5]$ & $243$ & $121$ & $0.4979$ \\
\hline
Ideal in $\Fb_2[\Zb_3^6]$ & $729$ & $365$ & $0.5007$ \\
\hline
Ideal in $\Fb_2[\Zb_3^7]$ & $2187$ & $1093$ & $0.4998$ \\
\hline
$\RM(5,11)$ & $2048$ & $1024$ & $0.5$ \\
\hline
\hline
\end{tabular}
\vspace{2mm}
\label{tbl:codes_in_threshold_fig}
\end{table}

We illustrate the sharp $0$-to-$1$ transition of the EXIT function (for increasing block length $N$) in Fig.~\ref{fig:threshold} using a sequence of four codes with rates close to $1/2$. These codes are ideals in $\Fb_2[\Zb_3^m]$, with $m=4,5,6,7$, respectively. We choose the zero-set of the code of length $3^m$ as the collection of all odd-weight vectors in $\Zb_3^m$.
Clearly, this zero-set is a union of weight classes, and satisfies the conditions required by Theorem~\ref{thm:capacity-abelian-codes}. The basic parameters of these four codes are summarized in Table~\ref{tbl:codes_in_threshold_fig}.
Fig.~\ref{fig:threshold} also shows the EXIT function of the RM code of rate $0.5$ and length $2048$. The rate and dimension of this RM code are similar to those of the abelian code with \mbox{$m=7$}.

\begin{table}[!t]
\renewcommand{\arraystretch}{1.3}
\centering
\caption{Parameters of the codes compared in Fig.~\ref{fig:berman_all} and~\ref{fig:dual_berman_all}.}
\begin{tabular}{|c|c|c|c|c|}
\hline
\hline
Code & Length & Dimension & Rate & Minimum Distance\\
\hline
\hline
$\Ds_{3}(5,7)$ & $2187$ & $576$ & $0.2634$ & $64$ \\
\hline
$\RM(4,11)$ & $2048$ & $562$ & $0.2744$ & $128$ \\
\hline
\hline
$\Cs_{3}(5,7)$ & $2187$ & $1611$ & $0.7366$ & $9$ \\
\hline
$\RM(6,11)$ & $2048$ & $1486$ & $0.7256$ & $32$ \\
\hline
\hline
\end{tabular}
\vspace{2mm}
\label{tbl:berman_and_dual_berman}
\end{table}

\begin{figure}[!t]
\centering
\includegraphics[width=3.75in]{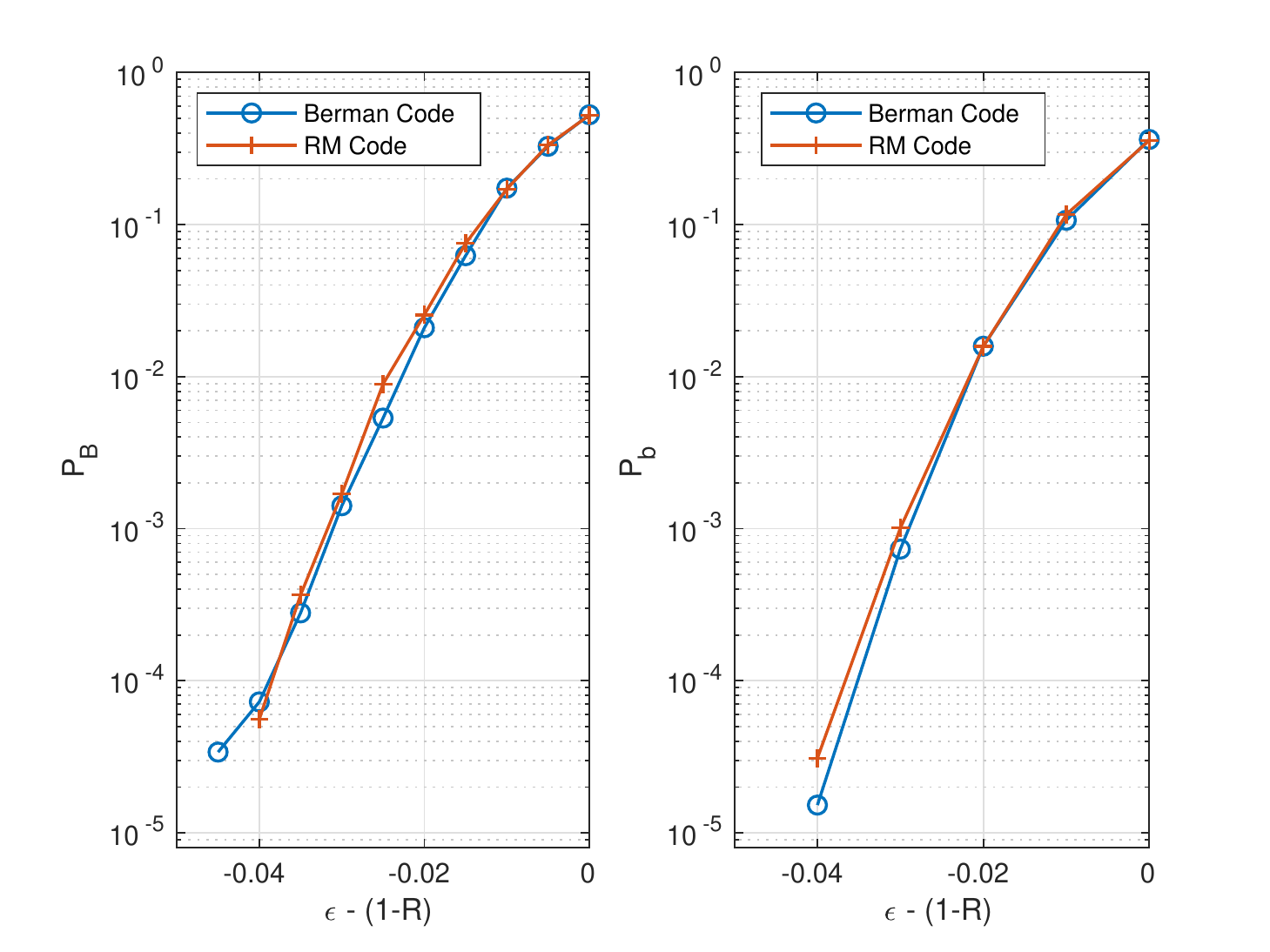}
\caption{The block erasure rate $P_B$ and bit erasure rate $P_b$ of $\Ds_{3}(5,7)$ and $\RM(4,11)$ in the BEC.}
\label{fig:berman_all}
\end{figure}

\begin{figure}[!t]
\centering
\includegraphics[width=3.75in]{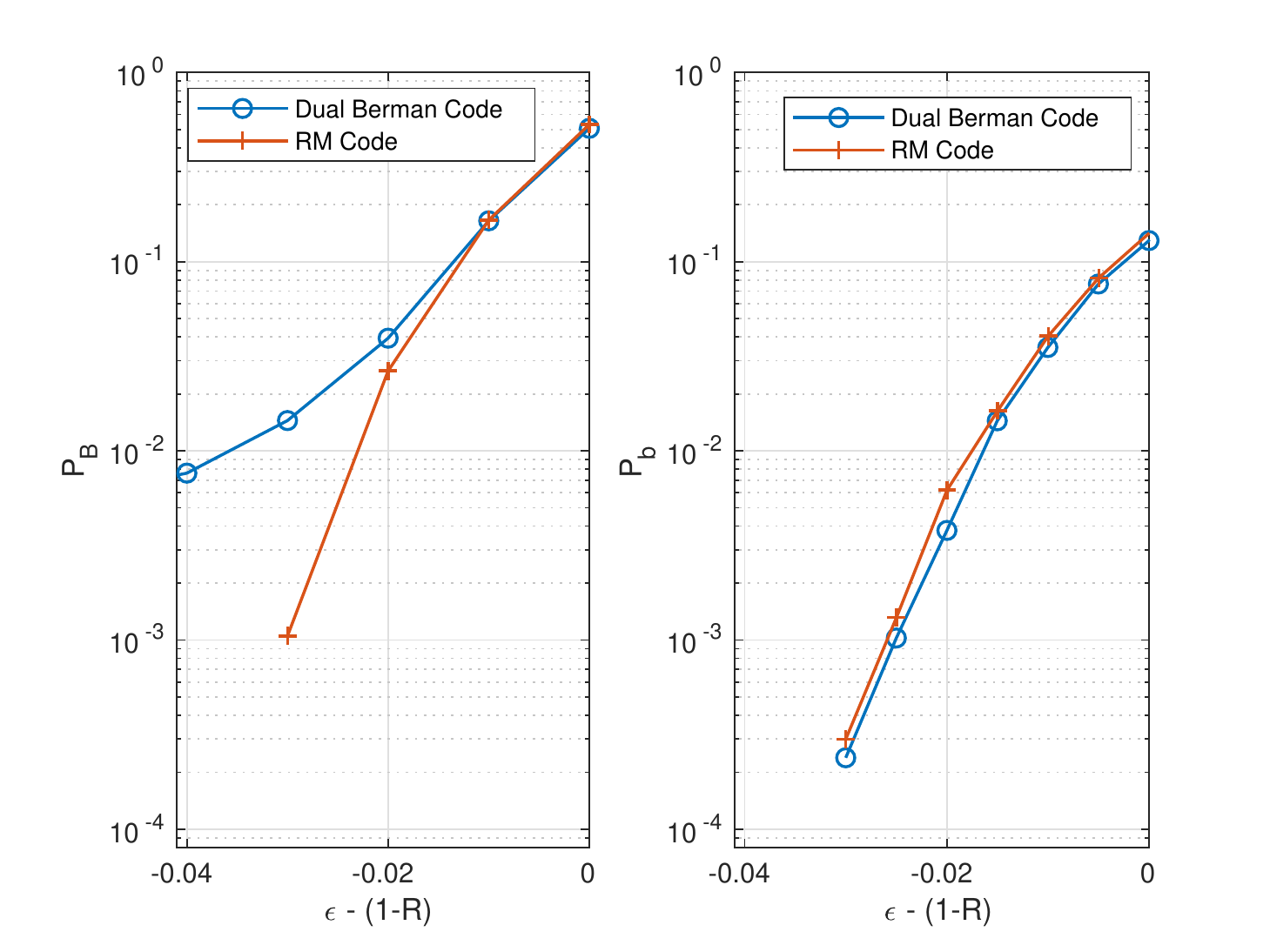}
\caption{The block erasure rate $P_B$ and bit erasure rate $P_b$ of $\Cs_{3}(5,7)$ and $\RM(6,11)$ in the BEC.}
\label{fig:dual_berman_all}
\end{figure}

We compare the block erasure rate $P_B$ (under block-MAP decoding) and bit erasure rate $P_b$ (under bit-MAP decoding) of $\Ds_3(5,7)$ and $\Cs_3(5,7)$ with RM codes in Fig.~\ref{fig:berman_all} and~\ref{fig:dual_berman_all}. The basic parameters of these codes are summarized in Table~\ref{tbl:berman_and_dual_berman}.
The code $\Ds_3(5,7)=\Ds_{\Zb_3}(5,7)$ belongs to a class of codes identified by Berman~\cite{Ber_Cybernetics_II_67}, and $\Cs_3(5,7)=\Cs_{\Zb_3}(5,7)$ is the dual of this Berman code.
In the simulation scenarios presented in Fig.~\ref{fig:berman_all} and~\ref{fig:dual_berman_all}, the bit erasure rates $P_b$ of $\Ds_3(5,7)$ and $\Cs_3(5,7)$ are similar to those of RM codes of similar rate and block length.
However, note the small minimum distance and the high floor in the block erasure rate of $\Cs_{3}(5,7)$ in Fig.~\ref{fig:dual_berman_all}.

\section{Discussion} \label{sec:discussion}

We identified a family of codes, that includes the RM codes~\cite{Reed_IRE_54,Muller_IRE_54} (\mbox{$n=2$}) and the Berman codes~\cite{Ber_Cybernetics_II_67,BlN_IT_01} ($n$ an odd prime) and whose properties are similar to RM codes.
When $n$ is odd, we used the semi-simple structure of $\Fb_2[G^m]$ and its associated DFT to identify a larger class of codes that achieve the BEC capacity.
While the similarity of $\Csn(r,m)$ and $\Dsn(r,m)$ to RM codes is striking, there are some key differences as well, especially in terms of the minimum distance and the automorphism group.
    Note that this family contains codes corresponding to a richer set of block lengths and dimensions than RM codes.
The code properties identified in Section~\ref{sec:capacity-related} and the simulation results presented in Section~\ref{sec:simulations} support the idea that these codes might enjoy a good performance in BMS channels when the block lengths are large.

The current work exposes some questions that seem to deserve investigation.
\begin{enumerate}
\item Our guarantees on capacity achievability in the BEC are in the sense of vanishing bit erasure probability $P_b$. It is not clear if these codes have vanishing block erasure probability $P_B$ under block-MAP decoding for rates close to capacity limit.
% % RO % %
\item It is possible that some of the differences of $\Csn(r,m)$ and $\Dsn(r,m)$ with RM codes might offer advantages.
{For large block lengths, the minimum distance of $\Csn(r,m)$ is significantly smaller than that of RM codes. This implies that its dual $\Dsn(r,m)$ has a parity-check matrix that is considerably sparser than the parity-check matrix of RM codes.
This sparsity might be useful in designing low complexity iterative decoders for $\Dsn(r,m)$, see~\cite{SHP_ISIT_18}.}
% % % % % % %
\end{enumerate}

\appendices

\section{Proofs} \label{app:proofs}

% \subsection{Proof of The Modulation Property of DFT}

% Let $a,b,c \in \Fb_2[G]$ be such that $c_i = a_ib_i$ for all $i \in G$. Then taking the DFT of $c$ we have
% \begin{align*}
% C_j &= \sum_{i \in G} \prod_{\ell \in \lset s \rset} \alpha_{\ell}^{i[\ell]\cdot j[\ell]} a_i b_i \\
% &=  \sum_{i \in G} \prod_{\ell \in \lset s \rset} \alpha_{\ell}^{i[\ell]\cdot j[\ell]} \left( \sum_{k \in G} \prod_{\ell' \in \lset s \rset} \alpha_{\ell}^{-i[\ell']\cdot k[\ell']}A_k \right) b_i \\
% &= \sum_{k \in G} A_k \sum_{i \in G} \prod_{\ell \in \lset s \rset} \alpha_{\ell}^{-i[\ell] \cdot (j[\ell] - k[\ell])} b_i \\
% &= \sum_{k \in G} A_kB_{j \ominus k}.
% \end{align*}
% Hence, $C=AB$ in $\Fb_q[G]$.

\subsection{Proof of Lemma~\ref{lem:rate-change:Berman}}

We first observe that $1-R_n(r,m) \geq 1 - R_n(r+k,m+k)$. To see this we use the fact that $R_n(r+k,m+k) - R_n(r,m)$ equals
\begin{equation*}
\Pr(X_0+\cdots+X_{m+k-1} \leq r+k) - \Pr(X_0+\cdots+X_{m-1} \leq r).
\end{equation*}

Using similar ideas as in the proof of Lemma~\ref{lem:rate-change:dual_Berman} we can show that the difference in the rates of $\Dsn(r,m)$ and \mbox{$\Dsn(r+k,m+k)$} is at the most
\begin{align*}
\frac{2\kappa}{\sqrt{m}} + \frac{1}{\sqrt{2\pi}} \left| \frac{r+k-(m+k)\mu}{\sigma\sqrt{m+k}} - \frac{r-m\mu}{\sigma\sqrt{m}} \right|.
\end{align*}
Using triangle inequality, we see that the second term in the above sum is at the most
\begin{align*}
% & \frac{1}{\sigma \sqrt{2 \pi m (m+k)}} \times \\ 
\frac{|(r+k)\sqrt{m} - r\sqrt{m+k}| + \mu\sqrt{m(m+k)}\,\,|\sqrt{m+k} - \sqrt{m}|}{\sigma \sqrt{2 \pi m (m+k)}}.
\end{align*}
Now using $\sqrt{m+k} - \sqrt{m} \leq \frac{k}{2\sqrt{m}}$ we can show that \mbox{$(r+k)\sqrt{m} - r\sqrt{m+k} \geq 0$}. Making use of the fact $k \geq 0$, we then have
\begin{equation*}
0 \leq (r+k)\sqrt{m} - r\sqrt{m+k} \leq k\sqrt{m}.
\end{equation*}
Straightforward algebraic manipulations then lead us to the following upper bound on the rate difference
\begin{align*}
\frac{2\kappa}{\sqrt{m}} + \frac{k}{\sqrt{m}}\frac{(\mu + 2)}{\sqrt{8\pi\sigma^2}}.
\end{align*}

\subsection{Proof of Lemma~\ref{lem:direct-sum-dual-berman}}

We first need the following basic results about the DFT.

\begin{lemma} \label{lem:dft-indicator-subgroup}
(DFT of the indicator function of a direct-product subgroup.)
If $a$ is the indicator function of \mbox{$G^{m-1} \times \{0\}$} then its DFT $A$ is the indicator function of $\{0\}^{m-1} \times G$.
\end{lemma}
\begin{IEEEproof}
We need to show that if \mbox{$A \in \Fb_q[G^m]$} is such that $A_{\p{j}} = 1$ if $(j_0,\dots,j_{m-2})=\p{0}$ and $A_{\p{j}}=0$ otherwise, then its inverse DFT is such that $a_{\p{i}} = 1$ if $i_{m-1}=0$ and $a_{\p{i}} = 0$ otherwise.
Taking the inverse DFT of $A$ we have
\begin{align*}
a_{\p{i}} &= \sum_{\substack{\p{j} \in G^{m}\\j_0,\dots,j_{m-2}=0}} \left( \prod_{k \in \lset m \rset } \prod_{\ell \in \lset s \rset} \alpha_{\ell}^{-i_k[\ell] \cdot j_k[\ell]} \right) \cdot 1 \\
&= \sum_{j_{m-1} \in G} \prod_{\ell \in \lset s \rset} \alpha_{\ell}^{-i_{m-1}[\ell] \cdot j_{m-1}[\ell]} \\
&= \sum_{n_0 \in \Zb_{m_0}} \cdots \sum_{n_{s-1} \in \Zb_{m_{s-1}}} \prod_{\ell \in \lset s \rset} \alpha_{\ell}^{-i_{m-1}[\ell] \cdot n_{\ell}} \\
&= \prod_{\ell \in \lset s \rset} \left( \sum_{n_{\ell} \in \Zb_{m_{\ell}}}  \alpha_{\ell}^{-i_{m-1}[\ell] \cdot n_{\ell}} \right).
\end{align*}
Since $\alpha_{\ell}$ is a primitive $m_{\ell}^\tth$ root of unity we know that
\begin{equation*}
\sum_{n_{\ell} \in \Zb_{m_{\ell}}} \!\!\! \alpha_{\ell}^{-i_{m-1}[\ell] \cdot n_{\ell}} = 1 + \alpha_{\ell}^{-i_{m-1}[\ell]} + \cdots + \alpha_{\ell}^{-i_{m-1}[\ell](m_{\ell}-1)}
\end{equation*}
is equal to $0$ if $i_{m-1}[\ell] \neq 0$ and is equal to $m_{\ell} \mod 2 = 1$ if $i_{m-1}[\ell] = 0$.
Thus we conclude that $a_{\p{i}} = 1$ if $i_{m-1}=0$ and $a_{\p{i}}=0$ otherwise.
\end{IEEEproof}

\begin{lemma} \label{lem:dft-periodic-sequence}
(DFT of a periodic sequence.)
Let $b \in \Fb_2[G^{m-1}]$, and let $a= (b~|~b~|~\cdots~|~b) \in \Fb_2[G^m]$.
The DFT of $a$ is related to that of $b$ as follows: $A_{\p{j}} = 0$ if $j_{m-1} \neq 0$ and $A_{\p{j}} = B_{(j_0,\dots,j_{m-2})}$ otherwise.
\end{lemma}
\begin{IEEEproof}
Let us assume that $b$ is given and $a$ is defined via its DFT as: $A_{\p{j}} = 0$ if $j_{m-1} \neq 0$ and $A_{\p{j}} = B_{(j_0,\dots,j_{m-2})}$ otherwise.
% Using the fact $B^2_{\p{j'}} = B_{2\p{j'}}$ for all $\p{j'} \in G^{m-1}$, it is straightforward to show that $A^2_{\p{j}} = A_{2\p{j}}$ for all $\p{j} \in G^m$, and hence, $a$ is indeed an element of $\Fb_2[G^m]$.
We now show that $a$ is the $m$-fold repetition of $b$.

Let $\p{i} \in G^{m}$ and $\p{i'} = (i_0,\dots,i_{m-2})$. From the inverse DFT, we have
\begin{align*}
a_{\p{i}} &= \sum_{\substack{\p{j} \in G^{m}\\j_{m-1}=0}} \left( \prod_{k \in \lset m \rset } \prod_{\ell \in \lset s \rset} \alpha_{\ell}^{-i_k[\ell] \cdot j_k[\ell]} \right) A_{\p{j}} \\
&= \sum_{\substack{\p{j} \in G^{m}\\j_{m-1}=0}} \left( \prod_{k \in \lset m-1 \rset } \prod_{\ell \in \lset s \rset} \alpha_{\ell}^{-i_k[\ell] \cdot j_k[\ell]} \right) A_{\p{j}} \\
&= \sum_{\substack{\p{j'} \in G^{m-1}}} \left( \prod_{k \in \lset m-1 \rset } \prod_{\ell \in \lset s \rset} \alpha_{\ell}^{-i_k[\ell] \cdot j'_k[\ell]} \right) A_{(\p{j'}|0)} \\
&= \sum_{\p{j'} \in G^{m-1}} \left( \prod_{k \in \lset m-1 \rset } \prod_{\ell \in \lset s \rset} \alpha_{\ell}^{-i'_k[\ell] \cdot j'_k[\ell]} \right) B_{\p{j'}} \\
&= b_{\p{i'}},
\end{align*}
where the second equality uses the fact $j_{m-1} = 0$, and the third equality uses the change of variables $\p{j} = (\p{j'}|0)$. Hence we have
\begin{equation*}
a = \sum_{\p{i'} \in G^{m-1}} ~ \sum_{\ell \in \lset \, |G| \, \rset} b_{\p{i'}} \, X^{(\p{i'}|g_{\ell})}.
\end{equation*}
This shows that $a = (b~|~\cdots~|~b)$.
\end{IEEEproof}

We are now ready to prove Lemma~\ref{lem:direct-sum-dual-berman}.
Let $\Gamma_{r,m-1}$ be the set of all vectors of weight $r$ in $G^{m-1}$.
Using Lemma~\ref{lem:conj_class_same_weight}, we observe that $\Gamma_{r,m-1}$ is a union of conjugacy classes. We define $\Cs(G^{m-1} \setminus \Gamma_{r,m-1})$ to be the ideal in $\Fb_2[G^{m-1}]$ whose non-zeros are $\Gamma_{r,m-1}$. Note that the dimension of \mbox{$\Cs(G^{m-1} \setminus \Gamma_{r,m-1})$} is $|\Gamma_{r,m-1}| = \binom{m-1}{r}(|G|-1)^r$.

In the first stage of the proof, we will show that $\Cs_G(r,m)$ is the direct sum of
\begin{align*}
\Cs_1 &= \left\{\, (b_0~|~\cdots~|~b_{|G|-1}) ~:~b_{\ell} \in \Cs_G(r-1,m-1) \, \right\}, \text{ and} \\
\Cs_2 &= \left\{\, (b~|~b~|~\cdots~|~b) ~:~ b \in  \Cs(G^{m-1}\setminus \Gamma_{r,m-1}) \, \right\}.
\end{align*}
We will complete the proof of this first stage by showing
\begin{enumerate}
\item $\Cs_1$ is a subcode of $\Cs_G(r,m)$,
\item $\Cs_2$ is a subcode of $\Cs_G(r,m)$,
\item $\Cs_1 \cap \Cs_2 = \{0\}$, and
\item $\dim(\Cs_1) + \dim(\Cs_2) = \dim(\Cs_G(r,m))$.
\end{enumerate}
We now provide the proofs of these results.

\subsubsection{$\Cs_1$ is a subcode of $\Cs_G(r,m)$}

Observe that it is sufficient to show that
\begin{equation*}
\{~(b~|~0~|~0~|~\cdots~|~0)~~:~~b \in \Cs_G(r-1,m-1)~\}
\end{equation*}
is a subcode of $\Cs_G(r,m)$.
The transitive property of $\Cs_G(r,m)$ (this follows from Lemma~\ref{lem:abelian-codes-transitive}) can then be leveraged to show that elements of the form $(0~|~b~|~0~|~\cdots~|~0)$ etc. also belong to $\Cs_G(r,m)$.

Now suppose $b \in \Cs_G(r-1,m-1)$.
Let us define $a \in \Fb_2[G^m]$ as $(b~|~\cdots~|~b)$.
Using the fact $B_{\p{j}} = 0$ if $\wt(\p{j}) \geq r$ and using Lemma~\ref{lem:dft-periodic-sequence} we deduce that $A_{\p{j}} = 0$ if $\wt(\p{j}) \geq r$.
Hence, $a \in \Cs_G(r-1,m)$.

Let $c \in \Fb_2[G^m]$ be the indicator function of $G^{m-1} \times \{0\}$, and $d \in \Fb_2[G^m]$ be defined as $d_{\p{i}} = c_{\p{i}}a_{\p{i}}$ for all $\p{i}$.
That is, $d=(b~|~0~|~\cdots~|~0)$.
From the modulation property of DFT we deduce that the DFT of $d$ is the convolution of $C$ and $A$. Hence,
\begin{equation*}
D = \sum_{\p{j} \in G^m} \sum_{\p{k} \in G^m} C_{\p{j}} A_{\p{k}} X^{\p{j} \oplus \p{k}}.
\end{equation*}
Now using Lemma~\ref{lem:dft-indicator-subgroup} with respect to $C$ we obtain
\begin{align*}
D &= \sum_{\substack{\p{j} \in G^m \\ j_0,\dots,j_{m-2} = 0}} \sum_{\p{k} \in G^m} A_{\p{k}} X^{\p{j} \oplus \p{k}} \\
&= \sum_{\substack{\p{j} \in G^m \\ j_0,\dots,j_{m-2} = 0}}~ \sum_{\substack{\p{k} \in G^m\\ \wt(\p{k}) \leq r-1}}A_{\p{k}} X^{\p{j} \oplus \p{k}},
\end{align*}
where in the last step we have used the fact $a \in \Cs_G(r-1,m)$.
If $j_0,\dots,j_{m-2}=0$ and $\wt(\p{k}) \leq r-1$, we have $\wt(\p{j} \oplus \p{k}) \leq \wt(\p{k}) + 1 \leq r$. Hence $d \in \Cs_G(r,m)$.

\subsubsection{$\Cs_2$ is a subcode of $\Cs_G(r,m)$}

Let $b \in \Cs(G^{m-1} \setminus \Gamma_{r,m-1})$, and $a$ be the $m$-fold repetition of $b$.
Using the fact $B_{\p{j'}} = 0$ if $\wt(\p{j'}) \neq r$, and using Lemma~\ref{lem:dft-periodic-sequence}, we deduce that $A_{\p{j}} = 0$ if $\wt(\p{j}) \geq r+1$.
Hence, $a \in \Cs_G(r,m)$.

\subsubsection{$\Cs_1 \cap \Cs_2 = \{0\}$}

Let $(b~|~b~|~\cdots~|~b) \in \Cs_1 \cap \Cs_2$. Then
\begin{equation*}
b \in \Cs_G(r-1,m-1) \cap \Cs(G^{m-1} \setminus \Gamma_{r,m-1}).
\end{equation*}
The non-zeros of $\Cs_G(r-1,m-1)$ are the vectors in $G^{m-1}$ of weight at the most $r-1$, and the non-zeros of \mbox{$\Cs(G^{m-1} \setminus \Gamma_{r,m-1})$} are the vectors of weight $r$. Hence, $B_{\p{j'}}=0$ for all $\p{j'} \in G^{m-1}$. We deduce that $b=0$, and therefore, $\Cs_1 \cap \Cs_2 = \{0\}$.

\subsubsection{Comparing the dimensions}

The dimension of $\Cs_1$ is $|G|\dim\left( \Cs_G(r-1,m-1) \right)$ and the dimension of $\Cs_2$ is equal to $\dim\left( \Cs(G^m \setminus \Gamma_{r,m-1}) \right)$. We have
$\dim(\Cs_1 + \Cs_2)$
\begin{align*}
 &=\dim(\Cs_1) + \dim(\Cs_2) \\
&= |G|\sum_{w=0}^{r-1}\binom{m-1}{w}(|G|-1)^w + |\Gamma_{r,m-1}| \\
&= |G|\sum_{w=0}^{r-1}\binom{m-1}{w}(|G|-1)^w + \binom{m-1}{r}(|G|-1)^r.
\end{align*}
Using $|G| = 1 + (|G|-1)$, we have $\dim(\Cs_1 + \Cs_2)$
\begin{align*}
&= \sum_{w=0}^{r}\binom{m-1}{w}(|G|-1)^w + \sum_{w=0}^{r-1}\binom{m-1}{w}(|G|-1)^{w+1} \\
&= \binom{m-1}{0} + \sum_{w=1}^{r}\left( \binom{m-1}{w} +\binom{m-1}{w-1} \right) (|G|-1)^w \\
&= \sum_{w=0}^{r} \binom{m}{w}(|G|-1)^w \\
&= \dim\left( \Cs_G(r,m) \right).
\end{align*}
This is the end of the first stage of the proof of Lemma~\ref{lem:direct-sum-dual-berman}.

To complete the proof we now apply a change of variables in the direct sum decomposition of $\Cs_G(r,m)$ into $\Cs_1$ and $\Cs_2$.
We define $u,u_0,\dots,u_{|G|-2}$ in terms of $b,b_0,\dots,b_{|G|-1}$ as follows
\begin{align*}
u = b_{|G|-1} + b, \text{ and } u_l = b_l + b_{|G|-1} \text{ for all } l \in \lset |G|-1 \rset.
\end{align*}
Hence, every element of $\Cs_G(r,m)$ is of the form 
\begin{equation*}
(u+u_0|\cdots|u + u_{|G|-2}|u).
\end{equation*}
Since $b \in \Cs(G^{m-1} \setminus \Gamma_{r,m-1})$ and $b_l \in \Cs_G(r-1,m-1)$, and since $\Cs_G(r,m-1)$ is the direct sum of $\Cs(G^{m-1} \setminus \Gamma_{r,m-1})$ and $\Cs_G(r-1,m-1)$, we deduce that
\begin{equation*}
u \in \Cs_G(r,m-1) \text{ and } u_0,\dots,u_{|G|-2} \in \Cs_G(r-1,m-1).
\end{equation*}
The last piece of argument required to show that $\Cs_G(r,m)$ is equal to
\begin{align} 
\{ &(u+u_0|u+u_1|\cdots|u+u_{|G|-2}|u)~: \nonumber \\
&~~~~~~~~u_l \in \Cs_G(r-1,m-1), u \in \Cs_G(r,m-1) \} \label{eq:app:berman-recursive:1}
\end{align}
is based on the dimension of these two codes.
Clearly, the dimension of the code in~\eqref{eq:app:berman-recursive:1} is
\begin{align*}
&=\dim(\Cs_G(r,m-1)) + (|G|-1)\dim(\Cs_G(r-1,m-1)) \\
&= \sum_{w=0}^{r}\binom{m-1}{w}(|G|-1)^w ~+~\sum_{w=0}^{r-1}\binom{m-1}{w}(|G|-1)^{w+1} \\
&= \sum_{w=0}^{r}\binom{m}{w}(|G|-1)^w \\
&= \dim(\Cs_G(r,m)).
\end{align*}
Thus, we conclude that the code in~\eqref{eq:app:berman-recursive:1} is indeed $\Cs_G(r,m)$.

\section{Relation to the Classical Codes of Berman and Blackmore-Norton} \label{app:classical_berman_codes}

Let $p$ be any odd prime, $G=\Zb_p$, and let $\Fb_q$ be a finite algebraic extension of $\Fb_2$ that contains a primitive $p^\tth$ root of unity $\alpha$.
Since $p$ is prime each of $\alpha,\alpha^2,\dots,\alpha^{p-1}$ is a primitive $p^\tth$ root of unity. For any $\ell \in \{1,\dots,p-1\}$, we have
\begin{equation*}
1 + \alpha^\ell + \alpha^{2\ell} + \cdots + \alpha^{(p-1)\ell} = 0.
\end{equation*}

The DFT and inverse DFT in $\Fb_2[\Zb_p^m]$ are
\begin{equation*}
A_{\p{j}} = \sum_{\p{i} \in \Zb_p^m} \alpha^{\p{i}\cdot\p{j}} a_{\p{i}} ~\text{ and }~ a_{\p{i}} = \sum_{\p{j} \in \Zb_p^m} \alpha^{-\p{i}\cdot\p{j}} A_{\p{j}},
\end{equation*}
where $\p{i} \cdot \p{j} = \sum_{k \in \lset m \rset} i_kj_k$ is the dot product in $\Zb_p^m$.

We will use the definition used by Blackmore and Norton~\cite{BlN_IT_01} as our point of reference. We will recall this definition using the notation followed in Section~\ref{sec:dft}.

For $k \in \lset m \rset$, let $\p{e}_k \in \Zb_p^m$ be the vector with a $1$ in the $k^\tth$ position and $0$ elsewhere. Define
\begin{equation*}
a^{(k)} = X^{\p{0}} + X^{\p{e}_k} + X^{2\p{e}_k} + \cdots + X^{(p-1)\p{e}_k} = \sum_{\beta \in \Zb_p} X^{\beta \p{e}_k}.
\end{equation*}
Note that $a^{(k)}$ is denoted as $e_p(X_k)$ in~\cite{BlN_IT_01}.
Let $A^{(k)}$ be the DFT of $a^{(k)}$. Then
\begin{align*}
A^{(k)}_{\p{j}} = \sum_{\beta \in \Zb_p} \alpha^{\beta \p{e}_k \cdot \p{j}} = \sum_{\beta \in \Zb_p} \alpha^{\beta j_k}.
\end{align*}
Thus, $A^{(k)}_{\p{j}} = 1$ if $j_k=0$ and $A^{(k)}_{\p{j}} = 0$ if $j_k \neq 0$.

Now consider $b^{(k)} = 1 + a^{(k)}$, where $1 = X^{\p{0}}$ is the multiplicative identity of $\Fb_2[\Zb_p^m]$. Since the DFT of $X^{\p{0}}$ is $\sum_{\p{j} \in \Zb_p^m} X^{\p{j}}$, we see that the DFT of $b^{(k)}$ is given by $B^{(k)}_{\p{j}} = 1 + A^{(k)}_{\p{j}}$ which equals $0$ if $j_k=0$ and equals $1$ if $j_k \neq 0$.

Blackmore and Norton define $E_p(r,m) \subset \Fb_2[\Zb_p^m]$ to be the set of all products of the form
\begin{equation} \label{eq:app:Eprm_element}
b^{(k_0)}\cdots b^{(k_{s-1})} \cdot a^{(k_{s})} \cdots a^{(k_{m-1})},
\end{equation}
where $\{k_0,\dots,k_{m-1}\} = \lset m \rset$ and $0 \leq s \leq r$.
The DFT of~\eqref{eq:app:Eprm_element} at $\p{j} \in \Zb_p^m$ is equal to $1$ if $j_{k_0},\dots,j_{k_{s-1}} \neq 0$ and $j_{k_s},\dots,j_{k_{m-1}}=0$, and the DFT at $\p{j}$ is equal to $0$ otherwise.
That is, the DFT of~\eqref{eq:app:Eprm_element} at $\p{j}$ is equal to $\mathbbm{1}\left( \supp(\p{j}) = \{k_0,\dots,k_{s-1}\} \right)$ where $\mathbbm{1}$ denotes the indicator function and $\supp$ is the support set of a vector.

Let $\langle E_p(r,m) \rangle$ be the ideal generated by the elements of $E_p(r,m)$ in $\Fb_2[\Zb_p^m]$.
Using the convolution property of the DFT we observe that $a \in \langle E_p(r,m) \rangle$ if and only if $A_{\p{j}}=0$ for all $\p{j}$ with $\wt(\p{j}) \geq r+1$.
That is, $\langle E_p(r,m) \rangle = \Cs_{\Zb_p}(r,m)$.

Blackmore and Norton show that the family of codes \mbox{$\langle E_p(r,m) \rangle =\Cs_{\Zb_p}(r,m)$} are the duals of the codes designed by Berman.
The code $\Cs_{\Zb_p}(r,m)$ is denoted as $\mathcal{B}_p(r,m)^\perp$ in~\cite{BlN_IT_01}.

\section{Generator Matrices of Abelian Codes from $\Fb_2[\Zb_p^m]$, with $2$ a primitive root modulo $p$} \label{app:gen_matrix_prime_group}

We utilize the fact that every abelian code in $\Fb_2[\Zb_p^m]$ is a direct sum of its irreducible subcodes~\cite{Wil_BellSys_70,Ber_Cybernetics_II_67}.
A generator matrix of an abelian code can therefore be obtained by juxtaposing the generator matrices of its irreducible subcodes.
Hence, we will now consider only the irreducible codes of $\Fb_2[\Zb_p^m]$.

The inverse DFT for $\Fb_2[\Zb_p^m]$ is given by 
\begin{equation*}
a_{\p{i}} = \sum_{\p{j} \in \Zb_p^m} \alpha^{-\p{i}\cdot \p{j}} A_{\p{j}}, 
\end{equation*}
where $\alpha$ is a primitive $p^\tth$ root of unity. Note that $\alpha,\alpha^2,\dots,\alpha^{p-1}$ are the roots of $x^{p-1}+x^{p-2}+\cdots+1$. Since $2$ is primitive in $\Zb_p$, $x^{p-1}+x^{p-2}+\cdots+1$ is irreducible. The field extension $\Fb_q = \Fb_2[\alpha]$ is of degree $(p-1)$ over $\Fb_2$, and $\{\alpha,\alpha^2,\dots,\alpha^{p-1}\}$ is an $\Fb_2$-basis of $\Fb_q$.

Let $\Gamma \subset \Zb_p^m$ be any conjugacy class and let $\p{j} \in \Gamma$. If $\p{j}=\p{0}$ then $\Gamma=\{\p{0}\}$ and the irreducible code corresponding to $\Gamma$ is the repetition code.
If $\p{j} \neq \p{0}$, then
\begin{equation*}
\Gamma = \{\p{j},2\p{j},2^2\p{j},\dots\} = \{\beta\p{j}~:~\beta \in \{1,2,\dots,p-1\}\},
\end{equation*}
where we have used the fact that $2$ is a primitive root in the finite field $\Zb_p$. Thus, $|\Gamma|=p-1$ and the irreducible code corresponding to $\Gamma$ has dimension $p-1$.
This code consists of all $a \in \Fb_2[\Zb_p^m]$ such that
\begin{equation*}
a_{\p{i}} = \sum_{t=0}^{p-2} \alpha^{-\p{i} \cdot 2^t \p{j}} A_{2^t\p{j}},
\end{equation*}
where $A_{\p{j}} \in \Fb_q$ and $A_{2^t\p{j}} = A_{\p{j}}^{2^t}$. Hence,
\begin{align*}
a_{\p{i}} = \sum_{t=0}^{p-2} \left( \alpha^{-\p{i}\cdot \p{j}} A_{\p{j}} \right)^{2^t}
= \Tr_{\Fb_q/\Fb_2}\left( \alpha^{-\p{i}\cdot \p{j}} A_{\p{j}} \right),
\end{align*}
where $\Tr_{\Fb_q/\Fb_2}$ is the trace function.
% Since $\alpha,\dots,\alpha^{p-1}$ is a $\Fb_2$-basis for $\Fb_q$,
We represent $A_{\p{j}}$ in the $\Fb_2$-basis $\{\alpha,\alpha^2,\dots,\alpha^{p-1}\}$ as $A_{\p{j}} = \sum_{k=1}^{p-1} m_k \alpha^k$, where $m_1,\dots,m_k \in \Fb_2$. Since the trace function is $\Fb_2$-linear we have
\begin{align*}
a_{\p{i}} = \sum_{k=1}^{p-1} m_k \, \Tr_{\Fb_q/\Fb_2}\left( \alpha^{k-\p{i}\cdot \p{j}} \right).
\end{align*}
Let us denote $k - \p{i} \cdot \p{j} \in \Zb_p$ by $l$.
If $l=0$, then $\Tr_{\Fb_q/\Fb_2}(\alpha^l) = 0$.
For $l \neq 0$, we have
\begin{align*}
\Tr_{\Fb_q/\Fb_2}(\alpha^l) &= \alpha^l + \alpha^{2l} + \alpha^{2^2 l} + \cdots + \alpha^{2^{p-2} l} \\
&= \alpha^l + \alpha^{2 l} + \alpha^{3 l} + \cdots + \alpha^{(p-1)l} \\
&= 1,
\end{align*}
since $\alpha^l$ is a root of $x^{p-1} + x^{p-2} + \cdots + x + 1$.
Hence,
\begin{equation*}
a_{\p{i}} = \sum_{k=1}^{p-1}m_k \, \mathbbm{1}(\p{i} \cdot \p{j} \neq k),
\end{equation*}
where $\mathbbm{1}$ is the indicator function.

We thus arrive at the following generator matrix $\p{G}$ for the irreducible code corresponding to the conjugacy class $\Gamma=\{\p{j},2\p{j},\dots\}$. Let $\p{G}$ be a $(p-1) \times p^m$ binary matrix whose rows are indexed by $k \in \{1,\dots,p-1\}$ and columns by $\p{i} \in \Zb_p^m$. The entry of $\p{G}$ in row $k$ and column $\p{i}$ is $\mathbbm{1}(\p{i} \cdot \p{j} \neq k)$.

% % % % % % % % % % % % % 
\section*{Acknowledgment}

The authors thank the two anonymous reviewers (from the submission of this work to the \emph{IEEE Transactions on Information Theory}) for their constructive comments and suggestions. 
% and Reviewer~1 for letting the authors know about the work~\cite{HeR_JAlgebra_13}.

% % bibliography % % %
% \bibliographystyle{IEEEtran}
% \bibliography{IEEEabrv,Berman_T-IT}


\end{document}